\documentclass[aps, prd, floatfix, twocolumn, superscriptaddress, nofootinbib]{revtex4-1}

\usepackage{latexsym}
\usepackage{amsmath}
\usepackage{amssymb}
\usepackage{amsfonts}

\usepackage{dsfont}
\usepackage{slashed}
\usepackage{bm}
\usepackage{urwchancal}

\usepackage[vcentermath]{youngtab}

\usepackage{color}
\definecolor{purple}{rgb}{0.5,0,0.5}
\definecolor{blue}{rgb}{0.0,0,0.9}
\definecolor{prdblue}{rgb}{0.133,0.118,0.498}

\usepackage{enumitem}
\usepackage{multirow}
\usepackage{subfigure}
\usepackage{textcomp}

\usepackage{supertabular} 
\usepackage{placeins}
\usepackage{epsfig}
\usepackage{graphicx}

\def\tstrut{\vrule height3.25ex depth0pt width0pt} 

\begin{document}


\title{Diffusion Monte Carlo calculations of fully-heavy (multiquark) bound states}

\author{M.C. Gordillo}
\email[]{cgorbar@upo.es}
\affiliation{Departamento de Sistemas F\'isicos, Qu\'imicos y Naturales, Universidad Pablo de Olavide, E-41013 Sevilla, Spain}

\author{F. De Soto}
\email[]{fcsotbor@upo.es}
\affiliation{Departamento de Sistemas F\'isicos, Qu\'imicos y Naturales, Universidad Pablo de Olavide, E-41013 Sevilla, Spain}

\author{J. Segovia}
\email[]{jsegovia@upo.es}
\affiliation{Departamento de Sistemas F\'isicos, Qu\'imicos y Naturales, Universidad Pablo de Olavide, E-41013 Sevilla, Spain}

\date{\today}

\begin{abstract}
We use a diffusion Monte Carlo method to solve the many-body Schr\"odinger equation describing fully-heavy tetraquark systems.
This approach allows to reduce the uncertainty of the numerical calculation at the percent level, accounts for multi-particle correlations in the physical observables, and avoids the usual quark-clustering assumed in other theoretical techniques applied to the same problem.
The interaction between particles was modeled by the most general and accepted potential, \emph{i.e.} a pairwise interaction including Coulomb, linear-confining and hyperfine spin-spin terms. This means that, in principle, our analysis should provide some rigorous statements about the mass location of the all-heavy tetraquark ground states, which is particularly timely due to the very recent observation made by the LHCb collaboration of some enhancements in the invariant mass spectra of $J/\psi$-pairs.
Our main results are:
(i) the $cc\bar c\bar c$, $cc\bar b\bar b$ ($bb\bar c\bar c$) and $bb\bar b \bar b$ lowest-lying states are located well above their corresponding meson-meson thresholds; 
(ii) the $J^{PC}=0^{++}$ $cc\bar c\bar c$ ground state with preferred quark-antiquark pair configurations is compatible with the enhancement(s) observed by the LHCb collaboration;
(iii) our results for the $cc\bar c\bar b$ and $bb\bar c\bar b$ sectors seem to indicate that the $0^+$ and $1^+$ ground states are almost degenerate with the $2^+$ located around $100\,\text{MeV}$ above them;
(iv) smaller mass splittings for the $cb\bar c\bar b$ system are predicted, with absolute mass values in reasonable agreement with other theoretical works;
(v) the $1^{++}$ $cb\bar c\bar b$ tetraquark ground state lies at its lowest $S$-wave meson-meson threshold and it is compatible with a molecular configuration.
\end{abstract}



\maketitle


\section{Introduction}
\label{sec:Introduction}

The $J/\psi$ signal was observed simultaneously at Brookhaven~\cite{Aubert:1974js} and SLAC~\cite{Augustin:1974xw} in 1974; it was a heavy resonance with a surprisingly small decay width. Three years later, an even heavier resonance but equally narrow, the so-called $\Upsilon$ state, was observed at Fermilab~\cite{Herb:1977ek, Innes:1977ae}. The interpretation of the $J/\psi$ and $\Upsilon$ as low-lying bound states of a heavy quark, $Q$, and its antiquark, $\bar{Q}$, with $Q$ either $c$- or $b$-quark, explained their narrow decay widths and, in fact, was proved to be crucial to establish Quantum Chromodynamics (QCD) as the strong-interaction sector of the Standard Model of Particle Physics~\cite{Appelquist:1974zd, DeRujula:1974rkb}. It is also worth mentioning that a related system, the $c\bar b$ bound state $(B_c^+)$, has also been found in nature~\cite{Abe:1998wi}; and that the heaviest of the quarks, \emph{i.e.} the top-quark, was discovered in $1995$ at Fermilab~\cite{Abe:1994xt}, with a mass around $175\,\text{GeV}$ and a large decay width that, due to weak interactions, forbids to form narrow $t\bar{t}$ resonances. 

Heavy quarkonia, \emph{viz.} mesons containing only a heavy valence quark-antiquark pair, opened the possibility to use a non-relativistic (NR) picture of QCD. They can indeed be classified in terms of the quantum numbers of a NR bound state, and the spacings between radial, orbital and spin excitations have a pattern similar to the ones observed in positronium, an $e^+e^-$ NR bound state well studied in Quantum Electrodynamics (QED)~\cite{Ore:1949te, Deutsch:1951zza}. Being baryonic analogues of heavy quarkonia, triply-heavy baryons may provide a complementary window in the understanding of the non-relativistic regime of QCD and the strong interaction between heavy quarks, without taking into account the usual light-quark complications. One can imagine that we could continue with the game of incorporating valence heavy quarks and/or antiquarks, and thus form tetraquark, pentaquark, hexaquark, etc., bound-state systems which would help to understand, at least, the generalization of the (heavy) quark-quark strong interaction to multi-body states. Note here that composite hadrons with four and more, light or heavy, quarks were already conjectured in 1964 by Gell-Mann~\cite{GellMann:1964nj} and Zweig~\cite{zweigcern} as a side product of explaining the observed spectrum of mesons and baryons from the quark model picture.

It is a fact that many precise experimental results are available for conventional heavy quarkonia~\cite{Tanabashi:2018oca}. In addition, tens of charmonium- and bottomonium-like states, the so-called XYZ states, which cannot fit the quark model picture, have been identified along the last two decades at B-factories (BaBar, Belle, and CLEO), $\tau$-charm facilities (CLEO-c and BES) and hadron-hadron colliders (CDF, D0, LHCb, ATLAS, and CMS). So far, there is no consensus about the nature of these exotic states (see Refs.~\cite{Brambilla:2010cs, Olsen:2014qna, Olsen:2017bmm, Brambilla:2019esw} for reviews of the experimental and theoretical status of the subject). Their analysis and new determinations will continue with the upgrade of experiments such as BES III~\cite{Asner:2008nq}, Belle II~\cite{Bevan:2014iga}, and HL- and HE-LHC~\cite{Cerri:2018ypt}. This will provide a sustained progress in the field as well as the breadth and depth necessary for a vibrant heavy quark research environment. 

The ultimate aim of theory is to describe the properties of the XYZ states from QCD's first principles. However, since this task is quite challenging, a more modest goal is to start with the development of QCD motivated phenomenological models that specify the colored constituents, how they are clustered and the forces between them. In that line, simultaneously to the experimental measurements, theorists have been proposing for the XYZ states different kinds of color-singlet clusters, made by quarks and gluons, which go beyond conventional mesons (quark-antiquark), baryons (three-quarks) and antibaryons (three-antiquarks); the most famous are glueballs, quark-gluon hybrids and multiquark systems (for a graphic picture of these kinds of hadrons see, for example, Figs.~1, 6, and 7 of Ref.~\cite{Olsen:2017bmm}). Related to the last ones, the first and best known exotic state is the $X(3872)$, which was observed in 2003 as an extremely narrow peak in the $B^+ \to K^+ (\pi^+ \pi^- J/\psi)$ channel and at exactly the $\bar D^{0} D^{\ast0}$ threshold~\cite{Choi:2003ue, Aubert:2004ns}. It is suspected to be a $c n \bar c \bar n$ ($n=u$ or $d$ quark) tetraquark state whose features resemble those of a molecule, but some experimental findings seem to point out the existence in its wavefunction of more compact components such as diquark-antidiquark and quark-antiquark~\cite{Acosta:2003zx, Abazov:2004kp, Aaij:2011sn, Chatrchyan:2013cld, CMS:2019vma, Aaij:2020qga}. 

Finally, fully-heavy tetraquarks have recently received considerable attention, both experimentally and theoretically. On the experimental side, it is thought that all-heavy tetraquark states will be very easy to spot because their masses should be far away from the typical mass regions populated by both conventional heavy mesons and the XYZ states discovered until now. A search for deeply bound $bb\bar b \bar b$ tetraquark states at the LHC was motivated by Eichten \emph{et al.} in Ref.~\cite{Eichten:2017ual}, and it was carried out by the LHCb collaboration~\cite{Aaij:2018zrb} determining that no significant excess is found in the $\mu^+\mu^-\Upsilon(1S)$ invariant-mass distribution. On the other hand, the LHCb collaboration has recently released in Ref.~\cite{1804391} a study of the $J/\psi$-pair invariant mass spectrum finding a narrow peak and a broad structure which could originate from hadron states consisting of four charm quarks. 

From the theoretical side, concerning the interaction between heavy quarks, chiral symmetry is explicitly broken and thus meson-exchange forces cannot exist in a fully heavy tetraquark system~\cite{Chen:2016jxd, Liu:2019zuc}, which would favor the formation of genuine tetraquark configurations rather than loosely bound hadronic molecules. This is interesting by itself, but also it simplifies the kind of quark--(anti\nobreakdash-)quark interactions to be taken into account, justifying the proliferation of theoretical works.

In the literature, we find fully-heavy tetraquark computations based on phenomenological mass formulae~\cite{Karliner:2016zzc, Berezhnoy:2011xn, Wu:2016vtq}, QCD sum rules~\cite{Chen:2016jxd, Wang:2017jtz, Wang:2018poa, Reinders:1984sr}, QCD motivated bag models~\cite{Heller:1985cb}, NR effective field theories~\cite{Anwar:2017toa, Esposito:2018cwh}, potential models~\cite{Ader:1981db, Zouzou:1986qh, Lloyd:2003yc, Barnea:2006sd, Richard:2018yrm, Richard:2017vry, Vijande:2009kj, Debastiani:2017msn, Liu:2019zuc, Chen:2019dvd, Chen:2019vrj, Chen:2020lgj, Wang:2019rdo, Yang:2020rih}, non-perturbative functional methods~\cite{Bedolla:2019zwg}, and even some exploratory lattice-QCD calculations~\cite{Hughes:2017xie}. Some works predict the existence of stable $QQ\bar Q\bar Q$ ($Q=c$ or $b$) bound states with masses slightly lower than the respective thresholds of quarkonium pairs (see, for instance, Refs.~\cite{Chen:2016jxd, Anwar:2017toa, Karliner:2016zzc, Berezhnoy:2011xn, Wang:2017jtz, Wang:2018poa, Debastiani:2017msn, Esposito:2018cwh}. In contrast, there are other studies that predict no stable $cc\bar c\bar c$ and $bb\bar b\bar b$ tetraquark bound states because their masses are larger than two-quarkonium thresholds (see, \emph{e.g.}, Refs.~\cite{Ader:1981db, Lloyd:2003yc, Richard:2018yrm, Wu:2016vtq, Hughes:2017xie}). To some extent, a better understanding of the  mass locations of fully-heavy tetraquark states would be desirable, if not crucial, for our comprehension of their underlying dynamics and their experimental hunting.

The goal of the present study is to achieve the most \emph{general} and \emph{accurate} prediction for the ground states of fully-heavy tetraquarks. In order to comply with the first feature, we are not going to assume any particular clustering between the valence quarks (antiquarks), the interaction between them is the most simple and accepted one: Coulomb$\,+\,$linear-confining$\,+\,$hyperfine spin-spin, and it will be implemented non-perturbativelly.\footnote{It is fair to notice that a similar calculation to the one presented herein has been recently released in Ref.~\cite{Bai:2016int}; however, important differences must be mentioned: (i) meson-meson and diquark-antidiquark clusters were assumed, (ii) the sextet-antisextet diquark-antidiquark configuration was fully neglected, and (iii) the hyperfine spin-spin interaction was computed perturbatively.} The second feature, accurateness, is fulfilled using a diffusion Monte Carlo (DMC) technique for solving the many-body Schr\"odinger equation wich, in contrast with variational methods, allows to reduce the uncertainty of the numerical calculation at the percent level, since the systematic one associated with the trial wave function is eliminated by the algorithm. Note, too, that it is possible to disentangle the uncertainty inherent of many-body techniques from the theoretical one coming from the model.

The manuscript is arranged as follows. In Sec.~\ref{sec:Theory} the theoretical framework is presented; we explain first the origin, features and implementation of the computational algorithm and, later, the quark model Hamiltonian and how their parameters are fixed. Section~\ref{sec:Results} is mostly devoted to the analysis and discussion of our theoretical results on fully-heavy tetraquarks; note here that we firstly study all-heavy mesons and baryons, comparing our results with those available from variational methods in order to confirm the validation of our approach. Finally, we summarize and give some prospects in Sec.~\ref{sec:Epilogue}.


\section{Theoretical framework}
\label{sec:Theory}


\subsection{Quark model}
\label{subsec:TeoMet}

The Hamiltonian which describes fully-heavy bound-state systems can be written as
\begin{equation}
H = \sum_{i=1}^{\text{n-part.}}\left( m_i+\frac{\vec{p}^{\,2}_i}{2m_i}\right) - T_{\text{CM}} + \sum_{j>i=1}^{\text{n-part.}} V(\vec{r}_{ij}) \,,
\label{eq:Hamiltonian}
\end{equation}
where $m_{i}$ is the quark mass, $\vec{p}_i$ is the momentum of the quark, and $T_{\text{CM}}$ is the center-of-mass kinetic energy. Since chiral symmetry is explicitly broken in the heavy quark sector, the two-body potential, $V(\vec{r}_{ij})$, can be deduced from the one-gluon exchange and confining interactions; \emph{i.e.}
\begin{equation}
V(\vec{r}_{ij}) = V_{\text{OGE}}(\vec{r}_{ij}) + V_{\text{CON}}(\vec{r}_{ij}) \,.
\end{equation}

It is important to highlight that the potentials above have tensor and spin-orbit contributions, which shall be neglected in this work. This is because our main purpose here is to get a first, unified and non-perturbative reliable description of fully-heavy hadron ground states (from 2- to 4-quark systems) within the DMC method. Moreover, this kind of interactions appeared not to be essential for a global description of baryons~\cite{SilvestreBrac:1996bg}, and beyond~\cite{Wang:2019rdo}.

The Coulomb and hyperfine terms are collected in the one-gluon exchange potential, and are given by
\begin{equation}
V_{\text{OGE}}(\vec{r}_{ij}) = \frac{1}{4} \alpha_{s} (\vec{\lambda}_{i}\cdot
\vec{\lambda}_{j}) \Bigg[\frac{1}{r_{ij}} - \frac{2\pi}{3m_{i}m_{j}} \delta^{(3)}(\vec{r}_{ij}) (\vec{\sigma}_{i}\cdot\vec{\sigma}_{j}) \Bigg] \,,
\end{equation}
where $\alpha_s$ is the strong coupling constant fixed in a phenomenological way, $\vec{\lambda}$ are the $SU(3)$-color Gell-Mann matrices, and the Pauli spin matrices are denoted by $\vec{\sigma}$. The Dirac delta function of the hyperfine term comes from the Fermi-Breit approximation of the one-gluon exchange interaction. In order to perform non-perturbative calculations, the $\delta^{(3)}(\vec{r}_{ij})$ is usually replaced by a smeared function that, in our case, reads as follows
\begin{equation}
\delta^{(3)}(\vec{r}_{ij}) \to \kappa \, \frac{e^{-r_{ij}^2/r_0^2}}{\pi^{3/2}r_{0}^3} \,,
\end{equation}
with $\kappa$ a quark model parameter, and $r_0=A\left(\frac{2m_im_j}{m_i+m_j}\right)^B$ a regulator which depends on the reduced mass of the quark--(anti-)quark pair.

Confinement is one of the crucial aspects of the strong interaction that is widely accepted and incorporated into any QCD based model. Studies of QCD on a lattice have demonstrated that multi-gluon exchanges produce an attractive linearly rising potential, which is proportional to the distance between infinitely heavy quarks~\cite{Bali:2005fu}. This phenomenological observation is usually modeled as 
\begin{equation}
V_{\text{CON}}(\vec{r}_{ij}) = (b\, r_{ij} + \Delta) (\vec{\lambda}_{i}\cdot
\vec{\lambda}_{j}) \,,
\end{equation}
where $b$ is the confinement strength and $\Delta$ is a global constant fixing the origin of energies.

\begin{table}[!t]
\caption{\label{tab:parameters} Quark model parameters used herein and taken from AL1 potential in Refs.~\cite{Semay:1994ht, SilvestreBrac:1996bg}.}
\begin{ruledtabular}
\begin{tabular}{llc}
Quark masses & $m_c$ (GeV) & 1.836 \\
             & $m_b$ (GeV) & 5.227 \\[2ex]
OGE          & $\alpha_s$        & 0.3802 \\
             & $\kappa$          & 3.6711 \\
             & $A$ (GeV)$^{B-1}$ & 1.6553 \\
             & $B$               & 0.2204 \\[2ex]
CON          & $b$ (GeV$^2$)  &  0.1653 \\
             & $\Delta$ (GeV) & -0.8321 \\
\end{tabular}
\end{ruledtabular}
\end{table}

Table~\ref{tab:parameters} shows the quark model parameters relevant for this work. Note here that we are using the so-called AL1 potential proposed by Silvestre-Brac and Semay in Ref.~\cite{Semay:1994ht}, and applied extensively to the baryon sector in Ref.~\cite{SilvestreBrac:1996bg}. It is worth emphasizing that the potential collects nicely the most important phenomenological features of QCD for heavy quarks, and that the parameters were constrained by a simultaneous fit of 36 mesons and 53 baryons with a remarkable agreement with data.


\subsection{Computational algorithm}
\label{subsec:NumMet}

Quantum Monte Carlo (QMC) methods have been successfully applied to many research areas but quantum chemistry and material science are the ones which have received more attention~\cite{Hammond:1994bk, Foulkes:2001zz, Nightingale:2014bk}. This is because QMC is a natural competitor of other methods where the uncorrelated or Hartree-Fock state does not provide a good description of the many-body ground state. Other applications of QMC algorithms are solid-state physics concerning the dynamics of condensed helium systems~\cite{Schmidt1992, Ceperley:1995zz}, and studies on the properties of both bosonic and fermionic ultracold quantum gases~\cite{Carlson:2003zz, Giorgini:2008zz, PhysRevA.89.023633, Carbonell_Coronado_2016}.

Since nuclear Hamiltonians induce strong correlations, QMC methods have appeared to be very valuable in the understanding of nuclei and nucleonic matter. Variational Monte Carlo (VMC) algorithms dealing with nuclear interactions were introduced in the early 1980s~\cite{Lomnitz-Adler:1981dmh}. Afterwards, methods based on Green Function Monte Carlo (GFMC) burst into nuclear physics in the late 1980s~\cite{Carlson:1987zz, Carlson:1988zz}, and were applied mostly to spin-isospin-dependent Hamiltonians. The GFMC technique is very accurate but becomes increasingly more complex when moving toward larger systems, being $^{12}$C the state-of-the-art studied one~\cite{Lovato:2013cua, Lovato:2014eva, Lovato:2015qka}. Diffusion Monte Carlo methods~\cite{Schmidt:1999lik} appear to be much more efficient at treating large systems; however, there are unsolved issues when dealing with spin-isospin dependent potentials.

The application of QMC methods to hadron physics has been scarce, basically because most known hadrons consist on 2- and 3-body relativistic bound states. However, the description from different perspectives of composite states with four and more, light or heavy, quarks (antiquarks) has recently received considerable attention since the discovery of the potentially non-relativistic, 4-quark, charmonium-like system $X(3872)$. It is in this context that QMC algorithms can contribute to shed some light to the study of tetraquark, pentaquark, hexaquark, etc., non-relativistic systems. 

Up to our knowledge, the first QMC study of mesons and baryons was performed by Carlson \emph{et al.} in Refs.~\cite{Carlson:1982xi, Carlson:1983rw}. The authors used a VMC method developed previously for nuclear physics problems and their results compared reasonably well with those of the well-known Isgur-Karl's quark model~\cite{Isgur:1978xj, Isgur:1978wd, Isgur:1979be, Capstick:1986bm}. The Diffusion Monte Carlo method has been applied recently to the fully-beautiful tetraquark system in Ref.~\cite{Bai:2016int}; in particular, the authors calculate the ground state energy of the $J^{PC}=0^{++}$ $bb\bar b\bar b$ system.

The central idea behind the DMC method is to write the Schr\"odinger equation for $n$-particles in imaginary time ($\hbar=c=1$):
\begin{equation}
-\frac{\partial \Psi_{\alpha'}(\bm{R},t)}{\partial t} = (H_{\alpha'\alpha}-E_s) \Psi_{\alpha}(\bm{R},t) \,,
\label{eq:Sch1}
\end{equation}
where $E_s$ is the usual energy shift used in DMC methods, $\bm{R}\equiv(\vec{r}_1,\ldots,\vec{r}_n)$ stands for the position of $n$ particles and $\alpha$ denotes each possible spin-color channel, with given quantum numbers, for the $n$-particles system. The function $\Psi_{\alpha}(\bm{R},t)$ can be expanded in terms of a complete set of the Hamiltonian's eigenfunctions as
\begin{equation}
\Psi_{\alpha}(\bm{R},t) = \sum_i c_{i,\alpha} \, e^{-(E_i-E_s)t} \, \phi_{i,\alpha}(\bm{R}) \,,
\end{equation}
where the $E_i$ are the eigenvalues of the system's Hamiltonian operator, $\hat{H}$. The ground state wave function, $\phi_{0,\alpha}(\bm{R})$, is obtained as the asymptotic solution of Eq.~\eqref{eq:Sch1} when $t\to \infty$, as long as there is overlap between $\Psi_{\alpha}(\bm{R},t=0)$ and $\phi_{0,\alpha}(\bm{R})$, for any $\alpha$-channel.

A crucial feature of QMC methods is the use of importance sampling techniques~\cite{PhysRevA.9.2178}, in order to reduce the statistical fluctuations to a manageable level. Given a Hamiltonian, $H=H_0+V$, of the form
\begin{equation}
H_{\alpha'\alpha} = -\frac{\nabla_{\bm{R}}^2}{2m} \, \delta_{\alpha'\alpha} + V_{\alpha'\alpha}(\bm{R}) \,,
\end{equation}
with $m$ the mass of each particle which composes the bound-state system, if one works with the function
\begin{equation}
f_{\alpha}(\bm{R},t) \equiv \psi(\bm{R}) \, \Psi_{\alpha}(\bm{R},t) \,,
\end{equation}
where $\psi(\bm{R})$ is the time-independent trial function:
\begin{equation}
\psi(\bm{R}) = \prod_{i<j} \phi(\vec{r}_{ij}) \,,
\end{equation}
where $i$, $j$ are indices which run over the number of particles. Within this work, we choose $\phi(\vec{r}_{ij})$ as the wave function solution of the 2-body Hamiltonian of the system at short interquark distances: $\phi(\vec{r}_{ij})=e^{-a_{ij}r_{ij}}$, where $a_{ij}$ are determined by the so-called cusp conditions.

Equation~\eqref{eq:Sch1} turns out to be
\begin{align}
&
-\frac{\partial f_{\alpha'}(\bm{R},t)}{\partial t} = - \frac{1}{2m} \nabla_{\bm{R}}^2 f_{\alpha'}(\bm{R},t) \nonumber \\
&
\hspace*{-0.00cm} + \frac{1}{2m} \nabla_{\bm{R}} \big[ F(\bm{R})f_{\alpha'}(\bm{R},t) \big] \nonumber \\
&
\hspace*{-0.00cm} + \big[ E_{L}(\bm{R})-E_s \big] f_{\alpha'}(\bm{R},t) \nonumber \\
&
\hspace*{-0.00cm} + V_{\alpha'\alpha}(\bm{R}) f_{\alpha}(\bm{R},t) \nonumber \\
&
\hspace*{-0.00cm} \equiv \Big[ A^{(1)}+A^{(2)}+A^{(3)} \Big] f_{\alpha'}(\bm{R},t) + V_{\alpha'\alpha}(\bm{R}) f_{\alpha}(\bm{R},t) \nonumber \\
&
\hspace*{-0.00cm} \equiv A f_{\alpha'}(\bm{R},t) + V_{\alpha'\alpha}(\bm{R}) f_{\alpha}(\bm{R},t) \,, 
\label{eq:Sch2}
\end{align}
where
\begin{align}
E_{L}(\bm{R}) &= \psi(\bm{R})^{-1} \, H_{0}\, \psi(\bm{R}) \,, \\
F(\bm{R}) &= 2\, \psi(\bm{R})^{-1}\, \nabla_{\bm{R}} \,\psi(\bm{R}) \,,
\end{align}
are the so-called local energy and drift force, respectively. The formal solution of Eq.~\eqref{eq:Sch2} is given by
\begin{equation}
f_{\alpha'}(\bm{R}',t+\Delta t) = \sum_{\alpha} \int\!\!d\bm{R}\,\, G_{\alpha'\alpha}(\bm{R}',\bm{R},\Delta t) f_{\alpha}(\bm{R},t) \,.
\label{eq:fsolver1}
\end{equation}
While algorithms based on GFMC implement the whole Green's function, DMC methods rely on reasonable approximations of $G_{\alpha'\alpha}(\bm{R}',\bm{R},\Delta t)$ for small values of the time step $\Delta t$ and iterates repeatedly to obtain the asymptotic solution $f_{\alpha}(\bm{R},t\to\infty)$. In our case, the Green's function is approximated by
\begin{equation}
G_{\alpha'\alpha}(\bm{R}',\bm{R},\Delta t) \approx \langle \bm{R}' | e^{-A \Delta t} | \bm{R} \rangle \, e^{- V_{\alpha'\alpha}(\bm{R}) \Delta t} \,,
\label{eq:Green}
\end{equation}
where for the first part we follow Ref.~\cite{PhysRevB.49.8920} and approximate it as
\begin{align}
\exp(-A \Delta t) &= \exp\left(-A^{(3)} \frac{\Delta t}{2}\right) \exp\left(-A^{(2)} \frac{\Delta t}{2}\right) \nonumber \\
&
\times \exp\left(-A^{(1)} \Delta t\right) \nonumber \\
&
\times \exp\left(-A^{(2)} \frac{\Delta t}{2}\right) \exp\left(-A^{(3)} \frac{\Delta t}{2}\right) \,,
\label{eq:fsolver2}
\end{align}
which is exact up to order $(\Delta t)^2$. With the expression above, Eq.~\eqref{eq:fsolver1} becomes
\begin{align}
f_{\alpha'}(\bm{R}',t+\Delta t) &= \int\!\!d\bm{R}\, \int d\bm{R}_1 \cdots d\bm{R}_4 \nonumber \\
&
\hspace*{-2.20cm} \times \Bigg[ G^{(3)}\Big( \bm{R}',\bm{R}_1,\frac{\Delta t}{2}\Big) \, G^{(2)}\Big(\bm{R}_1,\bm{R}_2,\frac{\Delta t}{2}\Big) \nonumber \\
&
\hspace*{-2.20cm} \times G^{(1)}\Big( \bm{R}_2,\bm{R}_3,\Delta t \Big) \nonumber \\
&
\hspace*{-2.20cm} \times G^{(2)}\Big( \bm{R}_3,\bm{R}_4,\frac{\Delta t}{2}\Big) \, G^{(3)}\Big(\bm{R}_4,\bm{R},\frac{\Delta t}{2}\Big) \Bigg] \nonumber \\
&
\hspace*{-2.20cm} \times \sum_{\alpha}\, e^{-V_{\alpha'\alpha}(\bm{R}) \Delta t} f_{\alpha}(\bm{R},t) \,,
\end{align}
with
\begin{widetext}
\begin{align}
&
G^{(1)}(\bm{R}',\bm{R},t) = \left(\frac{2\pi t}{m}\right)^{-\frac{3n}{2}} \, \exp\left(-\frac{m(\bm{R}'-\bm{R})^2}{2t}\right) \,, \\[1ex]
&
G^{(2)}(\bm{R}',\bm{R},t) = \delta(\bm{R}'-\bm{R}(t)) \,, \text{ where} \begin{cases} \bm{R}(0)=\bm{R} \,, \\[2ex] \frac{d\bm{R}(t)}{dt} = \frac{F(\bm{R}(t))}{2m} \,, \end{cases} \\[1ex]
&
G^{(3)}(\bm{R}',\bm{R},t) = \exp\big[-(E_{L}(\bm{R})-E_s)t \big] \, \delta(\bm{R}'-\bm{R}) \,.
\end{align}
\end{widetext}

Note herein that in the case of having more than one $\alpha$-channel, we follow the method proposed in Ref.~\cite{PhysRevA.98.053632} and propagate the quantity
\begin{equation}
{\cal F}(\bm{R},t) = \sum_\alpha f_{\alpha}(\bm{R},t) \,,
\end{equation}
such as
\begin{align}
{\cal F}(\bm{R}',t+\Delta t) &= \int dR \, \langle \bm{R}' | e^{-A \Delta t} | \bm{R} \rangle \, \sum_{\alpha'\alpha} e^{-V_{\alpha'\alpha}(\bm{R})\Delta t} \, f_\alpha(\bm{R},t) \nonumber \\ 
&
= \int dR \, \langle \bm{R}' | e^{-A \Delta t} | \bm{R} \rangle \, \omega(\bm{R},t) \, {\cal F}(\bm{R},t) \,,
\label{eq:TheF}
\end{align}
where we have introduced the weight factor
\begin{equation}
\omega(\bm{R},t) = \frac{\sum_{\alpha'\alpha}e^{-V_{\alpha'\alpha}(\bm{R})\Delta t}f_{\alpha}(\bm{R},t)}{\sum_\alpha f_{\alpha}(\bm{R},t)} \,.
\label{eq:omega}
\end{equation}

With either one chanel or more, the effective way of applying the DMC method defined by Eq.~\eqref{eq:TheF} is the following. Each walker is characterized by the positions, $\bm{R}$, and the coefficients of each channel, $c_\alpha$; all of them fixed by an initial guess. Then, for each walker,
\begin{itemize}
\item[(i)] Move it, under the drift force $F(\bm{R})$, during an interval $\Delta t/2$ with accuracy $(\Delta t)^2$.
\item[(ii)] Apply a displacement $\chi$, randomly drawn from the $3n$ Gaussian distribution $\exp\left[ -m\chi^2/(2\Delta t)\right]$
\item[(iii)] Repeat step (i), with the new drift force.
\item[(iv)] Randomly replicate the walker considering the product of branching ratios
\begin{equation}
{\cal B}_1 = e^{-\left[ \frac{E_L(\bm{R}')+E_L(\bm{R})}{2}-E_s\right] \Delta t} \,,
\end{equation}
and
\begin{equation}
{\cal B}_{2} = \frac{\sum_\alpha c'_\alpha}{\sum_\alpha c_\alpha} \,,
\end{equation}
where the last one is derived from Eq.~\eqref{eq:omega}, and therefore the coefficients are updated according to
\begin{align}
c'_\alpha &= \sum_{\alpha'}e^{-V_{\alpha\alpha'}(\bm{R})\Delta t} \, c_{\alpha'} \nonumber \\
&
\approx \sum_{\alpha'} \Big[ \delta_{\alpha\alpha'}-V_{\alpha\alpha'}(\bm{R})\Delta t \Big] c_{\alpha'} \,.
\end{align}
\end{itemize}
The procedure above must be repeated for each walker until the set of them is exhausted. The resulting set of walkers corresponds to the new positions and coefficients $\{\bm{R}',c_\alpha'\}$. Finally, the whole procedure must be repeated as many times as needed to reach the asymptotic limit $t\to \infty$.


\section{Results}
\label{sec:Results}

A detailed discussion about the particular features of our spectrum will be given in the following subsections. However, a comment is due here on the theoretical uncertainty of our results. There are two kind of theoretical errors: one is inherently connected to the statistical nature of the DMC algorithm and the other one is related with a shortcoming of the quark model approach and lies on the way to fix the model parameters. The statistical error is of the order of $1\,\text{MeV}$ and negligible with respect the systematic one related with the quark model. As mentioned above, the set of model parameters are fitted to reproduce a certain number of hadron observables within a determinate range of agreement with experiment. Therefore, it is difficult to assign an error to those parameters and, as a consequence, to the magnitudes calculated when using them. As the range of agreement between theory and experiment is around $10-20\%$, this value can be taken as an estimation of the model uncertainty for fully-heavy tetraquark systems.


\subsection{Fully-heavy mesons}
\label{subsec:Mesons}

Four fundamental degrees of freedom at the quark level: space, spin, flavor and color are generally accepted in QCD, and any hadron's wave function must be expressed as a product of these four terms: 
\begin{equation}
|\psi_{\text{hadron}}\rangle = |\phi_r\rangle |\chi_s \rangle \, |\chi_f\rangle \, |\chi_c\rangle \,.
\end{equation}
From now on, we shall drop out the \emph{trivial} flavor wave function because only fully-heavy hadrons are considered in this work and thus, due to the explicit breaking of chiral symmetry, the considered quark model Hamiltonian is blind to the heavy quark flavor.

Concerning the color degree of freedom, any hadron state must be a color singlet one, because no color charges has been observed in nature. The leading Fock state of a fully-heavy meson is constituted by a quark, $Q$, and antiquark, $\bar{Q}$, with $Q$ either $c$ or $b$-quark. Therefore, the $SU(3)_{\text{color}}$ wave function is constructed as follows: 
\begin{equation}
\label{eqn:young-Meson}
\left. \begin{array}{ccccccc}
\yng(1) & \otimes & \yng(1,1) & = & \yng(1,1,1) & \oplus & \yng(2,1)  \\
&&&&&&\\
{\bf 3}_{\rm c} & \otimes & \bar{\bf 3}_{\rm c} & = & {\bf 1}_{\rm c}  & \oplus & {\bf 8}_{\rm c}
\end{array} \right. \,,  
\end{equation}
where the singlet state is the physically interesting one in this work, and it is given by the well know, symmetric expression,
\begin{equation}
|\chi_c\rangle_{\text{meson}} = \frac{1}{\sqrt{3}} \Big( |r\bar r\rangle + |g\bar g\rangle + |b\bar b\rangle \Big) \,.
\end{equation}

Since a meson is made by distinguishable particles (a quark and an antiquark), there is no restriction in the spin wave functions due to Pauli principle and thus all
\begin{align}
|\chi_{S=0,S_z=0} \rangle &= \frac{1}{\sqrt{2}} \Big( |\!\uparrow\downarrow\rangle - |\!\downarrow\uparrow\rangle \Big) \,, \\[1ex]
|\chi_{S=1,S_z=+1}\rangle &= |\!\uparrow\uparrow\rangle \,, \\
|\chi_{S=1,S_z=0}\rangle  &= \frac{1}{\sqrt{2}} \Big( |\!\uparrow\downarrow\rangle + |\!\downarrow\uparrow\rangle \Big) \,, \\
|\chi_{S=1,S_z=-1}\rangle &= |\!\downarrow\downarrow\rangle \,,
\end{align}
possibilities can be considered. The $S=0$ state is antisymmetric respect the particle exchange $1\leftrightarrow2$, whereas the $S=1$ wave functions are all symmetric. 

One can guess that an excitation of a unit of angular momentum costs an energy around $500\,\text{MeV}$. This effect can be estimated from the experimental $M(L=1)-M(L=0)$ mass differences: $f_1(1285)-\omega(782) \approx 499\,\text{MeV}$, $a_1(1260)-\rho(770) \approx 460\,\text{MeV}$, $\chi_{c1}(1P)-J/\psi \approx 414$, and $\chi_{b1}(1P)-\Upsilon(1S) \approx 432\,\text{MeV}$, but also in the baryon sector as, for instance, $N(1535)-N(940) \approx 570\,\text{MeV}$. Therefore, we shall limit ourselves to analyze states with orbital angular momentum equal to zero. An important consequence is that the space-wave function will always represent an $S$-wave state, which is completely symmetric.

\begin{table}[!t]
\caption{\label{tab:Mesons} The mass spectra of the heavy quarkonia in units of MeV. The \emph{1st, 2nd and 3rd columns} refer to the name and quantum numbers of the considered hadron, \emph{4th column} is our result, \emph{5th column} is the same calculation but using a variational method~\cite{Wang:2019rdo}, and \emph{6th column} collects experimental data if exist.}
\begin{ruledtabular}
\begin{tabular}{lccccc}
& $n\,^{2S+1}\!L_J$ & $J^{PC}$ & DMC & VAR~\cite{Wang:2019rdo} & EXP~\cite{Tanabashi:2018oca} \\
\hline
\tstrut
$\eta_c$ & $1\,^{1}\!S_0$ & $0^{-+}$ & $3005$ & $3006.6$ & $2983.9\pm0.5$ \\
$J/\psi$ & $1\,^{3}\!S_1$ & $1^{--}$ & $3101$ & $3102.1$ & $3096.900\pm0.006$ \\[1.5ex]
$B_c$      & $1\,^{1}\!S_0$ & $0^{-+}$ & $6292$ & $6293.5$ & $6274.9\pm0.8$ \\
$B_c^\ast$ & $1\,^{3}\!S_1$ & $1^{--}$ & $6343$ & - & - \\[1.5ex]
$\eta_b$       & $1\,^{1}\!S_0$ & $0^{-+}$ & $9424$ & $9427.9$ & $9398.7\pm2.0$ \\
$\Upsilon(1S)$ & $1\,^{3}\!S_1$ & $1^{--}$ & $9462$ & $9470.4$ & $9460.30\pm0.26$ \\
\end{tabular}
\end{ruledtabular}
\end{table}

Once the (trial) wave function of the meson is constructed, we follow the DMC algorithm explained in the section above and obtain, as a proof of concept, the masses of the singlet and triplet $S$-wave ground states of heavy quarkonia. Our results are shown in Table~\ref{tab:Mesons}, they compare fairly well with the experimental data despite the simplicity of the quark model; this also supports our confidence on the mass prediction for fully-heavy tetraquark bound states. Our numerical technique is contrasted with the same calculation but using a variational approach~\cite{Wang:2019rdo}. As one can see in Table~\ref{tab:Mesons}, there are negligible differences between the two methods; as expected, our values are always below the ones reported in Ref.~\cite{Wang:2019rdo}. This validates our technique against the variational one, which should give an energy's upper limit quite close to the real eigenvalue in the case of meson systems. As we shall see later, the differences between the two numerical approaches will be larger as the number of particles increases.

\begin{figure*}
\includegraphics[width=0.32\textwidth]{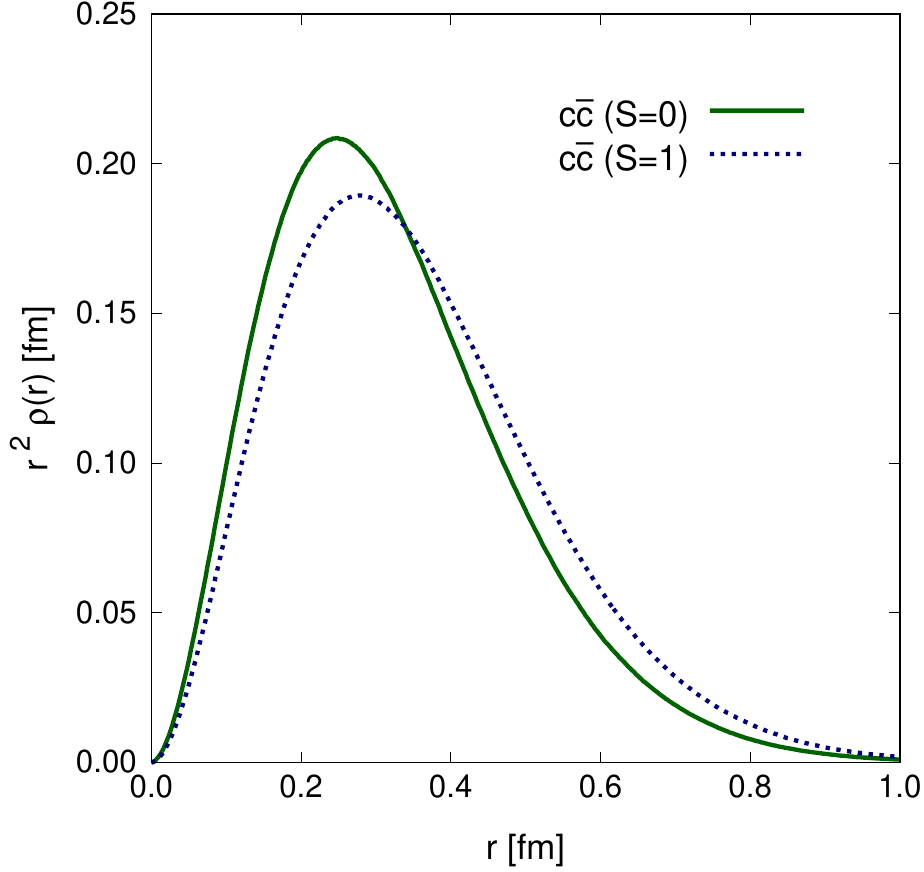}
\includegraphics[width=0.32\textwidth]{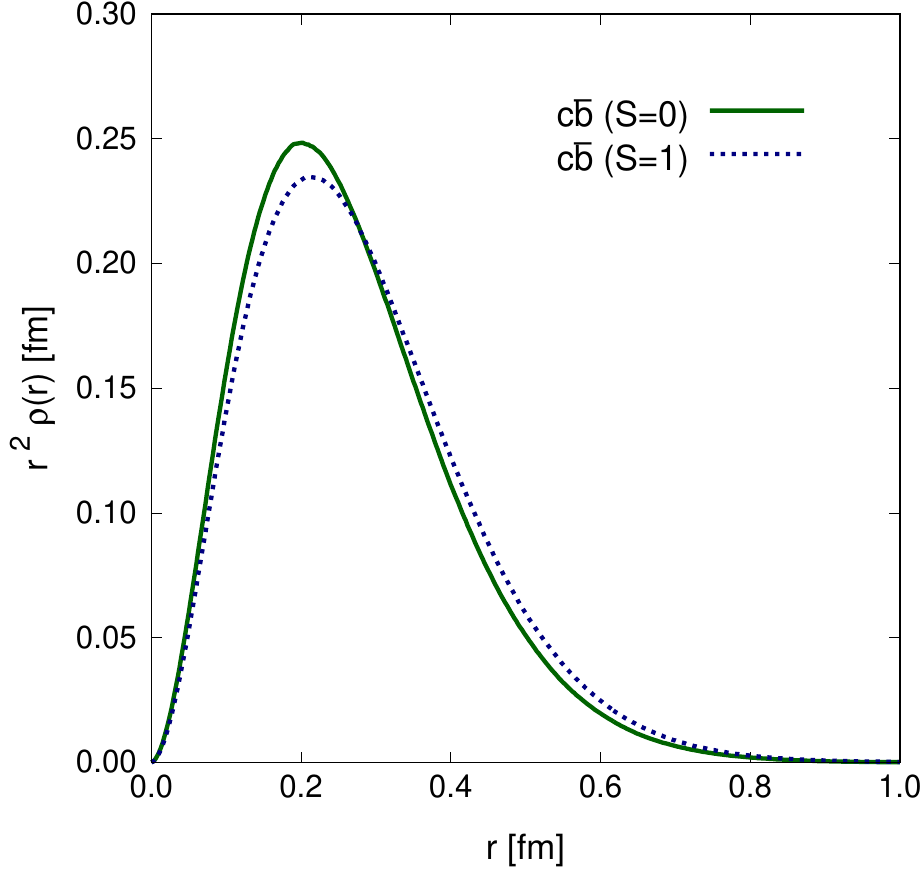}
\includegraphics[width=0.32\textwidth]{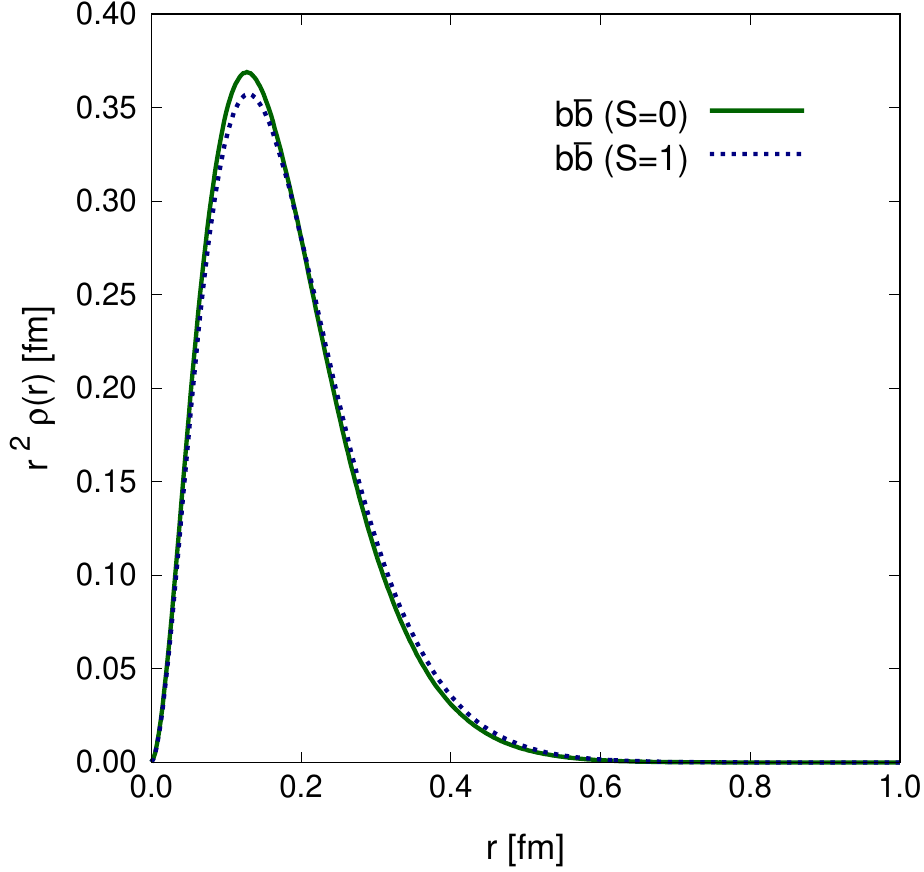}
\caption{\label{fig:Mesons} Radial distribution functions, $r^{2}\,\rho^{(2)}(r)$ with $r=|\vec{r}_2-\vec{r}_1|$ the relative coordinate between the 2 particles, for the studied charmonium (\emph{left panel}), $B_c$ (\emph{middle panel}), and bottomonium (\emph{right panel}) states.}
\end{figure*}

The concept of radial distribution function can be applied to multiquark systems and, in fact, can provide valuable information about the existence of interquark correlations; in particular, $2$-body correlations. If the $n$-particle wave function is defined as $\psi(\vec{r}_1,\ldots,\vec{r}_n)$, where spin, flavor and color degrees of freedom have been ignored for simplicity without lost of generalization, the probability of finding particle $1$ in position $\vec{r}_1$, particle $2$ in position $\vec{r}_2$, $\ldots$, particle $n$ in position $\vec{r}_n$ is:
\begin{equation}
P(\vec{r}_1,\ldots,\vec{r}_n) = \psi^{\ast}(\vec{r}_1,\ldots,\vec{r}_n) \psi(\vec{r}_1,\ldots,\vec{r}_n) \,,
\end{equation}
and it is normalize to one, \emph{i.e.}
\begin{equation}
1 = \int d\vec{r}_1\, \cdots d\vec{r}_n\, P(\vec{r}_1,\ldots,\vec{r}_n) \,.
\end{equation}
Therefore, one can define
\begin{equation}
\rho^{(2)}(\vec{r}_1,\vec{r}_2) = \int d\vec{r}_{3}\, \cdots d\vec{r}_{n}\,P(\vec{r}_1,\ldots,\vec{r}_n) \,,
\end{equation}
which expresses the probability of finding 2 particles in positions $\vec{r}_1$ and $\vec{r}_2$; and the radial distribution function as
\begin{equation}
\rho(r) = 4\pi r^2 \int d\vec{R}\, \rho^{(2)}(\vec{R}+\vec{r},\vec{R})
\end{equation}
where $r$ indicates now the distance between the two particles considered.

Figure~\ref{fig:Mesons} shows, for the studied mesons, the radial distribution functions which are pure estimators calculated within our DMC following Ref.~\cite{Casulleras_1995, Boronat:2002book}. Among the features one can observe, the following are of particular interest: (i) the $c\bar c$ states are the most extended objects and the system becomes more compact in going from the $c\bar c$ meson to the $b\bar b$ one, being the size of the $b\bar b$ system almost half of the $c\bar c$-meson's one; (ii) the $S=1$ state is slightly more extended than the $S=0$ state because the different sign of the spin-spin hyperfine interaction; and (iii) the structural differences between $S=0$ and $S=1$ states seems to blur as we go to heavier quarks, as expected since the hyperfine mass splitting gets smaller.

\begin{table}[!t]
\begin{center}
\caption{\label{tab:radiiMeson} Mean-square radii, $\langle r^2 \rangle$, of the studied quarkonium systems, in units of fm$^2$.}
\begin{tabular}{lc}
\hline
\hline
Meson & $\langle r^2 \rangle$ \\
\hline
\tstrut
$\eta_c$ & $0.131$ \\
$J/\psi$ & $0.158$ \\[1.5ex]
$B_c$      & $0.091$ \\
$B_c^\ast$ & $0.101$ \\[1.5ex]
$\eta_b$       & $0.041$ \\
$\Upsilon(1S)$ & $0.043$ \\
\hline
\hline
\end{tabular}
\end{center}
\end{table}

Finally, the typical mean-square radii, $\langle r^2 \rangle$, of the studied quarkonium systems are shown in Table~\ref{tab:radiiMeson}. They compare nicely with the results reported by a well-known non-relativistic QCD effective field theory for quarkonium~\cite{Pineda:2013lta, Peset:2015vvi}. Moreover, looking at the table, we can confirm that the interparticle distance of the $b\bar b$ system is almost half of the $c\bar c$ one; and the $B_c$'s one is closer to the bottomonium than to the charmonium.


\subsection{Fully-heavy baryons}
\label{subsec:Baryons}

We turn now our attention to the study of all-heavy ground state baryons and thus no orbital angular momentum excitations must be considered, \emph{i.e.} only $S$-wave symmetric states are under scrutiny.

Concerning the $SU(3)_{\text{color}}$ wave function, it is constructed as follows:
\begin{equation}
\label{eqn:young-Baryon}
\left. \begin{array}{ccccccccccccc}
\yng(1) & \otimes & \yng(1) & \otimes & \yng(1) & = & \yng(3) & \oplus & \yng(2,1) & \oplus & \yng(2,1) & \oplus & \yng(1,1,1) \\
&&&&&&\\
{\bf 3}_{\rm c} & \otimes & {\bf 3}_{\rm c} & \otimes & {\bf 3}_{\rm c} & = & {\bf 10}_{\rm c}  & \oplus & {\bf 8}_{\rm c} & \oplus & {\bf 8}_{\rm c} & \oplus & {\bf 1}_{\rm c}
\end{array} \right. \,,  
\end{equation}
where the colorless state is fully anti-symmetric and its color wave function is given by the textbook expression
\begin{align}
|\chi_c\rangle_{\text{baryon}} &= \frac{1}{\sqrt{6}} \Big( |rgb\rangle + |gbr\rangle + |brg\rangle \nonumber \\ 
& 
- |rbg\rangle - |grb\rangle - |bgr\rangle \Big) \,.
\end{align}

With three particles of spin $1/2$, one can construct the following spin states:
\begin{equation}
\label{eqn:young-Spin-Baryon}
\left. \begin{array}{ccccccccccc}
\yng(1) & \otimes & \yng(1) & \otimes & \yng(1) & = & \yng(3) & \oplus & \yng(2,1) & \oplus & \yng(2,1) \\
&&&&&&\\
{\bf 2}_{\rm s} & \otimes & {\bf 2}_{\rm s} & \otimes & {\bf 2}_{\rm s} & = & {\bf 3/2}_{\rm S}  & \oplus & {\bf 1/2}_{\rm MS} & \oplus & {\bf 1/2}_{\rm MA}
\end{array} \right. \,,  
\end{equation}
with
\begin{align}
|\chi_{S=3/2,S_z=+3/2}\rangle_{\text{S}} &= |\!\uparrow\uparrow\uparrow\rangle \,, \\
|\chi_{S=1/2,S_z=+1/2}\rangle_{\text{MS}} &= \frac{1}{\sqrt{6}} \Big( |\!\uparrow\downarrow\uparrow\rangle + |\!\downarrow\uparrow\uparrow\rangle \nonumber \\ & \hspace{1.00cm} - 2 |\!\uparrow\uparrow\downarrow\rangle \Big) \,, \\
|\chi_{S=1/2,S_z=+1/2}\rangle_{\text{MA}} &= \frac{1}{\sqrt{2}} \Big( |\!\uparrow\downarrow\uparrow\rangle - |\!\downarrow\uparrow\uparrow\rangle \Big) \,,
\end{align}
examples of the spin wave functions used herein, without lost of generality. For the baryons
\begin{equation}
\Omega_{ccc}^{++} \,, \hspace*{0.40cm}
\Omega_{ccb}^{+} \,, \hspace*{0.40cm}
\Omega_{cbb}^{0} \,, \hspace*{0.40cm}
\Omega_{bbb}^{-} \,, \hspace*{0.40cm}
\end{equation}
the spin wave function must comply Pauli statistics in the case that quarks are the same. Namely, the $S_{\text{sym.}} = 3/2_{\text S}$ wave function, which is completely symmetric, must be used for the $\Omega_{ccc}$ and $\Omega_{bbb}$ baryons; whereas the $S_{\text{sym.}} = 1/2_{\text{MS}}$ wave function, which is symmetric respect the two particles that are equal, must be used for the ground states of $\Omega_{ccb}$ and $\Omega_{cbb}$, because it is an allowed state with a lower energy than the $S=3/2$ case.

\begin{table}[!t]
\caption{\label{tab:Baryons} Masses, in MeV, of the ground states of fully-heavy baryons. The \emph{1st, 2nd and 3rd columns} refer to the name and quantum numbers of the considered hadron, \emph{4th column} is our result, and \emph{5th column} is the same calculation but using a variational method~\cite{SilvestreBrac:1996bg}. Note that, for comparison, the masses presented here include three-body-force corrections using the value of the constant $C$ reported in Table 1 of Ref.~\cite{SilvestreBrac:1996bg}.}
\begin{ruledtabular}
\begin{tabular}{cccrr}
Baryon & $^{2S+1}\!L_J$ & $J^P$ & DMC & Ref.~\cite{SilvestreBrac:1996bg} \\
\hline
\tstrut
$\Omega_{ccc}^{++}$ & $^{4}\!S_{3/2}$ & $3/2^+$ &  $4798$ &  $4799$ \\
$\Omega_{ccb}^{+}$  & $^{2}\!S_{1/2}$ & $1/2^+$ &  $8018$ &  $8019$ \\
$\Omega_{cbb}^{0}$  & $^{2}\!S_{1/2}$ & $1/2^+$ & $11215$ & $11217$ \\
$\Omega_{bbb}^{-}$  & $^{4}\!S_{3/2}$ & $3/2^+$ & $14398$ & $14398$ \\
\end{tabular}
\end{ruledtabular}
\end{table}

Table~\ref{tab:Baryons} shows the calculated masses for the ground states of the $\Omega_{QQQ}$ baryons ($Q=c$ or $b$) in each allowed spin channel, and compares them with those obtained in Ref.~\cite{SilvestreBrac:1996bg}. As one can see, there are negligible differences between the two numerical approaches. Unfortunately, there is no experimental data to compare with; therefore, we encourage the design of experimental set-ups at, for instance, the LHC@CERN facility able to detect this kind of particles because the reward could be high and, as mentioned above,  triply-heavy baryons are ideally suited to study QCD and, in particular, the heavy quark--(anti-)quark interaction, as it has been the case for heavy quarkonia.

\begin{table}[!t]
\caption{\label{tab:BarPro} The mass mean-square radii (in fm$^2$), charge mean-square radii (in e fm$^2$), and magnetic moments (in nuclear magnetons) for all the baryons listed in Table~\ref{tab:Baryons}.
}
\begin{ruledtabular}
\begin{tabular}{llcccc}
Observable & Approach & $\Omega_{ccc}^{++}$ & $\Omega_{ccb}^{+}$ & $\Omega_{cbb}^{0}$ & $\Omega_{bbb}^{-}$ \\
\hline
\tstrut
$\langle R_m^2 \rangle$ & DMC & $0.069$ & $0.040$ & $0.028$ & $0.021$ \\
& Ref.~\cite{SilvestreBrac:1996bg} & $0.069$ & $0.040$ & $0.028$ & $0.021$ \\[1.5ex]
$\langle R_c^2 \rangle$ & DMC & $0.138$ & $0.096$ & $0.034$ & $-0.021$\\
& Ref.~\cite{SilvestreBrac:1996bg} & $0.138$ & $0.097$ & $0.034$ & $-0.021$ \\[1.5ex]
$\langle \mu \rangle$ & DMC & $1.023$ & $0.475$ & $-0.193$ & $-0.180$ \\
& Ref.~\cite{SilvestreBrac:1996bg} & $1.023$ & $0.475$ & $-0.193$ & $-0.180$ \\[1.5ex]
\end{tabular}
\end{ruledtabular}
\end{table}

In order to check further the capabilities of the DMC method to describe heavy baryons, we calculate several structural (static) properties to compare them with the results obtained in Ref.~\cite{SilvestreBrac:1996bg}. In particular, the following ones:
\begin{align}
\langle R_m^2 \rangle &\equiv \langle \psi_{\text{hadron}} | \sum_{i=1}^{3} \frac{m_i}{M} (\vec{r}_i-\vec{R})^2 | \psi_{\text{hadron}} \rangle  \label{eq:stat1} \,, \\
\langle R_c^2 \rangle &\equiv \langle \psi_{\text{hadron}} | \sum_{i=1}^{3} e_i (\vec{r}_i-\vec{R})^2 | \psi_{\text{hadron}} \rangle \label{eq:stat2} \,, \\
\langle \mu \rangle &\equiv \langle \psi_{\text{hadron}} | \sum_{i=1}^{3} \frac{e_i}{2m_i} (\ell_z^i+2s_z^i)^2 | \psi_{\text{hadron}} \rangle \label{eq:stat3} \,.
\end{align}
They are presented in Table~\ref{tab:BarPro} and one can see that the values obtained in Ref.~\cite{SilvestreBrac:1996bg} are perfectly reproduced. Therefore, as a proof of concept, our numerical technique can be applied to study not only eigenenergies but also structure properties of the hadrons under consideration when the momentum of the prove is less or equal than the typical hadron scale, $1\,\text{GeV}$.

Equation~\eqref{eq:stat1} gives an idea of the physical size of the baryon, where
\begin{equation}
\vec{R} = \frac{m_1\vec{r}_1 + m_2\vec{r}_2 + m_3\vec{r}_3 }{M} \,,
\end{equation}
is the center-of-mass coordinate, with $M=m_1+m_2+m_3$ the total mass of the system. One can see in Table~\ref{tab:BarPro} that the results lie between $0.02$ and $0.07\,\text{fm}^{2}$. This means that the spatial extension of such baryons goes from $0.15$ to $0.26\,\text{fm}$ while for the mesons we had $\surd{\langle R_{m}^2 \rangle_{Q\bar{Q}}} = \surd{\langle r^2 \rangle}/2 \in [0.1,0.2]\,\text{fm}$, \emph{i.e.} triply-heavy baryons are objects only slightly less compact than their heavy quarkonia counterparts. This can be explained by the fact that the color quark-quark interaction is half as intense as the quark-antiquark one but, \emph{a priori}, one would expect a bigger effect and this could indicate that some kind of diquark correlations be at play.

Equation~\eqref{eq:stat2} defines the charge mean-square radii that can be deduced from the value of the baryon's electric form factor at the photon point extracted from lepton-baryon scattering experiments. A result that might be worth to emphasize is the electric charge radius of the positive charged baryon $\Omega_{ccb}^+$, which is $0.31\,\text{fm}$, three times smaller than the proton's radius $0.84\,\text{fm}$~\cite{Tanabashi:2018oca}.

If mesonic exchange currents and relativistic effects can be ignored, the magnetic moment operator is given by Eq.~\eqref{eq:stat3}, which is just the sum of the magnetic moments of each quark, with orbital and spin contributions. Our values for the positive-charged and neutral triply heavy baryons, $\Omega_{ccb}^+$ and $\Omega_{cbb}^0$, are respectively $0.475\mu_N$ and $-0.193\mu_N$. These can be compared with the values of the proton and neutron: $2.79\mu_N$ and $-1.91\mu_N$, collected in Ref.~\cite{Tanabashi:2018oca}. Namely, the triply-heavy baryon partner of the nucleon have a magnetic moment $5-10$ times smaller.

\begin{figure*}
\includegraphics[width=0.40\textwidth]{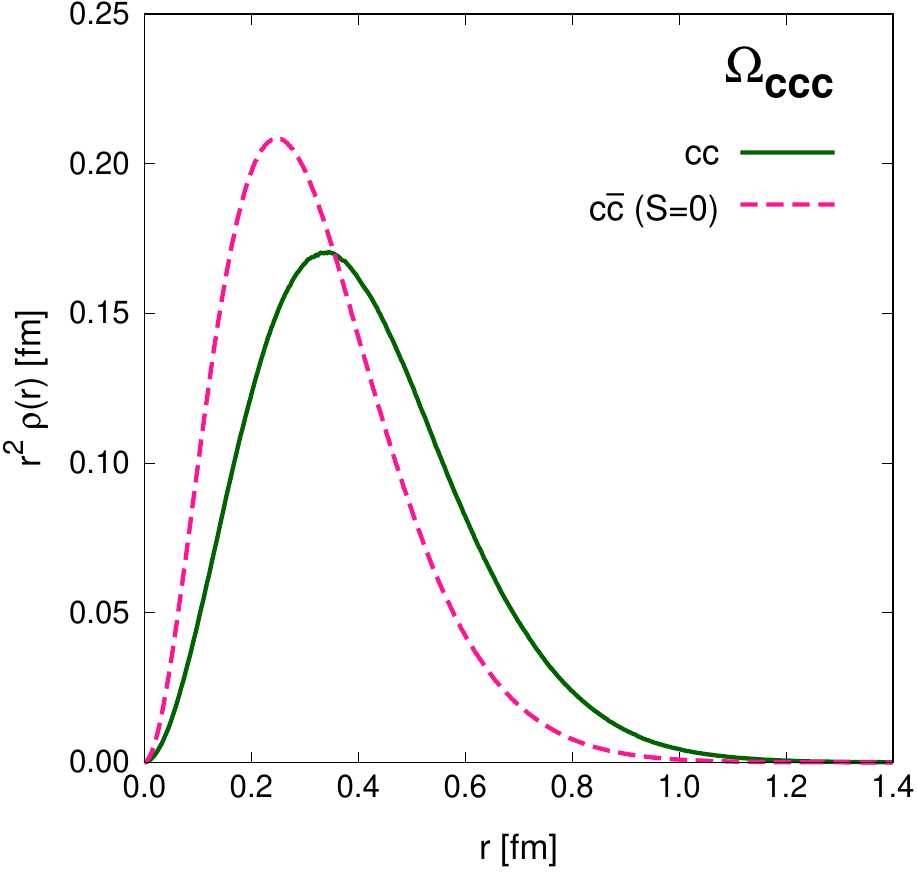}
\hspace*{0.50cm}
\includegraphics[width=0.40\textwidth]{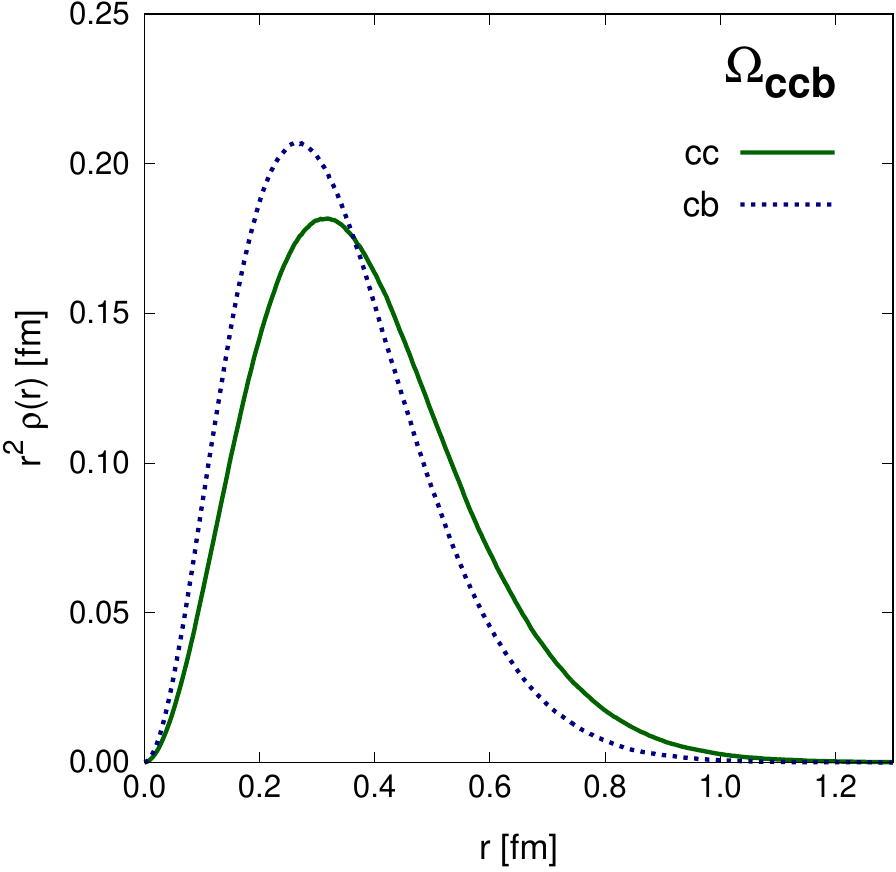} \\
\vspace*{0.50cm}
\includegraphics[width=0.40\textwidth]{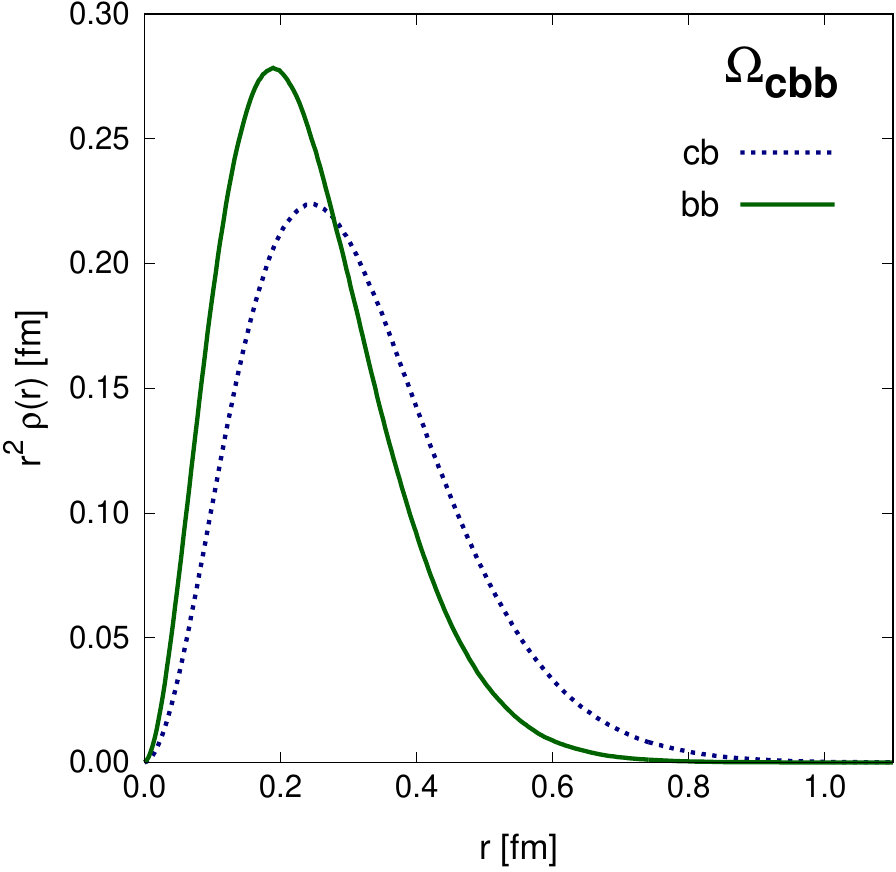}
\hspace*{0.50cm}
\includegraphics[width=0.40\textwidth]{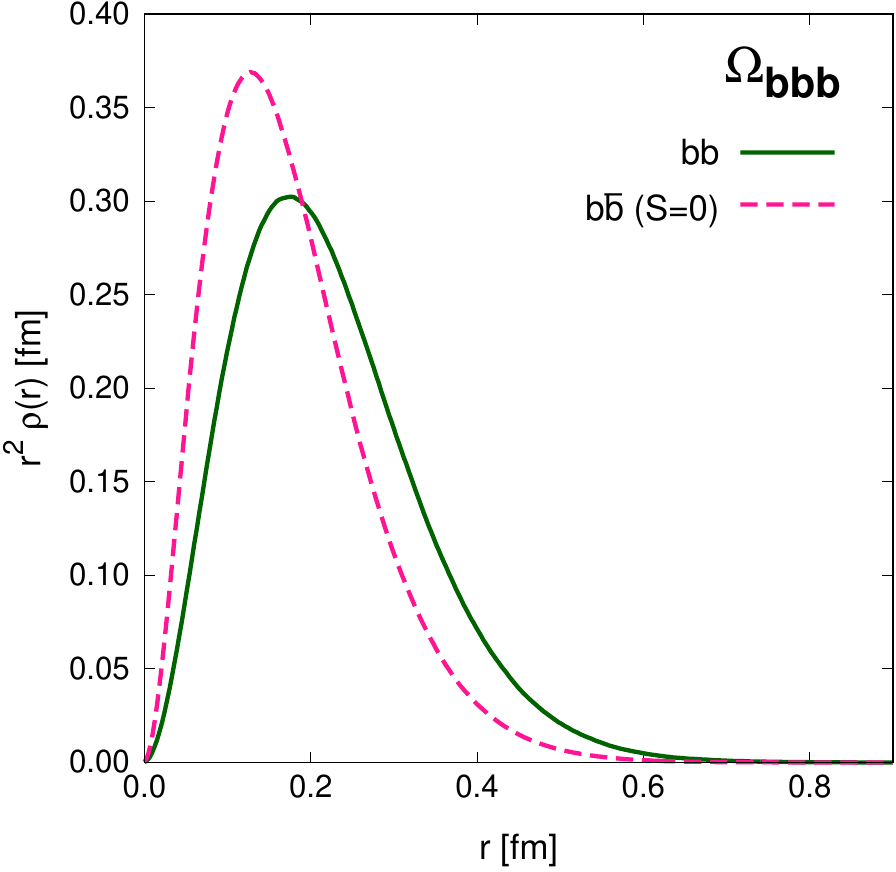}
\caption{\label{fig:Baryons} Radial distribution functions for the studied $\Omega_{ccc}$ (\emph{upper-left panel}), $\Omega_{ccb}$ (\emph{upper-right panel}), $\Omega_{cbb}$ (\emph{bottom-left panel}) and $\Omega_{bbb}$ (\emph{bottom-right panel}) baryons. The (pink) long-dashed curve plotted in the $\Omega_{ccc}$ and $\Omega_{bbb}$ panels is the same radial distribution function but for the $c\bar c$ and $b\bar b$ mesons, respectively.}
\end{figure*}

Figure~\ref{fig:Baryons} shows the relevant radial distribution functions. Among the observed features, the following are of particular interest. In the cases of $\Omega_{ccc}$ and $\Omega_{bbb}$ baryons, the quark-quark probability distribution function is broader than its quark-antiquark counterpart. There are two kinds of probability distributions in the cases of $\Omega_{ccb}$ and $\Omega_{cbb}$ baryons, one referring to the $QQ$-pair where the two heavy quarks are equal and the other one when the $QQ$-pair is made with different species. The same pattern as in heavy quarkonia is observed in triply-heavy baryons where the $QQ$-pair seems to be more compact as the heavy quark mass is larger, and thus this could facilitate the appearance of strong $(cb)$- and $(bb)$-diquark correlations inside the $\Omega_{ccb}$ and $\Omega_{cbb}$ baryons, respectively.

\begin{table}[!t]
\begin{center}
\caption{\label{tab:radiiBaryon} Mean-square radii, $\langle r_{ij}^2 \rangle$, of the studied triply-heavy baryons, in units of fm$^2$.}
\begin{tabular}{lccc}
\hline
\hline
Baryon & $\langle r_{cc}^2 \rangle$ & $\langle r_{cb}^2 \rangle$ & $\langle r_{bb}^2 \rangle$ \\
\hline
\tstrut
$\Omega_{ccc}^{++}$ & 0.206 & - & - \\
$\Omega_{ccb}^{+}$  & 0.182 & 0.136 & - \\
$\Omega_{cbb}^{0}$  & - & 0.117 & 0.073 \\
$\Omega_{bbb}^{-}$  & - & - & 0.062 \\
\hline
\hline
\end{tabular}
\end{center}
\end{table}

Finally, the typical mean-square radii, $\langle r_{ij}^2 \rangle$, of the studied triply-heavy baryons are collected in Table~\ref{tab:radiiBaryon}. Among other features, it highlights that the $cb$- and $bb$-pair are closer inside the $\Omega_{ccb}$ and $\Omega_{cbb}$ baryons, respectively.

For completeness, we have calculated the $J^P=3/2^+$ lowest-lying states of the $\Omega_{ccb}$ and $\Omega_{cbb}$ baryons. Their masses are $8046\,\text{MeV}$ and $11247\,\text{MeV}$, respectively. They compare well with the variational ones reported in Ref.~\cite{SilvestreBrac:1996bg}. We expect bigger differences when radial, angular and spin excitations will be compared; this goes beyond the scope of the present manuscript but indicates a possible next step to follow in the future. Finally, the predicted mass splittings $\Delta m = m(3/2^+)-m(1/2^+)$ are $28\,\text{MeV}$ and $32\,\text{MeV}$ for the $\Omega_{ccb}$ and $\Omega_{cbb}$ baryons; they are of the same order of magnitude than those collected in Ref.~\cite{Yang:2019lsg}, and references therein.


\subsection{Fully-heavy tetraquarks}
\label{subsec:tetraquarks}

Diffusion Monte Carlo methods are designed to compute non-relativistic bound-states of a few- to many-particles system. This technique has been successfully applied to the field of nuclear physics studying light and middle nuclei, as well as objects of high nuclear density such as neutron stars. However, as explained before, applications of DMC to hadron physics are scarce because most hadrons were understood as relativistic bound states of few (two or three) light quarks (antiquarks). Nowadays, many experimental signals point out the existence of tetra-~\cite{1804391}, penta-~\cite{Aaij:2015tga, lhcb:2019pc} and even hexa-quark~\cite{Clement:2016vnl} systems, mostly in heavy quark sectors where non-relativistic dynamics could be thought as a good assumption, and thus Monte Carlo techniques are becoming attractive approaches to apply in hadron physics. After the application of the method to mesons and baryons as a proof-of-concept, our next step is the study of all-heavy tetraquarks. Further studies of more complex multiquark systems shall be performed in the future but they go beyond the scope of this manuscript.

Multiquark systems present richer color structures than mesons and baryons. Without assuming any kind of clustering, the color algebra applied to tetraquark states leads to the following irreducible representations
\begin{align}
{\bf 3}_{\rm c} \,\otimes\, {\bf 3}_{\rm c} \,\otimes\, \bar{{\bf 3}}_{\rm c} \,\otimes\, \bar{{\bf 3}}_{\rm c} &\,=\, \Big( 2 \,\times\, {\bf 1}_{\rm c} \Big) \,\oplus\, \Big( 4 \,\times\, {\bf 8}_{\rm c} \Big) \nonumber \\
&
\,\oplus\, \Big( 2 \,\times\, {\bf 10}_{\rm c} \Big) \,\oplus\, {\bf 27}_{\rm c} \,,
\end{align}
where two color singlet states appear. They are usually known as, respectively, the $(\bar{3}_c\otimes3_c)$ and $(6_c\otimes\bar{6}_c)$ diquark-antidiquark configurations, \emph{i.e.} 
\begin{align}
{\bf 3}_{\rm c} \,\otimes\, {\bf 3}_{\rm c} \,\otimes\, \bar{{\bf 3}}_{\rm c} \,\otimes\, \bar{{\bf 3}}_{\rm c} &\,=\, \Big( \bar{{\bf 3}}_{\rm c} \,\oplus\, {\bf 6}_{\rm c} \Big) \,\otimes\, \Big( {\bf 3}_{\rm c} \,\oplus\, \bar{{\bf 6}}_{\rm c} \Big) \nonumber \\
&
\,=\, {\bf 1}_{(\bar{{\bf 3}}_{\rm c}\otimes{\bf 3}_{\rm c})} \oplus {\bf 1}_{({\bf 6}_{\rm c}\otimes\bar{{\bf 6}}_{\rm c})} \oplus \ldots
\end{align}
Knowing that $\bar{{\bf 3}}_c$ diquark and ${\bf 3}_c$ antidiquark representations are antisymmetric under the transposition of the two particles:
\begin{subequations}
\begin{align}
|\bar{3}_{12}^{\alpha}\rangle = \frac{1}{\sqrt{2}} \epsilon^{\alpha\beta\gamma}\, |Q^{\beta}(1)\rangle |Q^{\gamma}(2)\rangle \,, \\
|3_{12,\alpha}\rangle = \frac{1}{\sqrt{2}} \epsilon_{\alpha\beta\gamma}\, |\bar{Q}_{\beta}(1)\rangle |\bar{Q}_{\gamma}(2)\rangle \,,
\end{align}
\end{subequations}
where $\epsilon^{\alpha\beta\gamma}$ is the Levi-Civita tensor, with the greek letters going from $1$ to $3$; and that ${\bf 6}_c$ diquark and $\bar{{\bf 6}}_c$ antidiquark are symmetric under the same transposition:\footnote{The non-vanishing $d_{\alpha\beta\gamma}=d^{\alpha\beta\gamma}$ constants are $d_{111}=d_{222}=d_{333}=1$ and $d_{412}=d_{421}=d_{523}=d_{532}=d_{613}=d_{631}=1/\sqrt{2}$.}
\begin{subequations}
\begin{align}
|6_{12}^{\alpha}\rangle = d^{\alpha\beta\gamma}\, |Q^{\beta}(1)\rangle |Q^{\gamma}(2)\rangle \,, \\
|\bar{6}_{12,\alpha}\rangle = d_{\alpha\beta\gamma}\, |\bar{Q}^{\beta}(1)\rangle |\bar{Q}_{\gamma}(2)\rangle \,;
\end{align}
\end{subequations}
one can build two orthogonal singlet color tetraquark states
\begin{align}
|\bar{3}_{12} 3_{34}\rangle &= \frac{1}{\sqrt{12}} \epsilon^{\alpha\beta\gamma} \epsilon_{\alpha\lambda\sigma} |Q^{\beta}(1)\rangle |Q^{\gamma}(2)\rangle |\bar{Q}_{\lambda}(3)\rangle |\bar{Q}_{\sigma}(4)\rangle \,, \\
|6_{12} \bar{6}_{34}\rangle &= \frac{1}{\sqrt{6}} d^{\alpha\beta\gamma} d_{\alpha\lambda\sigma} |Q^{\beta}(1)\rangle |Q^{\gamma}(2)\rangle |\bar{Q}_{\lambda}(3)\rangle |\bar{Q}_{\sigma}(4)\rangle \,.
\end{align}
Their explicit expressions in terms of the familiar \emph{red}, \emph{green} and \emph{blue} degrees-of-freedom, and without explicitly clustering, can be written down as
\begin{align}
|\bar{3}_{12} 3_{34}\rangle = \frac{1}{\sqrt{12}} \Big( 
&
+|rg\bar{r}\bar{g}\rangle
+|gr\bar{g}\bar{r}\rangle
-|rg\bar{g}\bar{r}\rangle \nonumber \\
&
-|gr\bar{r}\bar{g}\rangle
+|rb\bar{r}\bar{b}\rangle
+|br\bar{b}\bar{r}\rangle \nonumber \\
&
-|rb\bar{b}\bar{r}\rangle
-|br\bar{r}\bar{b}\rangle
+|gb\bar{g}\bar{b}\rangle \nonumber \\
&
+|bg\bar{b}\bar{g}\rangle
-|gb\bar{b}\bar{g}\rangle
-|bg\bar{g}\bar{b}\rangle \Big) \,,
\end{align}
which is antisymmetric under the exchange of either both quarks or both antiquarks, and
\begin{align}
|6_{12} \bar{6}_{34}\rangle = \frac{1}{\sqrt{6}} \Big[
&
+|rr\bar{r}\bar{r}\rangle 
+|gg\bar{g}\bar{g}\rangle
+|bb\bar{b}\bar{b}\rangle \nonumber \\
&
+\frac{1}{2} \Big( 
+|rg\bar{r}\bar{g}\rangle 
+|gr\bar{g}\bar{r}\rangle 
+|rg\bar{g}\bar{r}\rangle \nonumber \\
& 
\hspace*{0.90cm} +|gr\bar{r}\bar{g}\rangle
+|rb\bar{r}\bar{b}\rangle
+|br\bar{b}\bar{r}\rangle \nonumber \\
&
\hspace*{0.90cm} +|rb\bar{b}\bar{r}\rangle
+|br\bar{r}\bar{b}\rangle
+|gb\bar{g}\bar{b}\rangle \nonumber \\
&
\hspace*{0.90cm} +|bg\bar{b}\bar{g}\rangle
+|gb\bar{b}\bar{g}\rangle
+|bg\bar{g}\bar{b}\rangle \Big) \Big] \,,
\end{align}
which is symmetric under the exchange of either both quarks or both antiquarks. 

It is also important to notice herein that the two color states defined above can be expressed in another set of color representations
\begin{align}
|\bar{3}_{12} 3_{34}\rangle &= + \sqrt{\frac{1}{3}} |1_{13} 1_{24} \rangle - \sqrt{\frac{2}{3}} |8_{13} 8_{24} \rangle \nonumber \\
&
= -\sqrt{\frac{1}{3}} |1_{14} 1_{23} \rangle + \sqrt{\frac{2}{3}} |8_{14} 8_{23} \rangle \,, \\[1.5ex]
|6_{12} \bar{6}_{34}\rangle &= + \sqrt{\frac{2}{3}} |1_{13} 1_{24} \rangle + \sqrt{\frac{1}{3}} |8_{13} 8_{24} \rangle \nonumber \\
&
= +\sqrt{\frac{2}{3}} |1_{14} 1_{23} \rangle + \sqrt{\frac{1}{3}} |8_{14} 8_{23} \rangle \,,
\end{align}
called meson-meson, $\bf{1}_c \otimes \bf{1}_c$, and color-hidden, $\bf{8}_c \otimes \bf{8}_c$, states.


\begin{table*}[!t]
\caption{\label{tab:Configurations} Spin-color configurations for fully-heavy tetraquark systems.
}
\begin{ruledtabular}
\begin{tabular}{llllll}
System & $J^{P(C)}$ & \multicolumn{4}{c}{Spin-Color configurations} \\
\hline
\tstrut
$\begin{matrix} cc\bar c\bar c, & bb\bar b\bar b \\ cc\bar b\bar b & (bb\bar c\bar c)\end{matrix}$ & $0^{+(+)}$ & $|\chi_{S=0}\rangle_{SS} \otimes |\bar{3}_c 3_c\rangle_{AA}$ & $|\chi_{S=0}\rangle_{AA} \otimes |6_c \bar{6}_c\rangle_{SS}$ & & \\[2ex]
& $1^{+(-)}$ & $|\chi_{S=1}\rangle_{SS} \otimes |\bar{3}_c 3_c\rangle_{AA}$ & & & \\[2ex]
& $2^{+(+)}$ & $|\chi_{S=2}\rangle_{SS} \otimes |\bar{3}_c 3_c\rangle_{AA}$ & & & \\[4ex]
$cc\bar c\bar b$, $bb\bar c\bar b$ & $0^{+}$ & $|\chi_{S=0}\rangle_{SS} \otimes |\bar{3}_c 3_c\rangle_{AA}$ & $|\chi_{S=0}\rangle_{AA} \otimes |6_c \bar{6}_c\rangle_{SS}$ & & \\[2ex]
& $1^{+}$ & $|\chi_{S=1}\rangle_{SS} \otimes |\bar{3}_c 3_c\rangle_{AA}$ & $|\chi_{S=1}\rangle_{SA} \otimes |\bar{3}_c 3_c\rangle_{AA}$ & $|\chi_{S=1}\rangle_{AS} \otimes |6_c \bar{6}_c\rangle_{SS}$ & \\[2ex]
& $2^{+}$ & $|\chi_{S=2}\rangle_{SS} \otimes |\bar{3}_c 3_c\rangle_{AA}$ & &  & \\[4ex]
$cb\bar c\bar b$ & $0^{+(+)}$ & $|\chi_{S=0}\rangle_{SS} \otimes |\bar{3}_c 3_c\rangle_{AA}$ & $|\chi_{S=0}\rangle_{SS} \otimes |6_c \bar{6}_c\rangle_{SS}$ & $|\chi_{S=0}\rangle_{AA} \otimes |\bar{3}_c 3_c\rangle_{AA}$ & $|\chi_{S=0}\rangle_{AA} \otimes |6_c \bar{6}_c\rangle_{SS}$ \\[3ex]
& $1^{+(-)}$ & \multicolumn{2}{c}{$|\chi_{S=1}\rangle_{SS} \otimes |\bar{3}_c 3_c\rangle_{AA}$} & \multicolumn{2}{c}{$|\chi_{S=1}\rangle_{SS} \otimes |6_c \bar{6}_c\rangle_{SS}$} \\
& & \multicolumn{2}{c}{$\frac{1}{\sqrt{2}} \big(|\chi_{S=1}\rangle_{SA} - |\chi_{S=1}\rangle_{AS} \big) \otimes |\bar{3}_c 3_c\rangle_{AA}$} & \multicolumn{2}{c}{$\frac{1}{\sqrt{2}} \big(|\chi_{S=1}\rangle_{SA} - |\chi_{S=1}\rangle_{AS} \big) \otimes |6_c \bar{6}_c\rangle_{SS}$} \\[3ex]
& $1^{+(+)}$ & \multicolumn{2}{c}{$\frac{1}{\sqrt{2}} \big(|\chi_{S=1}\rangle_{SA} + |\chi_{S=1}\rangle_{AS} \big) \otimes |\bar{3}_c 3_c\rangle_{AA}$} & \multicolumn{2}{c}{$\frac{1}{\sqrt{2}} \big(|\chi_{S=1}\rangle_{SA} + |\chi_{S=1}\rangle_{AS} \big) \otimes |6_c \bar{6}_c\rangle_{SS}$} \\[3ex]
& $2^{+(+)}$ & $|\chi_{S=2}\rangle_{SS} \otimes |\bar{3}_c 3_c\rangle_{AA}$ & $|\chi_{S=2}\rangle_{SS} \otimes |6_c \bar{6}_c\rangle_{SS}$ & & \\
\end{tabular}
\end{ruledtabular}
\end{table*}


If we turn now our attention to the spin degree-of-freedom, the $QQ\bar Q\bar Q$ ($Q=c$ or $b$) system, made by fermions of spin $1/2$, can have total spin $S=0$, $1$ and $2$. There are two linearly independent $S=0$ wave functions that can be written as
\begin{subequations}
\begin{align}
|\chi_{S=0,S_z=0}\rangle_{\text{SS}} &= \frac{1}{\sqrt{12}} \Big( +2\,|\downarrow\downarrow\uparrow\uparrow\rangle +2\,|\uparrow\uparrow\downarrow\downarrow\rangle \nonumber \\
& 
\hspace*{1.45cm} -|\downarrow\uparrow\uparrow\downarrow\rangle -|\uparrow\downarrow\downarrow\uparrow\rangle \nonumber \\
&
\hspace*{1.45cm} +|\downarrow\uparrow\downarrow\uparrow\rangle -|\uparrow\downarrow\uparrow\downarrow\rangle \Big) \,, \\
|\chi_{S=0,S_z=0}\rangle_{\text{AA}} &= \frac{1}{2} \Big( -|\downarrow\uparrow\downarrow\uparrow \rangle - |\downarrow\uparrow\uparrow\downarrow \rangle \nonumber \\
&
\hspace*{1.00cm} - |\uparrow\downarrow\downarrow\uparrow \rangle + |\uparrow\downarrow\uparrow\downarrow\rangle \Big) \,.
\end{align}
\end{subequations}
They are, respectively, symmetric ($S$) and antisymmetric ($A$) under the exchange of both quarks and both antiquarks. The linearly independent $S=1$ wave functions are given by ($S_z=S$ is assumed without lost of generality):
\begin{subequations}
\begin{align}
|\chi_{S=1,S_z=+1}\rangle_{\text{SA}} &= \frac{1}{\sqrt{2}} \Big( |\uparrow\uparrow\uparrow\downarrow \rangle - |\uparrow\uparrow\downarrow\uparrow\rangle \Big) \,, \\
|\chi_{S=1,S_z=+1}\rangle_{\text{AS}}&= \frac{1}{\sqrt{2}} \Big( |\uparrow\downarrow\uparrow\uparrow \rangle - |\downarrow\uparrow\uparrow\uparrow\rangle \Big)\,, \\
|\chi_{S=1,S_z=+1}\rangle_{\text{SS}} &= \frac{1}{2} \Big( + |\uparrow\uparrow\uparrow\downarrow \rangle + |\uparrow\uparrow\downarrow\uparrow\rangle \nonumber \\
&
\hspace*{1.00cm} - |\uparrow\downarrow\uparrow\uparrow \rangle - |\downarrow\uparrow\uparrow\uparrow\rangle \Big) \,, 
\end{align}
\end{subequations}
which correspond to the symmetric$\,-\,$antisymmetric, antisymmetric$\,-\,$symmetric and symmetric$\,-\,$symmetric exchange of quarks and antiquarks. Finally, the $S=2$ spin wave function can be written as 
\begin{equation}
|\chi_{S=2,S_z=+2}\rangle_{\text{SS}} = |\!\uparrow\uparrow\uparrow\uparrow\rangle \,,
\end{equation}
where $S_z=S$ is again assumed without lost of generality. Note that this spin wave function is fully symmetric under the exchange of both quarks and both antiquarks.

We shall focus our attention to the ground states of $QQ\bar Q\bar Q$, with $Q$ either $c$- or $b$-quark in any possible combination. This implies that the space wave function is totally symmetric and, in order to fulfill Fermi-Dirac statistics, the wave-function configurations of Table~\ref{tab:Configurations} must be taken into account for each tetraquark sector and $J^{P(C)}$ quantum numbers. We perform a coupled-channels calculation, based on the DMC algorithm explained above, for those sectors shown in Table~\ref{tab:Configurations} which have more than one spin-color configuration. It is worth highlighting that the treatment of spin-dependent potentials and coupled-channels calculations are challenging features for Monte Carlo methods which have avoided their extended application to hadron physics, both are considered herein.

Now, let us proceed to describe in detail our theoretical findings for each sector of fully-heavy tetraquarks.


\begin{table*}[!t]
\caption{\label{tab:ccccmass} Predicted masses, in MeV, for the $cc\bar{c}\bar{c}$ system computed with the Diffusion Monte Carlo technique and compared with those obtained with the variational approach~\cite{Wang:2019rdo}. For completeness, we also compare our results with those of other frameworks.
}
\begin{ruledtabular}
\begin{tabular}{ccc|cccccccccccc}
$J^{PC}$ & DMC & VAR~\cite{Wang:2019rdo} & \cite{Liu:2019zuc} & \cite{Wu:2016vtq} & \cite{Lloyd:2003yc} & \cite{Chen:2016jxd} & \cite{Ader:1981db} & \cite{Iwasaki:1975pv} & \cite{Karliner:2016zzc} & \cite{Barnea:2006sd} & \cite{Wang:2017jtz,Wang:2018poa} & \cite{Debastiani:2017msn} & \cite{Berezhnoy:2011xn} & \cite{Anwar:2017toa} \\
\hline
\tstrut
$0^{++}$ & $6351$ & $6371$ & 6487 & 6797 & 6477 & 6460-6470 & 6437 & 6200 & 6192 & 6038-6115 & 5990 & 5969 & 5966 & $< 6140$ \\
$1^{+-}$ & $6441$ & $6450$ & 6500 & 6899 & 6528 & 6370-6510 & 6437 & - & - & 6101-6176 & 6050 & 6021 & 6051 & - \\
$2^{++}$ & $6471$ & $6479$ & 6524 & 6956 & 6573 & 6370-6510 & 6437 & - & - & 6172-6216 & 6090 & 6115 & 6223 & - \\
\end{tabular}
\end{ruledtabular}
\end{table*}

\begin{figure*}
\includegraphics[width=0.31\textwidth]{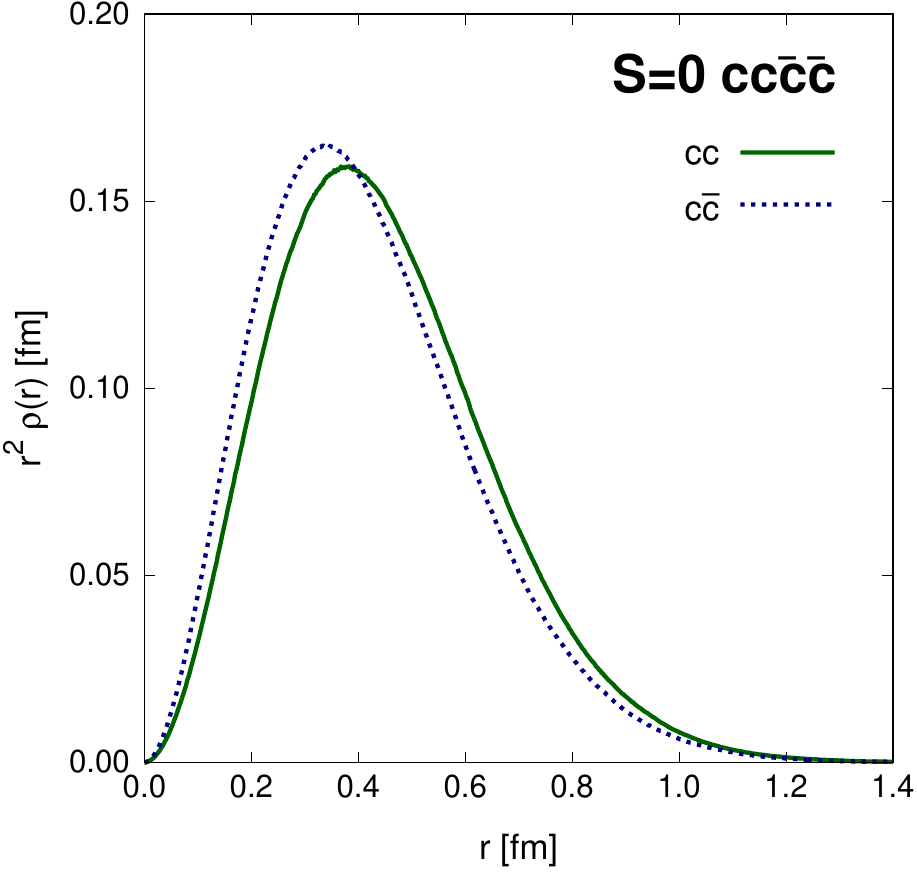}
\includegraphics[width=0.31\textwidth]{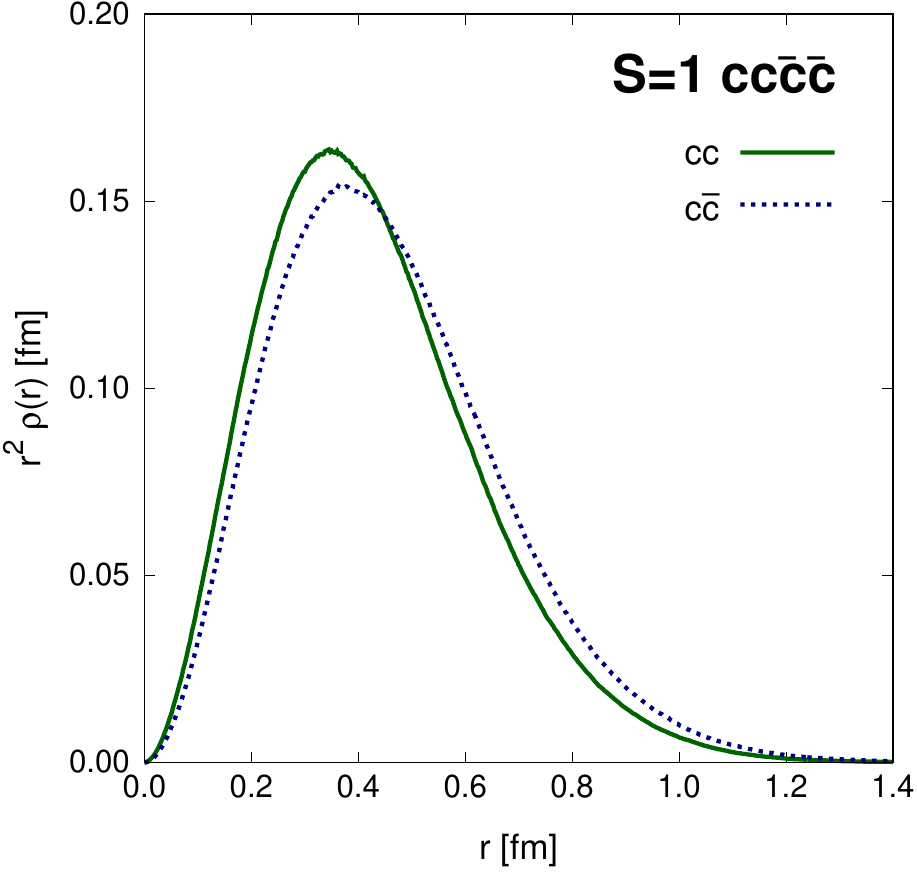}
\includegraphics[width=0.31\textwidth]{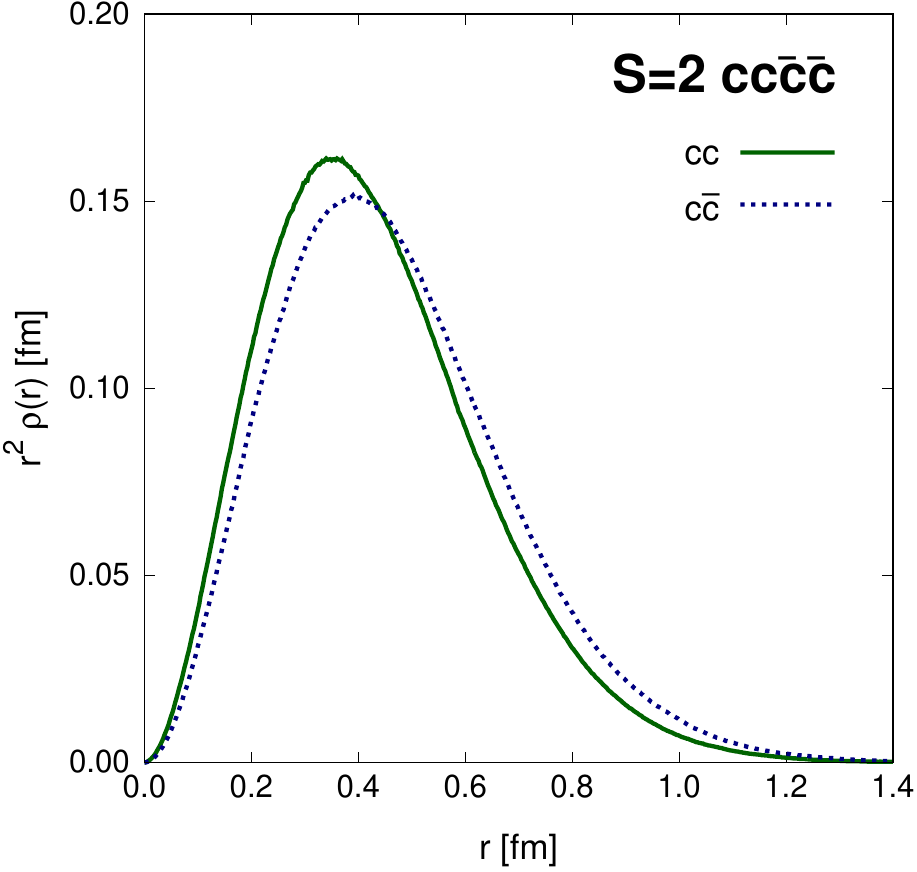}
\caption{\label{fig:cccc} Relevant radial distribution functions for the studied $cc\bar c\bar c$ tetraquark ground states.}
\end{figure*}

{\bf The $\mathbf{cc\bar c\bar c}$ tetraquark ground states.} The masses of the $S=0$, $1$ and $2$ ground states of $cc\bar c\bar c$ tetraquarks are shown in Table~\ref{tab:ccccmass}. We compare first our results with those obtained when solving the same Hamiltonian but using a Rayleigh-Ritz variational method. One can see that both approaches are compatible; however, the differences are larger than the ones shown in the meson and baryon sectors because the variational methods begin to have difficulties as the number of particles grow. Nevertheless, the disagreement between both approaches is between $8$ and $20\,\text{MeV}$, which is smaller than the usual uncertainty assigned to any quark model. Note, too, that the variational method only provides an upper limit of the eigenenergy and it depends on the goodness of the trial wave function; DMC results depend less on such details and, in principle, provides an exact estimate of the eigenenergy.

Table~\ref{tab:ccccmass} provides also a comparison with numerous works that reported results on $cc\bar c\bar c$ tetraquarks using a large variety of techniques. It seems that our results $6.35\,\text{GeV}$, $6.44\,\text{GeV}$ and $6.47\,\text{GeV}$ for the ground states with quantum numbers $J^{PC}=0^{++}$, $1^{+-}$ and $2^{++}$, respectively, are located just in the middle of the ranges covered by all approaches which are $[5.97-6.80]\,\text{GeV}$, $[6.05-6.90]\,\text{GeV}$ and $[6.09-6.96]\,\text{GeV}$. These mass ranges are in agreement with the recently observed structures in the invariant mass distribution of $J/\psi$-pairs~\cite{1804391}. The $cc\bar c\bar c$ ground state with quantum numbers $J^{PC}=0^{++}$ is about $400-500\,\text{MeV}$ above the $\eta_c\eta_c$ and $J/\psi J/\psi$ thresholds. This suggests that the $J^{PC}=0^{++}$ state is unstable and can decay into $\eta_c\eta_c$ and $J/\psi J/\psi$ final states through quark rearrangements. The $J^{PC}=1^{+-}$ state lies about $400\,\text{MeV}$ above the mass threshold of $\eta_c J/\psi$, while $J^{PC}=2^{++}$ is about $300\,\text{MeV}$ above the mass threshold of $J/\psi J/\psi$, they can also easily decay into such charmonium pairs through quark rearrangement.

Figure~\ref{fig:cccc} shows for the $cc\bar c\bar c$ ground states, the relevant radial distribution functions. It is very interesting to observe that quark-antiquark pairs are slightly closer than quark-quark ones in the $S=0$ ($J^{PC}=0^{++}$) state whereas the situation gradually changes as spin increases. In any case, the $cc$ and $c\bar c$ radial distribution functions are very similar with a mean value close to $\sim0.5\,\text{fm}$, indicating that our tetraquark structures are compact objects and not meson-meson molecular states with typical inter-meson distances of $\gtrsim1\,\text{fm}$. 

In order to quantify our observations, we have computed the mass mean-square radii of the $J^{PC}=0^{++}$, $1^{+-}$ and $2^{++}$ $cc\bar c\bar c$ ground states. They are, respectively, $0.087\,\text{fm}^{2}$, $0.092\,\text{fm}^{2}$ and $0.095\,\text{fm}^{2}$; i.e. they are comparable with those obtained for mesons and baryons. Moreover, Table~\ref{tab:radiicccc} shows the typical mean-square radii for each (anti-)quark--quark pair indicating that no interquark distance is very different from the rest and thus all pairs play a similar role with distances of the order of $\sim0.5\,\text{fm}$.

\begin{table}[!t]
\begin{center}
\caption{\label{tab:radiicccc} Mean-square radii, $\langle r_{ij}^2 \rangle$, of the studied fully-charmed tetraquarks, in units of fm$^2$.}
\begin{tabular}{lccc}
\hline
\hline
\tstrut
$J^{PC}$ & $\langle r_{cc}^2 \rangle$ & $\langle r_{c\bar c}^2 \rangle$ & $\langle r_{\bar c\bar c}^2 \rangle$ \\
\hline
\tstrut
$0^{++}$ & 0.246 & 0.216 & 0.246 \\
$1^{+-}$ & 0.225 & 0.255 & 0.223 \\
$2^{++}$ & 0.229 & 0.266 & 0.231 \\
\hline
\hline
\end{tabular}
\end{center}
\end{table}


\begin{table*}[!t]
\caption{\label{tab:bbbbmass} Predicted masses, in MeV, for the $bb\bar{b}\bar{b}$ system computed with the Diffusion Monte Carlo technique and compared with those obtained with the variational approach~\cite{Wang:2019rdo}. For completeness, we also compare our results with those of other frameworks.}
\begin{ruledtabular}
\begin{tabular}{ccc|cccccccccc}
$J^{PC}$ & DMC & VAR~\cite{Wang:2019rdo} & \cite{Liu:2019zuc} & \cite{Wu:2016vtq} & \cite{Wang:2017jtz,Wang:2018poa} & \cite{Karliner:2016zzc} & \cite{Berezhnoy:2011xn} & \cite{Anwar:2017toa} & \cite{Bai:2016int} & \cite{Chen:2016jxd} & \cite{Hughes:2017xie} & \cite{Anwar:2017toa} \\
\hline
\tstrut
$0^{++}$ & $19199$ & $19243$ & 19322 & 20155 & 18840 & 18826 & 18754 & 18720 & 18690 & 18460-18490 & 18798 & $< 18890$ \\
$1^{+-}$ & $19276$ & $19311$ & 19329 & 20212 & 18840 & - & 18808 & - & - & 18320-18540 & - & - \\
$2^{++}$ & $19289$ & $19325$ & 19341 & 20243 & 18850 & - & 18916 & - & - & 18320-18530 & - & - \\
\end{tabular}
\end{ruledtabular}
\end{table*}

\begin{figure*}
\includegraphics[width=0.31\textwidth]{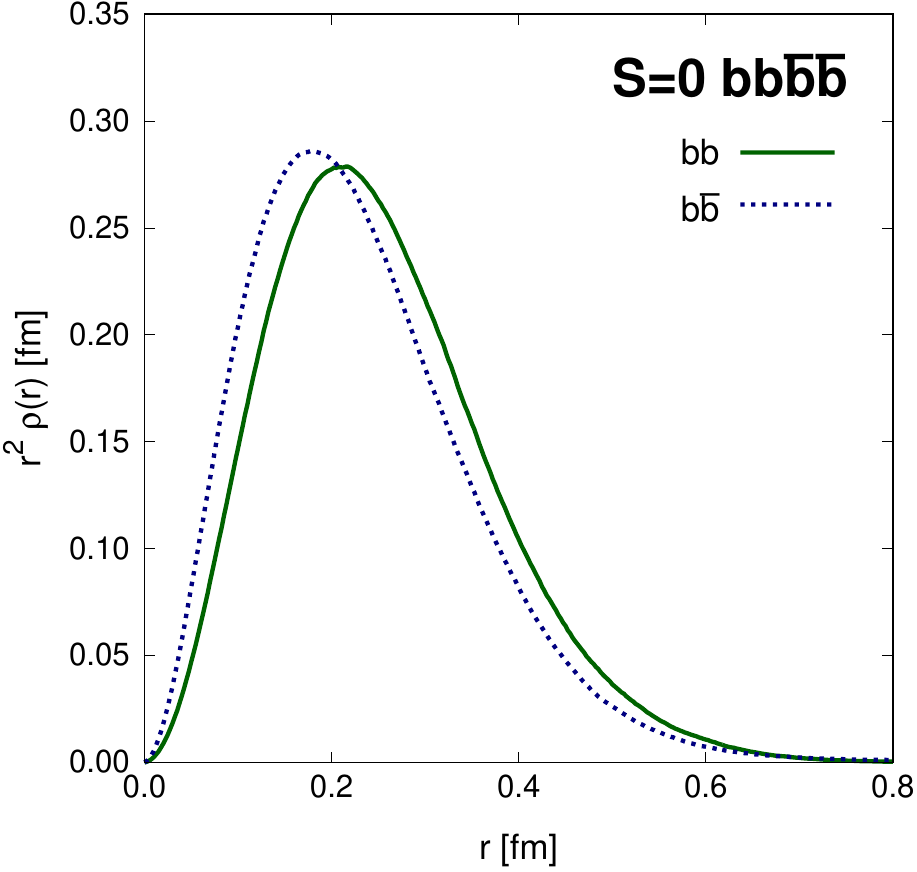}
\includegraphics[width=0.31\textwidth]{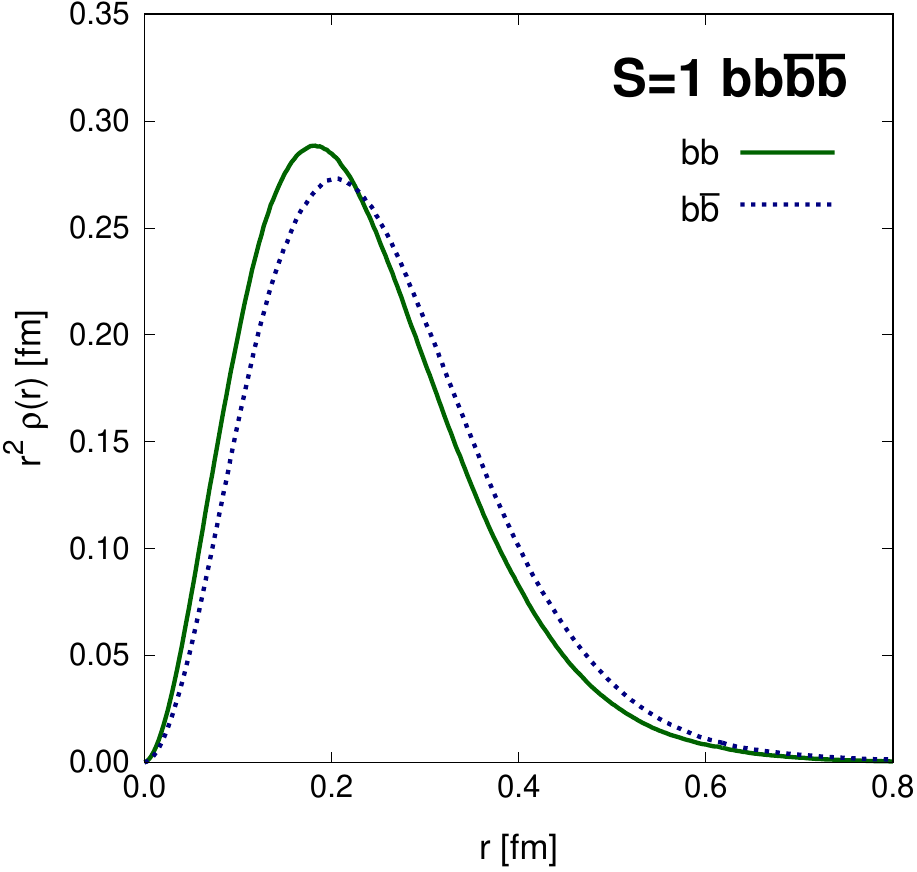}
\includegraphics[width=0.31\textwidth]{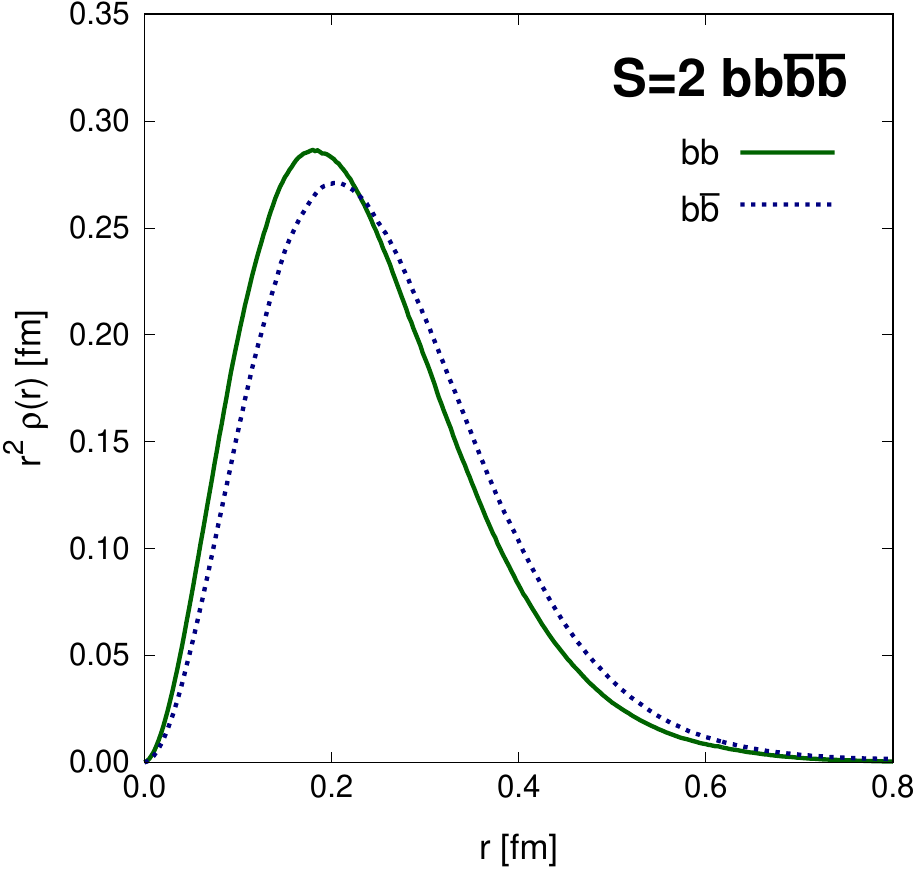}
\caption{\label{fig:bbbb} Relevant radial distribution functions for the studied $bb\bar b\bar b$ tetraquark ground states.}
\end{figure*}

{\bf The $\mathbf{bb\bar b\bar b}$ tetraquark ground states.} Let us now turn our attention to the $J^{PC}=0^{++}$, $1^{+-}$ and $2^{++}$ ground states of $bb\bar b\bar b$ tetraquarks shown in Table~\ref{tab:bbbbmass}. We compare first our results with those obtained when solving the same Hamiltonian but using a Rayleigh-Ritz variational method. One can see that even larger differences, between $30$ and $50\,\text{MeV}$, are found in the $bb\bar b\bar b$ sector, with our results always below the ones reported in Ref.~\cite{Wang:2019rdo}, as expected. From the dynamical point of view, we would expect better agreement in tetraquark sectors where heavier quark masses are involved and thus the issue might be related with the oscillating parameters of the Gaussian basis used in Ref.~\cite{Wang:2019rdo} for the $bb\bar b\bar b$ sector. Another possibility is that, as it will be shown later, interquark distances are smaller in these tetraquarks and thus multi-body correlations could play a more important role. Table~\ref{tab:bbbbmass} provides also a comparison with numerous works that reported results on $bb\bar b\bar b$ tetraquarks using a large variety of theoretical techniques. Again, our results $19.20\,\text{GeV}$, $19.28\,\text{GeV}$ and $19.29\,\text{GeV}$ for the ground states with quantum numbers $J^{PC}=0^{++}$, $1^{+-}$ and $2^{++}$, respectively, are located just in the middle of the ranges covered by all approaches which are $[18.46-20.16]\,\text{GeV}$, $[18.32-20.21]\,\text{GeV}$ and $[18.32-20.24]\,\text{GeV}$.

Figure~\ref{fig:bbbb} shows the relevant radial distribution functions for the $bb\bar b\bar b$ ground states. As in the case of fully-charm tetraquarks, the quark-antiquark pairs are closer than quark-quark ones in the $S=0$ ($J^{PC}=0^{++}$) state. The situation, however, gradually changes as spin increases and thus the diquark (antidiquark) distance seems to be slightly smaller for $J^{PC}=1^{+-}$ and $2^{++}$ $bb\bar b\bar b$ tetraquarks. The theoretical fact that the $J^{PC}=0^{++}$ $bb\bar b\bar b$ ground state could prefer to be in quark-antiquark pairs, together with a predicted mass which is $300-400\,\text{MeV}$ above the $\eta_b\eta_b$ and $\Upsilon(1S)\Upsilon(1S)$ thresholds, could provide founded reasons to keep looking for (reasonably wide) bumps in the invariant mass of bottomonium-pairs at the LHCb experiment.

We report in Table~\ref{tab:radiibbbb} the typical mean-square radii of the studied $bb\bar b\bar b$ tetraquarks. Such table highlights again the change of configuration picture from quark-antiquark pairs to diquark-antidiquark ones, when going from $S=0$ to $S=2$ ground states of $bb\bar b\bar b$ tetraquarks. Finally, the mass mean-square radii for the $J^{PC}=0^{++}$, $1^{+-}$ and $2^{++}$ ground states are, respectively, $0.026\,\text{fm}^{2}$, $0.028\,\text{fm}^{2}$ and $0.029\,\text{fm}^{2}$. This indicates that these states are very compact and far away from the picture of meson-meson molecules.

\begin{table}[!t]
\begin{center}
\caption{\label{tab:radiibbbb} Mean-square radii, $\langle r_{ij}^2 \rangle$, of the studied fully-bottom tetraquarks, in units of fm$^2$.}
\begin{tabular}{lccc}
\hline
\hline
\tstrut
$J^{PC}$ & $\langle r_{bb}^2 \rangle$ & $\langle r_{b\bar b}^2 \rangle$ & $\langle r_{\bar b\bar b}^2 \rangle$ \\
\hline
\tstrut
$0^{++}$ & 0.079 & 0.067 & 0.079 \\
$1^{+-}$ & 0.068 & 0.078 & 0.068 \\
$2^{++}$ & 0.068 & 0.080 & 0.068 \\
\hline
\hline
\end{tabular}
\end{center}
\end{table}


{\bf The $\mathbf{cc\bar b\bar b}$ ($\mathbf{bb\bar c\bar c}$) tetraquark ground states.} A similar theoretical calculation must be performed for computing the ground states of the $J^{P}=0^{+}$, $1^{+}$ and $2^{+}$ $cc\bar b\bar b$ ($bb\bar c\bar c$) tetraquarks. We obtain the masses:
\begin{subequations}
\begin{align}
m(J^P=0^+) &= 12865 \,\text{MeV} \,, \\
m(J^P=1^+) &= 12908 \,\text{MeV} \,, \\
m(J^P=2^+) &= 12926 \,\text{MeV} \,,
\end{align}
\end{subequations}
which compare as follows 
\begin{subequations}
\begin{align}
m(J^P=0^+) &= 12886 \,\text{MeV} \,, \\
m(J^P=1^+) &= 12924 \,\text{MeV} \,, \\
m(J^P=2^+) &= 12940 \,\text{MeV} \,,
\end{align}
\end{subequations}
with the variational calculation~\cite{Wang:2019rdo}. As one can see, the differences between the two numerical methods are around $20\,\text{MeV}$, and our values are always below those obtained by the variational method. Note, too, that our results are in reasonable agreement with the scarce ones reported by other theoretical methods~\cite{Liu:2019zuc}. Moreover, our predictions are above their lowest open-flavor decay channels for about $240-280\,\text{MeV}$ and thus they should appear as resonances with relatively large widths in the invariant mass of $B_cB_c$, $B_c B_c^\ast$, or $B_c^\ast B_c^\ast$ final states.

\begin{figure}
\includegraphics[width=0.45\textwidth]{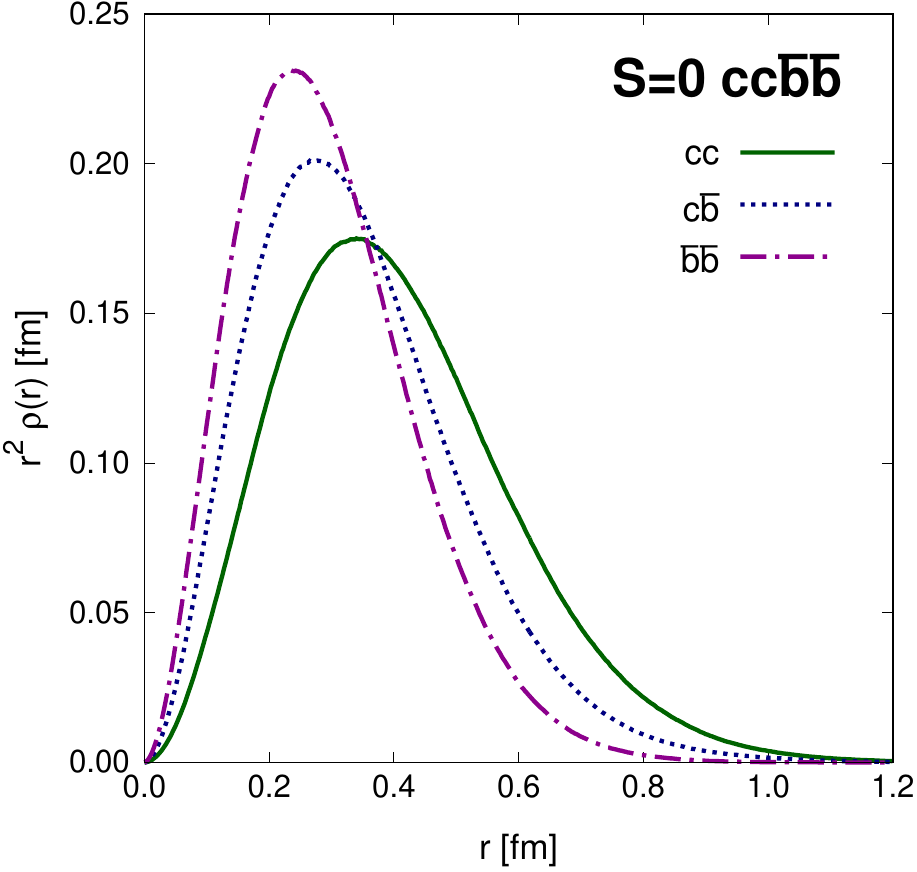}
\caption{\label{fig:ccbb} Radial distribution functions for the $J^P=0^+$ ground state of $cc\bar b\bar b$ tetraquark. The case $bb\bar c\bar c$ is obviously equal. No further information is obtained for the $J^P=1^+$ and $2^+$ cases.}
\end{figure}

An interesting feature distinguishes this sector with respect the two already discussed, \emph{i.e.} the $cc\bar c\bar c$ and $bb\bar b\bar b$ sectors. The expected pattern of interquark distances for $cc$, $c\bar b$ and $\bar b\bar b$ (from larger to shorter) is conserved when changing the total spin $S=0$, $1$ and $2$. An example of the output is drawn in Fig.~\ref{fig:ccbb} for the $J^P=0^+$ $cc\bar b\bar b$ ground state. We report, too, the mass mean-square radii for the $J^{P}=0^{+}$, $1^{+}$ and $2^{+}$ ground states, which are $0.046\,\text{fm}^{2}$, $0.047\,\text{fm}^{2}$, $0.048\,\text{fm}^{2}$, respectively. These values lie between the ones reported above for the $cc\bar c\bar c$ and $bb\bar b\bar b$ tetraquarks

\begin{figure*}
\includegraphics[width=0.31\textwidth]{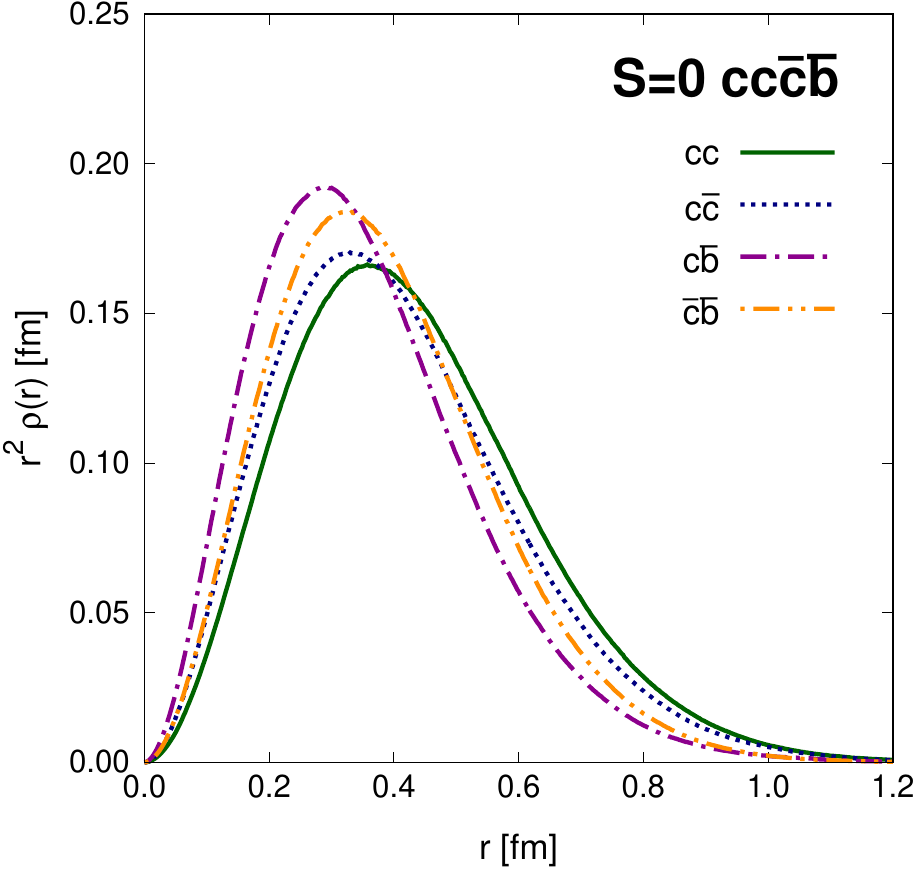}
\includegraphics[width=0.31\textwidth]{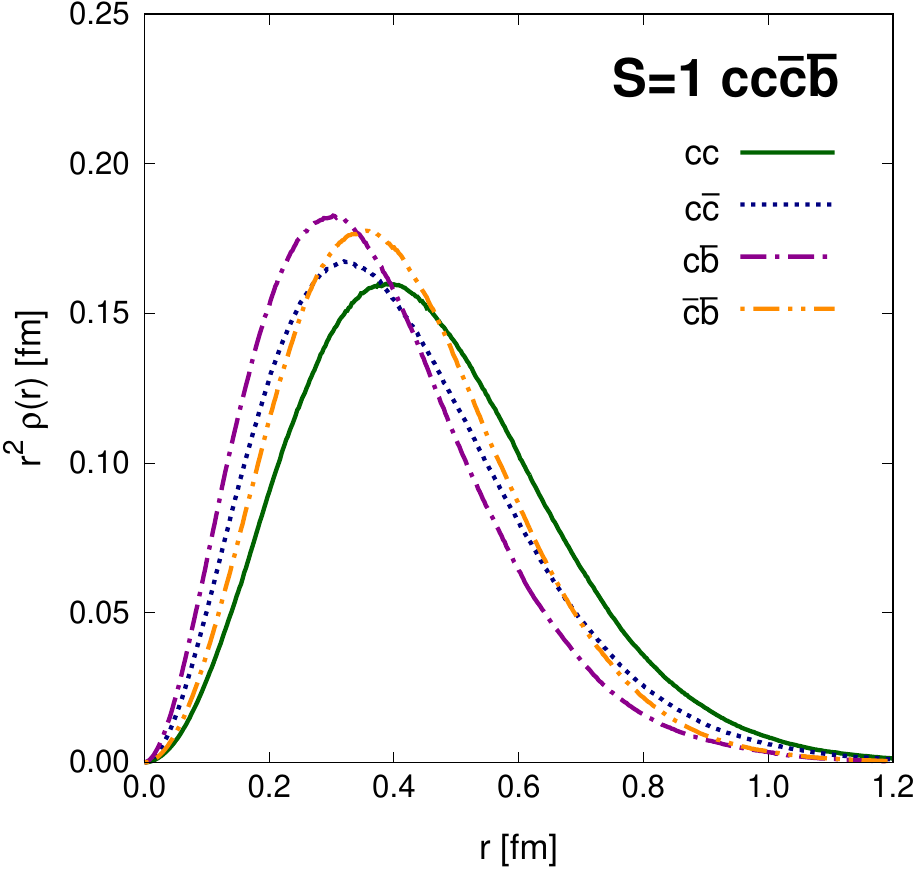}
\includegraphics[width=0.31\textwidth]{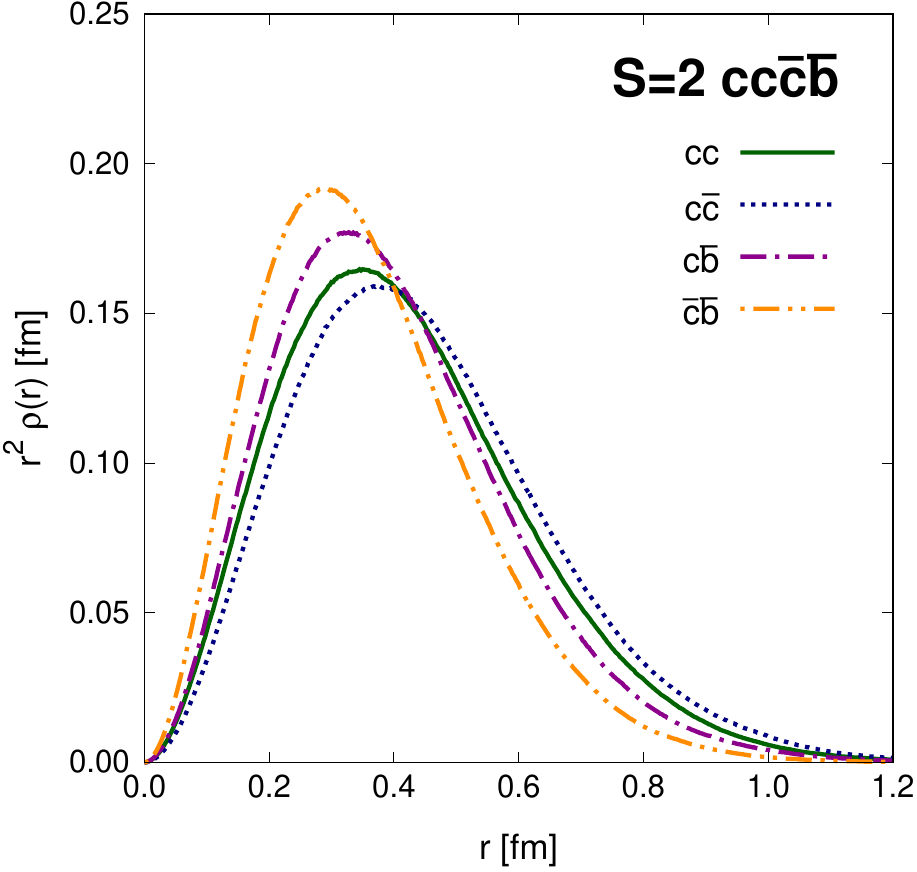} \\
\includegraphics[width=0.31\textwidth]{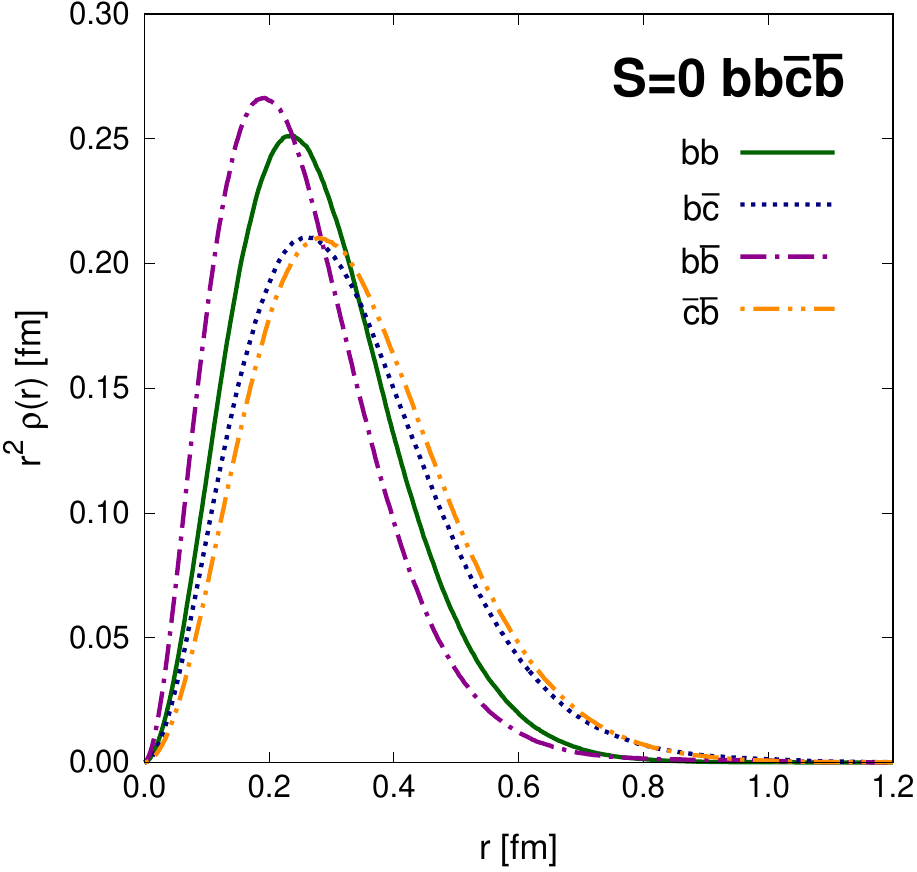}
\includegraphics[width=0.31\textwidth]{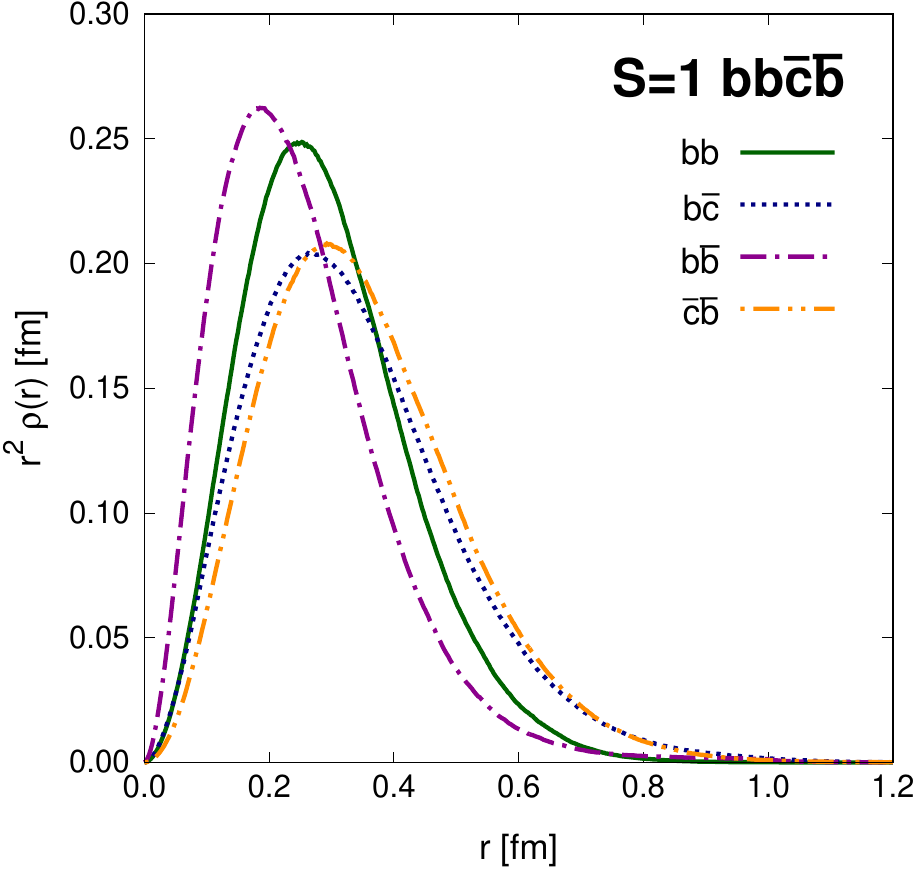}
\includegraphics[width=0.31\textwidth]{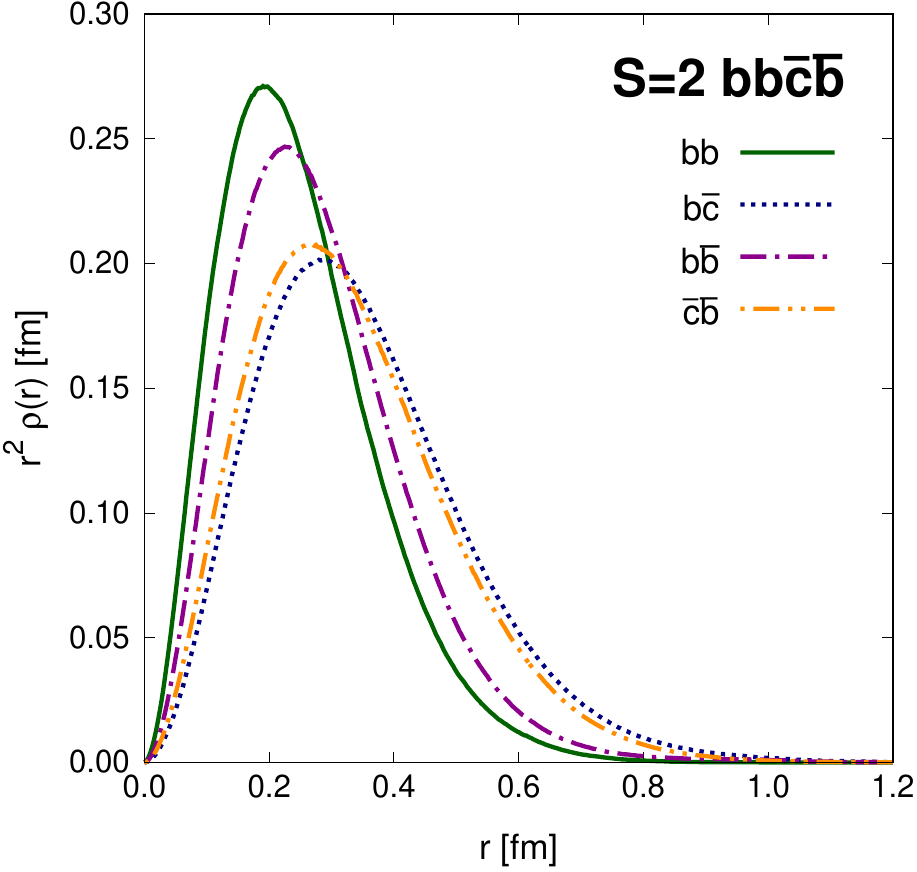} \\
\caption{\label{fig:cccb} Relevant radial distribution functions for the $J^P=0^+$, $1^+$ and $2^+$ ground states of the $cc\bar c\bar b$ (upper panels) and $bb\bar c\bar b$ (lower panels) systems.}
\end{figure*}

{\bf The $\mathbf{cc\bar c\bar b}$ and $\mathbf{bb\bar c\bar b}$ tetraquark ground states.} The $cc\bar c\bar b$, $bb\bar c\bar b$ (and $cb\bar c\bar b$) systems have not been studied before with neither the Hamiltonian used herein nor the variational method. Therefore, we cannot analyze in these cases the goodness of our numerical framework, diffusion Monte Carlo, with respect the variational one. We, however, shall compare our results with those obtained by other theoretical approaches, and numerical techniques, but such works are scarce because the complexity of the coupled-channels calculation needed to study the $cc\bar c\bar b$, $bb\bar c\bar b$ and $cb\bar c\bar b$ tetraquarks.

\begin{table}[!t]
\caption{\label{tab:cccbmass} Predicted masses, in MeV, for the $cc\bar{c}\bar{b}$ and $bb\bar{c}\bar{b}$ systems computed with the diffusion Monte Carlo technique and compared with those obtained by Refs.~\cite{Liu:2019zuc, Wu:2016vtq}.}
\begin{ruledtabular}
\begin{tabular}{cccc|ccc}
$J^{P}$ & DMC & \cite{Liu:2019zuc} & \cite{Wu:2016vtq} & DMC & \cite{Liu:2019zuc} & \cite{Wu:2016vtq} \\
\hline
\tstrut
$0^{+}$ & $9615$ & 9715 & 10144 & $16040$ & 16141 & 16823 \\
$1^{+}$ & $9610$ & 9727 & 10174 & $16013$ & 16148 & 16840 \\
$2^{+}$ & $9719$ & 9768 & 10273 & $16129$ & 16176 & 16917 \\
\end{tabular}
\end{ruledtabular}
\end{table}

The $cc\bar c\bar b$ and $bb\bar c\bar b$ systems share some common features in terms of heavy quark symmetry and thus they will be discussed together. Table~\ref{tab:cccbmass} shows our spectrum of ground states with quantum numbers $J^{P}=0^+$, $1^+$ and $2^+$. As one can see, in both tetraquark sectors, the $0^+$ and $1^+$ states are almost degenerate, with the $2^+$ ground state lying around $100\,\text{MeV}$ above them. The same picture is drawn by Ref.~\cite{Wu:2016vtq} while smaller mass splittings are predicted in Ref.~\cite{Liu:2019zuc}. It is interesting to observe, too, that our absolute figures are in disagreement with the other two cases reported in Table~\ref{tab:cccbmass}, with results of Ref.~\cite{Wu:2016vtq} much higher than ours and those of Ref.~\cite{Liu:2019zuc}; the latter being of the same order of magnitude than ours. Another feature needed to be mentioned is that the lowest strong-decay threshold is $\eta_c B_c$ ($\eta_b B_c$) for the $cc\bar c\bar b$ ($bb\bar c\bar b$) tetraquark sector and it is around $400\,\text{MeV}$ ($400\,\text{MeV}$) below the lowest-lying bound state; therefore, bound states of the $cc\bar c\bar b$ and $bb\bar c\bar b$ systems with narrow widths are not favored.

Figure~\ref{fig:cccb} shows the radial distribution functions for the $J^P=0^+$, $1^+$ and $2^+$ ground states of the $cc\bar c\bar b$ (upper panels) and $bb\bar c\bar b$ (lower panels) systems. The interested reader can observe how the different quark--(anti-)quark rearrangements are taking over as the total spin is getting higher. For instance, in both $cc\bar c\bar b$ and $bb\bar c\bar b$ tetraquark sectors, the (purple) dot-dashed curve which represents a quark-antiquark pair is the one that least extends in space for $S=0$ and $1$ indicating that quark-antiquark-pairs configuration could be important in such cases whereas the $S=2$ prefers the diquark-antidiquark-pairs.

The typical mean-square radii, $\langle r_{ij}^2 \rangle$, of the studied $cc\bar c\bar b$ (upper rows) and $bb\bar c\bar b$ (lower rows) tetraquarks are collected in Table~\ref{tab:radiiccbb}. For both tetraquark sectors, one can observe that the closest quark--(anti-)quark pair is the $c\bar b$ for $S=0$ and $S=1$ whereas is the $\bar c\bar b$ for $S=2$; indicating the change between quark-antiquark-pairs configuration and the one related with diquark-antidiquark pairs when the total spin grows. Finally, the mass mean-square radii for the $J^{P}=0^{+}$, $1^{+}$ and $2^{+}$ ground states are, respectively, $0.059\,\text{fm}^{2}$, $0.064\,\text{fm}^{2}$ and $0.066\,\text{fm}^{2}$ for $cc\bar c\bar b$ system, and $0.035\,\text{fm}^{2}$, $0.036\,\text{fm}^{2}$ and $0.037\,\text{fm}^{2}$ for $bb\bar c\bar b$ system.

\begin{table}[!t]
\begin{center}
\caption{\label{tab:radiiccbb} Mean-square radii, $\langle r_{ij}^2 \rangle$, of the studied $cc\bar c\bar b$ (upper rows) and $bb\bar c\bar b$ (lower rows) tetraquarks, in units of fm$^2$.}
\begin{tabular}{lcccc}
\hline
\hline
\tstrut
$J^{P}$ & $\langle r_{cc}^2 \rangle$ & $\langle r_{c\bar c}^2 \rangle$ & $\langle r_{c\bar b}^2 \rangle$ & $\langle r_{\bar c\bar b}^2 \rangle$ \\
\hline
\tstrut
$0^{+}$ & 0.225 & 0.202 & 0.155 & 0.181 \\
$1^{+}$ & 0.253 & 0.205 & 0.166 & 0.208 \\
$2^{+}$ & 0.217 & 0.241 & 0.191 & 0.159 \\
\hline
\hline
\tstrut
$J^{P}$ & $\langle r_{bb}^2 \rangle$ & $\langle r_{b\bar c}^2 \rangle$ & $\langle r_{b\bar b}^2 \rangle$ & $\langle r_{\bar c\bar b}^2 \rangle$ \\
\hline
\tstrut
$0^{+}$ & 0.097 & 0.130 & 0.076 & 0.097 \\
$1^{+}$ & 0.108 & 0.136 & 0.074 & 0.153 \\
$2^{+}$ & 0.076 & 0.148 & 0.095 & 0.136 \\
\hline
\hline
\end{tabular}
\end{center}
\end{table}


{\bf The $\mathbf{cb\bar c\bar b}$ tetraquark ground states.} The $cb\bar c\bar b$ system has no constraints from the Pauli principle. Therefore, looking at Table~\ref{tab:Configurations}, there are four spin-color configurations for the $J^{PC}=0^{++}$ ground state, another four in the case of the $J^{PC}=1^{+-}$, only two for the $J^{PC}=1^{++}$ channel, and two more for the $J^{PC}=2^{++}$ ground state. The predicted ground-state masses are listed in Table~\ref{tab:cbcbmass} and compared with the available results reported by other theoretical approaches~\cite{Liu:2019zuc, Berezhnoy:2011xn, Wu:2016vtq}. Our results, with masses at around $12.5\,\text{GeV}$, are in reasonable agreement with those of Ref.~\cite{Berezhnoy:2011xn}; however, both approaches predict figures which are slightly lower than the ones reported in Ref.~\cite{Liu:2019zuc}. The results published in Ref.~\cite{Wu:2016vtq} are systematically higher than others because the confinement term is ignored completely. As in the cases of $cc\bar c\bar b$ and $bb\bar c\bar b$ tetraquarks, the $J=0$ and the lowest $J=1$ states are almost degenerate with even the $J^{PC}=1^{+-}$ $cb\bar c\bar b$ ground state located below the $J^{PC}=0^{++}$ one. At this stage of our work, it could be necessary to remind that our calculation is parameter-free once the model is fitted to the meson and baryon sector. 

The lowest $S$-wave meson-meson thresholds in the $cb\bar c\bar b$ tetraquark sector are the $\eta_c\eta_b$ $(12429\,\text{MeV})$ for the $J^{PC}=0^{++}$ channel, $\eta_c\Upsilon(1S)$ $(12467\,\text{MeV})$ for the $J^{PC}=1^{+-}$ channel, and $J/\psi\Upsilon(1S)$ $(12563\,\text{MeV})$ for the $J^{PC}=1^{++}$ and $2^{++}$ channels. We are predicting tetraquark ground state masses which lie $\lesssim100\,\text{MeV}$ above their lowest open-flavor $S$-wave meson-meson threshold, \emph{i.e.} they could appear as resonance candidates in the invariant mass of hteir corresponding meson-meson channel. It is worth emphasizing that the $J^{PC}=1^{++}$ ground state is located at exactly, within theoretical uncertainty, its lowest $S$-wave meson-meson threshold.

\begin{table}[!t]
\caption{\label{tab:cbcbmass} Predicted masses, in MeV, for the $cb\bar{c}\bar{b}$ system computed with the Diffusion Monte Carlo technique and compared with those obtained by other theoretical frameworks.}
\begin{ruledtabular}
\begin{tabular}{ccccc}
$J^{PC}$ & DMC & \cite{Liu:2019zuc} & \cite{Berezhnoy:2011xn} & \cite{Wu:2016vtq} \\
\hline
$0^{++}$ & $12534$ & 12854 & 12359 & 13396 \\
$1^{+-}$ & $12510$ & 12881 & 12424 & 13478 \\
$1^{++}$ & $12569$ & 12933 & 12485 & 13510 \\
$2^{++}$ & $12582$ & 12933 & 12566 & 13590 \\
\end{tabular}
\end{ruledtabular}
\end{table}

\begin{figure*}
\includegraphics[width=0.45\textwidth]{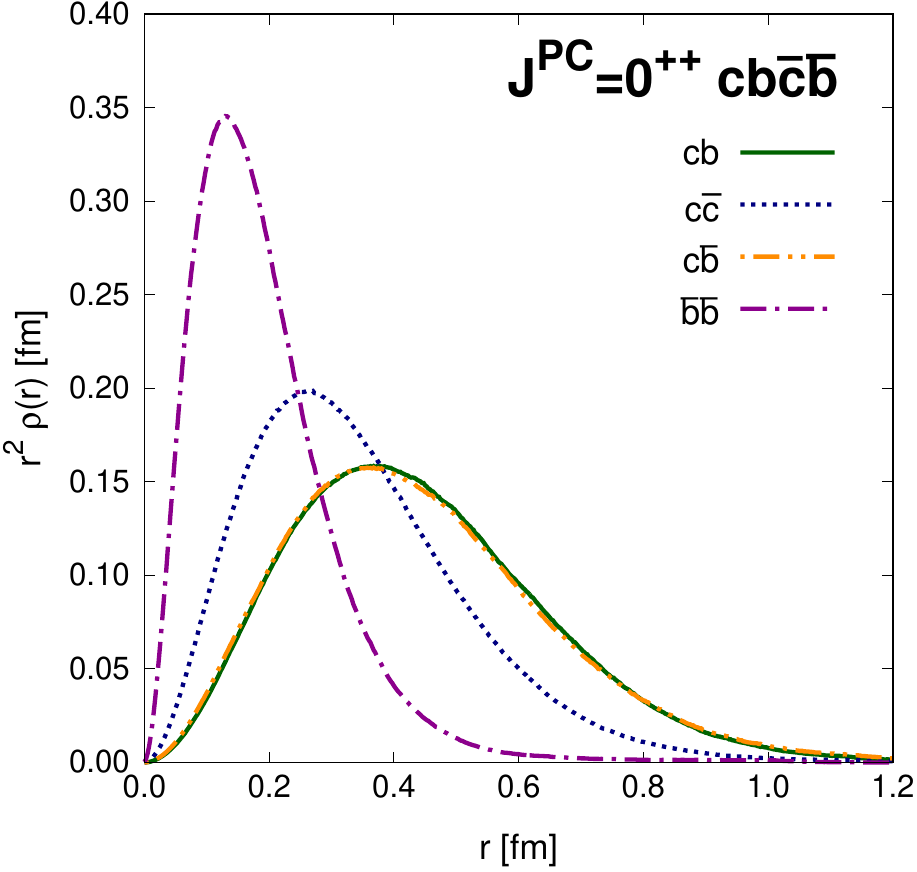}
\includegraphics[width=0.45\textwidth]{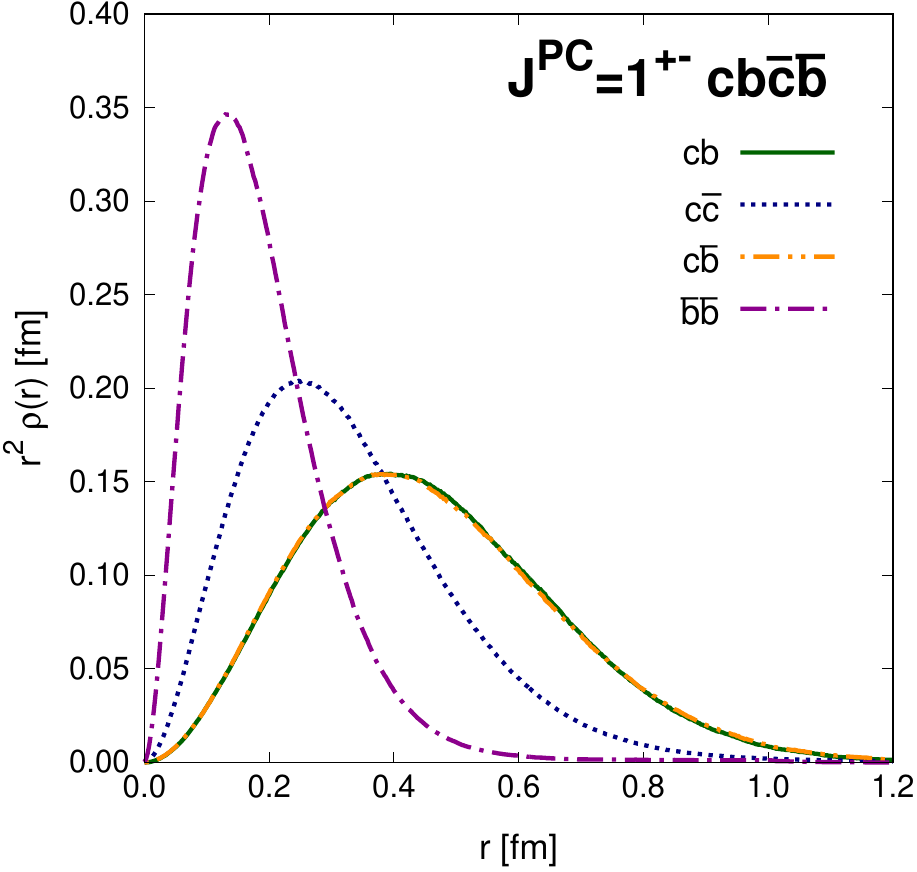} \\
\includegraphics[width=0.45\textwidth]{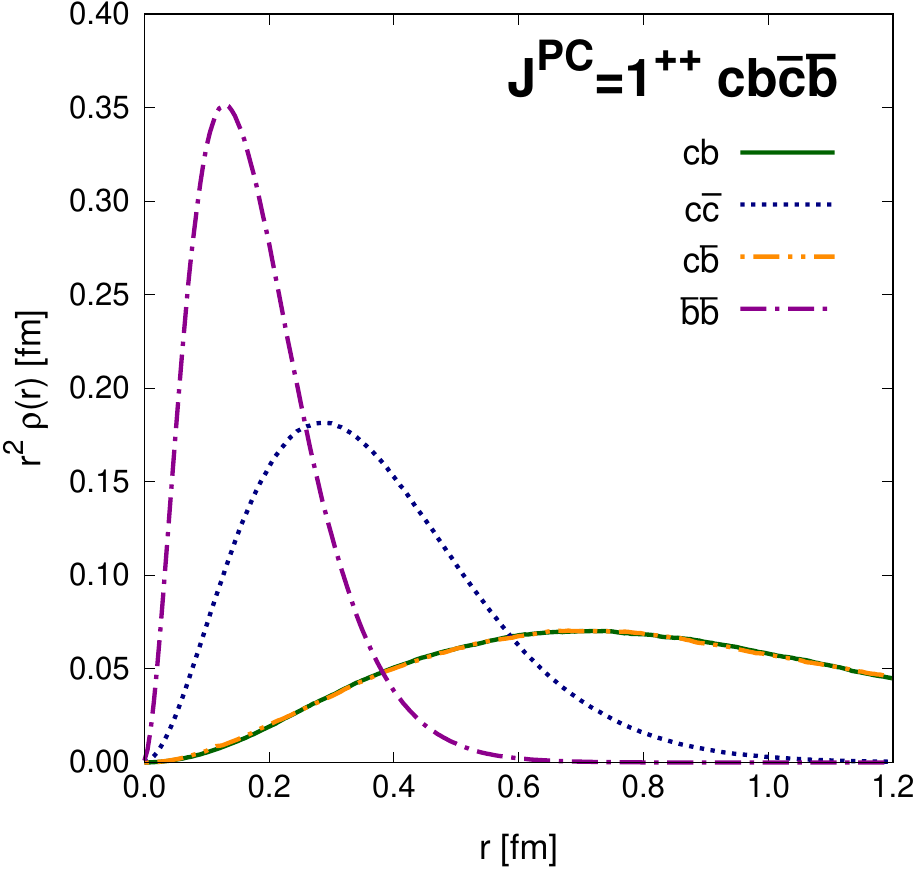}
\includegraphics[width=0.45\textwidth]{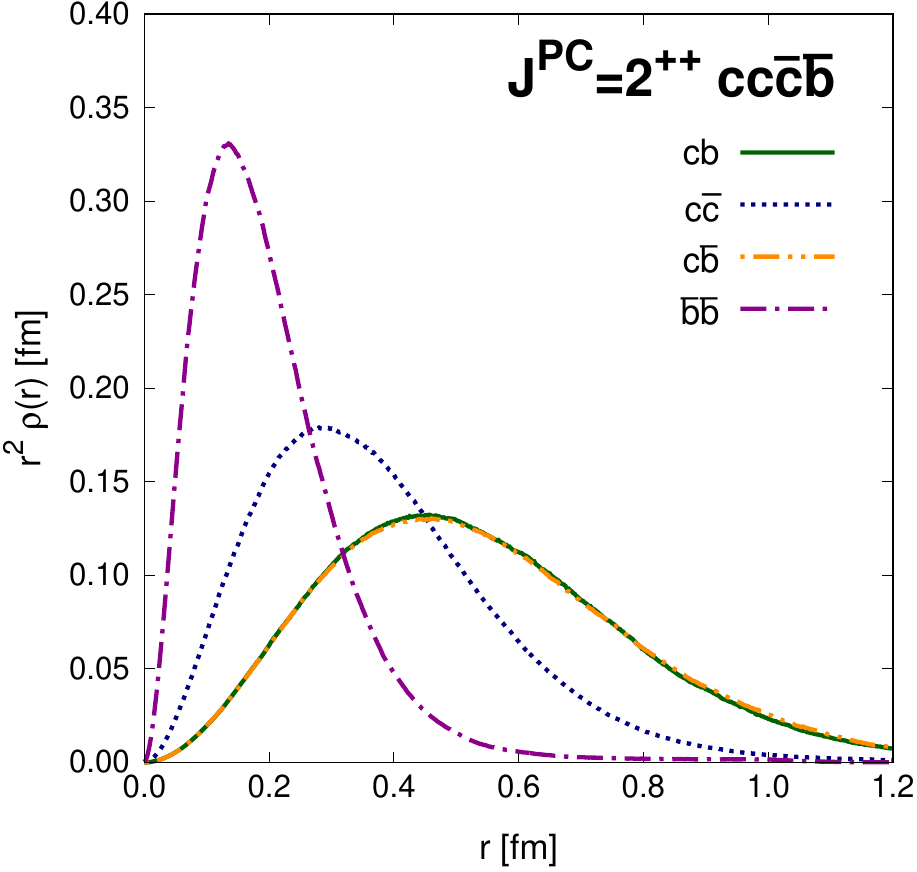}
\caption{\label{fig:cbcb} Relevant radial distribution functions for the $J^{PC}=0^{++}$, $1^{+-}$, $1^{++}$ and $2^{++}$ ground states of the $cb\bar c\bar b$ system.}
\end{figure*}

Figure~\ref{fig:cbcb} shows the relevant radial distribution functions for the $J^{PC}=0^{++}$, $1^{+-}$, $1^{++}$ and $2^{++}$ ground states of the $cb\bar c\bar b$ system. One can see that the radial distribution functions follow the same pattern for all ground states, \emph{i.e.} the interquark distance of the $b\bar b$-pair is the shortest one, it is followed by the one of the $c\bar c$-pair, and finally the distances between the $c$-quark and either the $\bar b$- or $b$-quark are the same and the largest. However, the left bottom panel of Fig.~\ref{fig:cbcb} shows that the $J^{PC}=1^{++}$ $cb\bar c\bar b$ ground state is particularly different with respect the others, with a more extended radial distribution for the $cb$ and $c\bar b$ pairs. In fact, such distribution is indicating that the interquark distance between the $c$- and $b$-quarks is of the order of $1\,\text{fm}$, whereas the $c\bar c$- and $b\bar b$-pairs are clustered within a distance of $0.5\,\text{fm}$ or less. Therefore, it seems that the $J^{PC}=1^{++}$ $cb\bar c\bar b$ tetraquark ground state prefers to be in a meson-meson configuration.

\begin{table}[!t]
\begin{center}
\caption{\label{tab:radiicbcb} Mean-square radii, $\langle r_{ij}^2 \rangle$, of the studied $cb\bar c\bar b$ tetraquarks, in units of fm$^2$.}
\begin{tabular}{lcccc}
\hline
\hline
\tstrut
$J^{PC}$ & $\langle r_{cb}^2 \rangle$ & $\langle r_{c\bar c}^2 \rangle$ & $\langle r_{c\bar b}^2 \rangle$ & $\langle r_{b\bar b}^2 \rangle$ \\
\hline
\tstrut
$0^{++}$ & 0.233 & 0.142 & 0.233 & 0.045 \\
$1^{+-}$ & 0.245 & 0.134 & 0.245 & 0.045 \\
$1^{++}$ & 1.397 & 0.171 & 1.397 & 0.044 \\
$2^{++}$ & 0.345 & 0.172 & 0.344 & 0.050 \\
\hline
\hline
\end{tabular}
\end{center}
\end{table}

In order to quantify our statements above, Table~\ref{tab:radiiccbb} shows the typical mean-square radii, $\langle r_{ij}^2 \rangle$, of the studied $cb\bar c\bar b$ tetraquark ground states. In the case of the $J^{PC}=1^{++}$ ground state, a mean-square radii larger than $1\,\text{fm}^2$ is obtained for the $cb$ and $c\bar b$ pairs, whereas the others are $\sim0.1\,\text{fm}^{2}$. This may indicate, in contrast with the other cases studied, that $c\bar c$ and $b\bar b$-pairs tend to form separated clusters within a distance of $1\,\text{fm}$. Furthermore, we report herein the mass mean-square radii for the $J^{PC}=0^{++}$, $1^{+-}$, $1^{++}$ and $2^{++}$ ground states which are, respectively, $0.053\,\text{fm}^{2}$, $0.055\,\text{fm}^{2}$, $0.3\,\text{fm}^{2}$ and $0.076\,\text{fm}^{2}$. One can see again that the $J^{PC}=1^{++}$ $cb\bar c\bar b$ ground state is twice more extended than the others.

Our results suggest that the $J^{PC}=1^{++}$ ground state is remarkably different with respect to the other cases studied because, on one hand, its mass lies exactly at its lowest $S$-wave meson-meson threshold and, on the other hand, clusters of $Q\bar Q$-pairs (with Q either $c$ or $b$) appear with a separation between them of the order of $1\,\text{fm}$. We believe that both features are intimately related and further studies shall be performed, which go beyond the scope of this manuscript. 


\section{Summary and outlook}
\label{sec:Epilogue}

Fully-heavy tetraquarks have recently received considerable attention from experiment. The most significant example is the observation made by the LHCb collaboration of some enhancements in the $J/\psi$-pair invariant mass spectrum whose origin could be linked to hadron states consisting of four charm quarks. Moreover, one should expect that the searching of doubly hidden-bottom and -charm tetraquark states will probably become one of the most attractive experimental goals with the future running of BES~III, Belle~II, and LHC.

From the theoretical side, several approaches have been proposed to calculate the spectrum of tetraquark systems made up only by heavy quarks. Their main goal was to established theoretically the existence of fully-heavy tetraquarks with narrow widths, \emph{i.e.} stable. Mixed results have been obtained with some theoretical studies claiming that some lowest-lying all-heavy tetraquarks could be located slightly lower than their respective thresholds of quarkonium pairs and others asserting that fully-heavy tetraquarks are located much higher in mass than their lowest possible meson-meson strong decay channel.

In order to contribute on a better understanding of the multiquark dynamics, we have used a diffusion Monte Carlo method to solve the many-body Schr\"odinger equation that describes the fully-heavy tetraquark systems. This approach allows to reduce the uncertainty of the numerical calculation, accounts for multi-particle correlations in the physical observables, and avoids the usual quark-clustering assumed in other theoretical techniques applied to the same problem. Moreover, Quantum Monte Carlo computations shall enable to scale the same bound-state problem to other multiquark systems such as pentaquarks, hexaquarks, etc.

The used quark model Hamiltonian has a pairwise interaction which is the most general and accepted one: Coulomb$\,+\,$linear-confining$\,+\,$hyperfine spin-spin; therefore, our analysis should provide some rigorous statements about the mass location of the all-heavy tetraquark ground states. Note, too, that such conclusions are parameter-free because the model parameters were constrained by a simultaneous fit of $36$ mesons and $53$ baryons, with a range of agreement between theory and experiment around $10-20\%$, which can be taken as an estimation of the model uncertainty for fully-heavy tetraquarks.

The $cc\bar c\bar c$, $cc\bar b\bar b$ ($bb\bar c\bar c$) and $bb\bar b \bar b$ lowest-lying states are located just in the middle of the mass ranges predicted by other theoretical approaches. All states appear above their corresponding meson-meson thresholds and thus the existence of stable $cc\bar c\bar c$, $cc\bar b\bar b$ ($bb\bar c\bar c$) and $bb\bar b \bar b$ systems with very narrow widths is disfavored; nevertheless, this does not forbid to have resonances in these tetraquark sectors which can be experimentally observed in the near future. Interestingly too is the observation that there is a transition between quark-antiquark pairs and diquark-antidiquark ones when going from $S=0$ to $S=2$ in the $QQ\bar Q\bar Q$ with all $Q$ either $c$- or $b$-quark, but not in the $cc\bar b\bar b$ ($bb\bar c\bar c$) sector. However, it is important to clarify that these states are compact ones, with sizes in the order of a typical hadron. At last, the $J^{PC}=0^{++}$ $cc\bar c\bar c$ ground state is predicted to have a mass compatible with the enhancements observed by the LHCb collaboration.

Theoretical studies of the $cc\bar c\bar b$, $bb\bar c\bar b$ and $cb\bar c\bar b$ systems are scarce because the complexity of the needed coupled-channels calculation. For the $cc\bar c\bar b$ and $bb\bar c\bar b$ sectors, our results seem to indicate that the $0^+$ and $1^+$ ground states are almost degenerate, with the $2^+$ lowest-lying state located around $100\,\text{MeV}$ above them. For the $cb\bar c\bar b$ system, we predict small mass splittings between the studied bound-states and absolute mass values located in between those predicted by other theoretical works. Moreover, we found clear evidence that the $J^{PC}=1^{++}$ $cb\bar c\bar b$ ground state has a meson-meson molecular configuration which deserves to be investigated further.

Finally, the diffusion Monte Carlo method is, in principle, applicable to a wide range of related problems and open questions. For instance, some natural extensions of the work presented herein could be the analysis of excited states and the exploration of other multiquark systems such as pentaquarks, hexaquarks, etc. On the other hand, answering some complex questions related with few- and many-body hadron physics appears scientifically interesting such as what would be the binding energy per quark in a many-body bound-state system?, is there any limit in the number of quarks and antiquarks that a hadron could host?, or could other constituents play a role in the stability of exotic hadrons?


\begin{acknowledgments}
We thank A. Lovato, J.M. Morgado, C.D. Roberts, J. Rodr\'iguez-Quintero for constructive discussions and continuous support. 
This work has been partially funded by the Ministerio Espa\~nol de Ciencia e Innovaci\'on under grant No. PID2019-107844GB-C22 and FIS2017-84114-C2-2-P; the Junta de Andaluc\'ia under contract No. Operativo FEDER Andaluc\'ia 2014-2020 UHU-1264517; but also PAIDI FQM-205 and -370. The authors acknowledges, too, the use of the computer facilities of C3UPO at the Universidad Pablo de Olavide, de Sevilla.
\end{acknowledgments}


\bibliography{MultiquarksDMC}

\begin{thebibliography}{95}%
\makeatletter
\providecommand \@ifxundefined [1]{%
 \@ifx{#1\undefined}
}%
\providecommand \@ifnum [1]{%
 \ifnum #1\expandafter \@firstoftwo
 \else \expandafter \@secondoftwo
 \fi
}%
\providecommand \@ifx [1]{%
 \ifx #1\expandafter \@firstoftwo
 \else \expandafter \@secondoftwo
 \fi
}%
\providecommand \natexlab [1]{#1}%
\providecommand \enquote  [1]{``#1''}%
\providecommand \bibnamefont  [1]{#1}%
\providecommand \bibfnamefont [1]{#1}%
\providecommand \citenamefont [1]{#1}%
\providecommand \href@noop [0]{\@secondoftwo}%
\providecommand \href [0]{\begingroup \@sanitize@url \@href}%
\providecommand \@href[1]{\@@startlink{#1}\@@href}%
\providecommand \@@href[1]{\endgroup#1\@@endlink}%
\providecommand \@sanitize@url [0]{\catcode `\\12\catcode `\$12\catcode
  `\&12\catcode `\#12\catcode `\^12\catcode `\_12\catcode `\%12\relax}%
\providecommand \@@startlink[1]{}%
\providecommand \@@endlink[0]{}%
\providecommand \url  [0]{\begingroup\@sanitize@url \@url }%
\providecommand \@url [1]{\endgroup\@href {#1}{\urlprefix }}%
\providecommand \urlprefix  [0]{URL }%
\providecommand \Eprint [0]{\href }%
\providecommand \doibase [0]{http://dx.doi.org/}%
\providecommand \selectlanguage [0]{\@gobble}%
\providecommand \bibinfo  [0]{\@secondoftwo}%
\providecommand \bibfield  [0]{\@secondoftwo}%
\providecommand \translation [1]{[#1]}%
\providecommand \BibitemOpen [0]{}%
\providecommand \bibitemStop [0]{}%
\providecommand \bibitemNoStop [0]{.\EOS\space}%
\providecommand \EOS [0]{\spacefactor3000\relax}%
\providecommand \BibitemShut  [1]{\csname bibitem#1\endcsname}%
\let\auto@bib@innerbib\@empty
\bibitem [{\citenamefont {Aubert}\ \emph {et~al.}(1974)\citenamefont {Aubert}
  \emph {et~al.}}]{Aubert:1974js}%
  \BibitemOpen
  \bibfield  {author} {\bibinfo {author} {\bibfnamefont {J.}~\bibnamefont
  {Aubert}} \emph {et~al.} (\bibinfo {collaboration} {E598}),\ }\href {\doibase
  10.1103/PhysRevLett.33.1404} {\bibfield  {journal} {\bibinfo  {journal}
  {Phys. Rev. Lett.}\ }\textbf {\bibinfo {volume} {33}},\ \bibinfo {pages}
  {1404} (\bibinfo {year} {1974})}\BibitemShut {NoStop}%
\bibitem [{\citenamefont {Augustin}\ \emph {et~al.}(1974)\citenamefont
  {Augustin} \emph {et~al.}}]{Augustin:1974xw}%
  \BibitemOpen
  \bibfield  {author} {\bibinfo {author} {\bibfnamefont {J.}~\bibnamefont
  {Augustin}} \emph {et~al.} (\bibinfo {collaboration} {SLAC-SP-017}),\ }\href
  {\doibase 10.1103/PhysRevLett.33.1406} {\bibfield  {journal} {\bibinfo
  {journal} {Phys. Rev. Lett.}\ }\textbf {\bibinfo {volume} {33}},\ \bibinfo
  {pages} {1406} (\bibinfo {year} {1974})}\BibitemShut {NoStop}%
\bibitem [{\citenamefont {Herb}\ \emph {et~al.}(1977)\citenamefont {Herb},
  \citenamefont {Hom}, \citenamefont {Lederman}, \citenamefont {Sens},
  \citenamefont {Snyder} \emph {et~al.}}]{Herb:1977ek}%
  \BibitemOpen
  \bibfield  {author} {\bibinfo {author} {\bibfnamefont {S.}~\bibnamefont
  {Herb}}, \bibinfo {author} {\bibfnamefont {D.}~\bibnamefont {Hom}}, \bibinfo
  {author} {\bibfnamefont {L.}~\bibnamefont {Lederman}}, \bibinfo {author}
  {\bibfnamefont {J.}~\bibnamefont {Sens}}, \bibinfo {author} {\bibfnamefont
  {H.}~\bibnamefont {Snyder}},  \emph {et~al.},\ }\href {\doibase
  10.1103/PhysRevLett.39.252} {\bibfield  {journal} {\bibinfo  {journal} {Phys.
  Rev. Lett.}\ }\textbf {\bibinfo {volume} {39}},\ \bibinfo {pages} {252}
  (\bibinfo {year} {1977})}\BibitemShut {NoStop}%
\bibitem [{\citenamefont {Innes}\ \emph {et~al.}(1977)\citenamefont {Innes},
  \citenamefont {Appel}, \citenamefont {Brown}, \citenamefont {Brown},
  \citenamefont {Ueno} \emph {et~al.}}]{Innes:1977ae}%
  \BibitemOpen
  \bibfield  {author} {\bibinfo {author} {\bibfnamefont {W.~R.}\ \bibnamefont
  {Innes}}, \bibinfo {author} {\bibfnamefont {J.}~\bibnamefont {Appel}},
  \bibinfo {author} {\bibfnamefont {B.}~\bibnamefont {Brown}}, \bibinfo
  {author} {\bibfnamefont {C.}~\bibnamefont {Brown}}, \bibinfo {author}
  {\bibfnamefont {K.}~\bibnamefont {Ueno}},  \emph {et~al.},\ }\href {\doibase
  10.1103/PhysRevLett.39.1240} {\bibfield  {journal} {\bibinfo  {journal}
  {Phys. Rev. Lett.}\ }\textbf {\bibinfo {volume} {39}},\ \bibinfo {pages}
  {1240} (\bibinfo {year} {1977})}\BibitemShut {NoStop}%
\bibitem [{\citenamefont {Appelquist}\ and\ \citenamefont
  {Politzer}(1975)}]{Appelquist:1974zd}%
  \BibitemOpen
  \bibfield  {author} {\bibinfo {author} {\bibfnamefont {T.}~\bibnamefont
  {Appelquist}}\ and\ \bibinfo {author} {\bibfnamefont {H.}~\bibnamefont
  {Politzer}},\ }\href {\doibase 10.1103/PhysRevLett.34.43} {\bibfield
  {journal} {\bibinfo  {journal} {Phys. Rev. Lett.}\ }\textbf {\bibinfo
  {volume} {34}},\ \bibinfo {pages} {43} (\bibinfo {year} {1975})}\BibitemShut
  {NoStop}%
\bibitem [{\citenamefont {De~Rujula}\ and\ \citenamefont
  {Glashow}(1975)}]{DeRujula:1974rkb}%
  \BibitemOpen
  \bibfield  {author} {\bibinfo {author} {\bibfnamefont {A.}~\bibnamefont
  {De~Rujula}}\ and\ \bibinfo {author} {\bibfnamefont {S.}~\bibnamefont
  {Glashow}},\ }\href {\doibase 10.1103/PhysRevLett.34.46} {\bibfield
  {journal} {\bibinfo  {journal} {Phys. Rev. Lett.}\ }\textbf {\bibinfo
  {volume} {34}},\ \bibinfo {pages} {46} (\bibinfo {year} {1975})}\BibitemShut
  {NoStop}%
\bibitem [{\citenamefont {Abe}\ \emph {et~al.}(1998)\citenamefont {Abe} \emph
  {et~al.}}]{Abe:1998wi}%
  \BibitemOpen
  \bibfield  {author} {\bibinfo {author} {\bibfnamefont {F.}~\bibnamefont
  {Abe}} \emph {et~al.} (\bibinfo {collaboration} {CDF}),\ }\href {\doibase
  10.1103/PhysRevLett.81.2432} {\bibfield  {journal} {\bibinfo  {journal}
  {Phys. Rev. Lett.}\ }\textbf {\bibinfo {volume} {81}},\ \bibinfo {pages}
  {2432} (\bibinfo {year} {1998})},\ \Eprint
  {http://arxiv.org/abs/hep-ex/9805034} {arXiv:hep-ex/9805034} \BibitemShut
  {NoStop}%
\bibitem [{\citenamefont {Abe}\ \emph {et~al.}(1994)\citenamefont {Abe} \emph
  {et~al.}}]{Abe:1994xt}%
  \BibitemOpen
  \bibfield  {author} {\bibinfo {author} {\bibfnamefont {F.}~\bibnamefont
  {Abe}} \emph {et~al.} (\bibinfo {collaboration} {CDF}),\ }\href {\doibase
  10.1103/PhysRevLett.73.225} {\bibfield  {journal} {\bibinfo  {journal} {Phys.
  Rev. Lett.}\ }\textbf {\bibinfo {volume} {73}},\ \bibinfo {pages} {225}
  (\bibinfo {year} {1994})},\ \Eprint {http://arxiv.org/abs/hep-ex/9405005}
  {arXiv:hep-ex/9405005} \BibitemShut {NoStop}%
\bibitem [{\citenamefont {Ore}\ and\ \citenamefont
  {Powell}(1949)}]{Ore:1949te}%
  \BibitemOpen
  \bibfield  {author} {\bibinfo {author} {\bibfnamefont {A.}~\bibnamefont
  {Ore}}\ and\ \bibinfo {author} {\bibfnamefont {J.}~\bibnamefont {Powell}},\
  }\href {\doibase 10.1103/PhysRev.75.1696} {\bibfield  {journal} {\bibinfo
  {journal} {Phys. Rev.}\ }\textbf {\bibinfo {volume} {75}},\ \bibinfo {pages}
  {1696} (\bibinfo {year} {1949})}\BibitemShut {NoStop}%
\bibitem [{\citenamefont {Deutsch}(1951)}]{Deutsch:1951zza}%
  \BibitemOpen
  \bibfield  {author} {\bibinfo {author} {\bibfnamefont {M.}~\bibnamefont
  {Deutsch}},\ }\href {\doibase 10.1103/PhysRev.82.455} {\bibfield  {journal}
  {\bibinfo  {journal} {Phys. Rev.}\ }\textbf {\bibinfo {volume} {82}},\
  \bibinfo {pages} {455} (\bibinfo {year} {1951})}\BibitemShut {NoStop}%
\bibitem [{\citenamefont {Gell-Mann}(1964)}]{GellMann:1964nj}%
  \BibitemOpen
  \bibfield  {author} {\bibinfo {author} {\bibfnamefont {M.}~\bibnamefont
  {Gell-Mann}},\ }\href {\doibase 10.1016/S0031-9163(64)92001-3} {\bibfield
  {journal} {\bibinfo  {journal} {Phys.\ Lett.}\ }\textbf {\bibinfo {volume}
  {8}},\ \bibinfo {pages} {214} (\bibinfo {year} {1964})}\BibitemShut {NoStop}%
\bibitem [{\citenamefont {Zweig}(1964)}]{zweigcern}%
  \BibitemOpen
  \bibfield  {author} {\bibinfo {author} {\bibfnamefont {G.}~\bibnamefont
  {Zweig}},\ }\href@noop {} {\bibfield  {journal} {\bibinfo  {journal}
  {CERN-TH-412, NP-14146}\ } (\bibinfo {year} {1964})}\BibitemShut {NoStop}%
\bibitem [{\citenamefont {Tanabashi}\ \emph {et~al.}(2018)\citenamefont
  {Tanabashi} \emph {et~al.}}]{Tanabashi:2018oca}%
  \BibitemOpen
  \bibfield  {author} {\bibinfo {author} {\bibfnamefont {M.}~\bibnamefont
  {Tanabashi}} \emph {et~al.} (\bibinfo {collaboration} {Particle Data
  Group}),\ }\href {\doibase 10.1103/PhysRevD.98.030001} {\bibfield  {journal}
  {\bibinfo  {journal} {Phys. Rev. D}\ }\textbf {\bibinfo {volume} {98}},\
  \bibinfo {pages} {030001} (\bibinfo {year} {2018})}\BibitemShut {NoStop}%
\bibitem [{\citenamefont {Brambilla}\ \emph {et~al.}(2011)\citenamefont
  {Brambilla} \emph {et~al.}}]{Brambilla:2010cs}%
  \BibitemOpen
  \bibfield  {author} {\bibinfo {author} {\bibfnamefont {N.}~\bibnamefont
  {Brambilla}} \emph {et~al.},\ }\href {\doibase
  10.1140/epjc/s10052-010-1534-9} {\bibfield  {journal} {\bibinfo  {journal}
  {Eur. Phys. J. C}\ }\textbf {\bibinfo {volume} {71}},\ \bibinfo {pages}
  {1534} (\bibinfo {year} {2011})},\ \Eprint {http://arxiv.org/abs/1010.5827}
  {arXiv:1010.5827 [hep-ph]} \BibitemShut {NoStop}%
\bibitem [{\citenamefont {Olsen}(2015)}]{Olsen:2014qna}%
  \BibitemOpen
  \bibfield  {author} {\bibinfo {author} {\bibfnamefont {S.~L.}\ \bibnamefont
  {Olsen}},\ }\href {\doibase 10.1007/S11467-014-0449-6} {\bibfield  {journal}
  {\bibinfo  {journal} {Front. Phys. (Beijing)}\ }\textbf {\bibinfo {volume}
  {10}},\ \bibinfo {pages} {121} (\bibinfo {year} {2015})},\ \Eprint
  {http://arxiv.org/abs/1411.7738} {arXiv:1411.7738 [hep-ex]} \BibitemShut
  {NoStop}%
\bibitem [{\citenamefont {Olsen}\ \emph {et~al.}(2018)\citenamefont {Olsen},
  \citenamefont {Skwarnicki},\ and\ \citenamefont {Zieminska}}]{Olsen:2017bmm}%
  \BibitemOpen
  \bibfield  {author} {\bibinfo {author} {\bibfnamefont {S.~L.}\ \bibnamefont
  {Olsen}}, \bibinfo {author} {\bibfnamefont {T.}~\bibnamefont {Skwarnicki}}, \
  and\ \bibinfo {author} {\bibfnamefont {D.}~\bibnamefont {Zieminska}},\ }\href
  {\doibase 10.1103/RevModPhys.90.015003} {\bibfield  {journal} {\bibinfo
  {journal} {Rev. Mod. Phys.}\ }\textbf {\bibinfo {volume} {90}},\ \bibinfo
  {pages} {015003} (\bibinfo {year} {2018})},\ \Eprint
  {http://arxiv.org/abs/1708.04012} {arXiv:1708.04012 [hep-ph]} \BibitemShut
  {NoStop}%
\bibitem [{\citenamefont {Brambilla}\ \emph {et~al.}(2019)\citenamefont
  {Brambilla}, \citenamefont {Eidelman}, \citenamefont {Hanhart}, \citenamefont
  {Nefediev}, \citenamefont {Shen}, \citenamefont {Thomas}, \citenamefont
  {Vairo},\ and\ \citenamefont {Yuan}}]{Brambilla:2019esw}%
  \BibitemOpen
  \bibfield  {author} {\bibinfo {author} {\bibfnamefont {N.}~\bibnamefont
  {Brambilla}}, \bibinfo {author} {\bibfnamefont {S.}~\bibnamefont {Eidelman}},
  \bibinfo {author} {\bibfnamefont {C.}~\bibnamefont {Hanhart}}, \bibinfo
  {author} {\bibfnamefont {A.}~\bibnamefont {Nefediev}}, \bibinfo {author}
  {\bibfnamefont {C.-P.}\ \bibnamefont {Shen}}, \bibinfo {author}
  {\bibfnamefont {C.~E.}\ \bibnamefont {Thomas}}, \bibinfo {author}
  {\bibfnamefont {A.}~\bibnamefont {Vairo}}, \ and\ \bibinfo {author}
  {\bibfnamefont {C.-Z.}\ \bibnamefont {Yuan}},\ }\href@noop {} {\  (\bibinfo
  {year} {2019})},\ \Eprint {http://arxiv.org/abs/1907.07583} {arXiv:1907.07583
  [hep-ex]} \BibitemShut {NoStop}%
\bibitem [{\citenamefont {Asner}\ \emph {et~al.}(2009)\citenamefont {Asner}
  \emph {et~al.}}]{Asner:2008nq}%
  \BibitemOpen
  \bibfield  {author} {\bibinfo {author} {\bibfnamefont {D.}~\bibnamefont
  {Asner}} \emph {et~al.},\ }\href@noop {} {\bibfield  {journal} {\bibinfo
  {journal} {Int. J. Mod. Phys. A}\ }\textbf {\bibinfo {volume} {24}},\
  \bibinfo {pages} {S1} (\bibinfo {year} {2009})},\ \Eprint
  {http://arxiv.org/abs/0809.1869} {arXiv:0809.1869 [hep-ex]} \BibitemShut
  {NoStop}%
\bibitem [{\citenamefont {Bevan}\ \emph {et~al.}(2014)\citenamefont {Bevan}
  \emph {et~al.}}]{Bevan:2014iga}%
  \BibitemOpen
  \bibfield  {author} {\bibinfo {author} {\bibfnamefont {A.}~\bibnamefont
  {Bevan}} \emph {et~al.} (\bibinfo {collaboration} {BaBar, Belle}),\ }\href
  {\doibase 10.1140/epjc/s10052-014-3026-9} {\bibfield  {journal} {\bibinfo
  {journal} {Eur. Phys. J. C}\ }\textbf {\bibinfo {volume} {74}},\ \bibinfo
  {pages} {3026} (\bibinfo {year} {2014})},\ \Eprint
  {http://arxiv.org/abs/1406.6311} {arXiv:1406.6311 [hep-ex]} \BibitemShut
  {NoStop}%
\bibitem [{\citenamefont {Cerri}\ \emph {et~al.}(2019)\citenamefont {Cerri}
  \emph {et~al.}}]{Cerri:2018ypt}%
  \BibitemOpen
  \bibfield  {author} {\bibinfo {author} {\bibfnamefont {A.}~\bibnamefont
  {Cerri}} \emph {et~al.},\ }\enquote {\bibinfo {title} {{Report from Working
  Group 4}: {Opportunities in Flavour Physics at the HL-LHC and HE-LHC}},}\ in\
  \href {\doibase 10.23731/CYRM-2019-007.867} {\emph {\bibinfo {booktitle}
  {{Report on the Physics at the HL-LHC,and Perspectives for the HE-LHC}}}},\
  Vol.~\bibinfo {volume} {7},\ \bibinfo {editor} {edited by\ \bibinfo {editor}
  {\bibfnamefont {A.}~\bibnamefont {Dainese}}, \bibinfo {editor} {\bibfnamefont
  {M.}~\bibnamefont {Mangano}}, \bibinfo {editor} {\bibfnamefont {A.~B.}\
  \bibnamefont {Meyer}}, \bibinfo {editor} {\bibfnamefont {A.}~\bibnamefont
  {Nisati}}, \bibinfo {editor} {\bibfnamefont {G.}~\bibnamefont {Salam}}, \
  and\ \bibinfo {editor} {\bibfnamefont {M.~A.}\ \bibnamefont {Vesterinen}}}\
  (\bibinfo {year} {2019})\ pp.\ \bibinfo {pages} {867--1158},\ \Eprint
  {http://arxiv.org/abs/1812.07638} {arXiv:1812.07638 [hep-ph]} \BibitemShut
  {NoStop}%
\bibitem [{\citenamefont {Choi}\ \emph {et~al.}(2003)\citenamefont {Choi} \emph
  {et~al.}}]{Choi:2003ue}%
  \BibitemOpen
  \bibfield  {author} {\bibinfo {author} {\bibfnamefont {S.-K.}\ \bibnamefont
  {Choi}} \emph {et~al.} (\bibinfo {collaboration} {Belle}),\ }\href {\doibase
  10.1103/PhysRevLett.91.262001} {\bibfield  {journal} {\bibinfo  {journal}
  {Phys.Rev.Lett.}\ }\textbf {\bibinfo {volume} {91}},\ \bibinfo {pages}
  {262001} (\bibinfo {year} {2003})},\ \Eprint
  {http://arxiv.org/abs/hep-ex/0309032} {arXiv:hep-ex/0309032} \BibitemShut
  {NoStop}%
\bibitem [{\citenamefont {Aubert}\ \emph {et~al.}(2005)\citenamefont {Aubert}
  \emph {et~al.}}]{Aubert:2004ns}%
  \BibitemOpen
  \bibfield  {author} {\bibinfo {author} {\bibfnamefont {B.}~\bibnamefont
  {Aubert}} \emph {et~al.} (\bibinfo {collaboration} {Babar}),\ }\href
  {\doibase 10.1103/PhysRevD.71.071103} {\bibfield  {journal} {\bibinfo
  {journal} {Phys.Rev.}\ }\textbf {\bibinfo {volume} {D71}},\ \bibinfo {pages}
  {071103} (\bibinfo {year} {2005})},\ \Eprint
  {http://arxiv.org/abs/hep-ex/0406022} {arXiv:hep-ex/0406022} \BibitemShut
  {NoStop}%
\bibitem [{\citenamefont {Acosta}\ \emph {et~al.}(2004)\citenamefont {Acosta}
  \emph {et~al.}}]{Acosta:2003zx}%
  \BibitemOpen
  \bibfield  {author} {\bibinfo {author} {\bibfnamefont {D.}~\bibnamefont
  {Acosta}} \emph {et~al.} (\bibinfo {collaboration} {CDF}),\ }\href {\doibase
  10.1103/PhysRevLett.93.072001} {\bibfield  {journal} {\bibinfo  {journal}
  {Phys.Rev.Lett.}\ }\textbf {\bibinfo {volume} {93}},\ \bibinfo {pages}
  {072001} (\bibinfo {year} {2004})},\ \Eprint
  {http://arxiv.org/abs/hep-ex/0312021} {arXiv:hep-ex/0312021 [hep-ex]}
  \BibitemShut {NoStop}%
\bibitem [{\citenamefont {Abazov}\ \emph {et~al.}(2004)\citenamefont {Abazov}
  \emph {et~al.}}]{Abazov:2004kp}%
  \BibitemOpen
  \bibfield  {author} {\bibinfo {author} {\bibfnamefont {V.}~\bibnamefont
  {Abazov}} \emph {et~al.},\ }\href {\doibase 10.1103/PhysRevLett.93.162002}
  {\bibfield  {journal} {\bibinfo  {journal} {Phys.Rev.Lett.}\ }\textbf
  {\bibinfo {volume} {93}},\ \bibinfo {pages} {162002} (\bibinfo {year}
  {2004})},\ \Eprint {http://arxiv.org/abs/hep-ex/0405004}
  {arXiv:hep-ex/0405004 [hep-ex]} \BibitemShut {NoStop}%
\bibitem [{\citenamefont {Aaij}\ \emph {et~al.}(2012)\citenamefont {Aaij} \emph
  {et~al.}}]{Aaij:2011sn}%
  \BibitemOpen
  \bibfield  {author} {\bibinfo {author} {\bibfnamefont {R.}~\bibnamefont
  {Aaij}} \emph {et~al.} (\bibinfo {collaboration} {LHCb}),\ }\href {\doibase
  10.1140/epjc/s10052-012-1972-7} {\bibfield  {journal} {\bibinfo  {journal}
  {Eur.Phys.J.}\ }\textbf {\bibinfo {volume} {C72}},\ \bibinfo {pages} {1972}
  (\bibinfo {year} {2012})},\ \Eprint {http://arxiv.org/abs/1112.5310}
  {arXiv:1112.5310 [hep-ex]} \BibitemShut {NoStop}%
\bibitem [{\citenamefont {Chatrchyan}\ \emph {et~al.}(2013)\citenamefont
  {Chatrchyan} \emph {et~al.}}]{Chatrchyan:2013cld}%
  \BibitemOpen
  \bibfield  {author} {\bibinfo {author} {\bibfnamefont {S.}~\bibnamefont
  {Chatrchyan}} \emph {et~al.} (\bibinfo {collaboration} {CMS}),\ }\href
  {\doibase 10.1007/JHEP04(2013)154} {\bibfield  {journal} {\bibinfo  {journal}
  {JHEP}\ }\textbf {\bibinfo {volume} {1304}},\ \bibinfo {pages} {154}
  (\bibinfo {year} {2013})},\ \Eprint {http://arxiv.org/abs/1302.3968}
  {arXiv:1302.3968 [hep-ex]} \BibitemShut {NoStop}%
\bibitem [{\citenamefont {Collaboration}(2019)}]{CMS:2019vma}%
  \BibitemOpen
  \bibfield  {author} {\bibinfo {author} {\bibfnamefont {C.}~\bibnamefont
  {Collaboration}} (\bibinfo {collaboration} {CMS}),\ }\href@noop {} {\enquote
  {\bibinfo {title} {{Evidence for $\chi_{c1}$(3872) in PbPb collisions and
  studies of its prompt production at
  $\sqrt{\smash[b]{s_{_{\mathrm{NN}}}}}=5.02$ TeV}},}\ } (\bibinfo {year}
  {2019})\BibitemShut {NoStop}%
\bibitem [{\citenamefont {Aaij}\ \emph
  {et~al.}(2020{\natexlab{a}})\citenamefont {Aaij} \emph
  {et~al.}}]{Aaij:2020qga}%
  \BibitemOpen
  \bibfield  {author} {\bibinfo {author} {\bibfnamefont {R.}~\bibnamefont
  {Aaij}} \emph {et~al.} (\bibinfo {collaboration} {LHCb}),\ }\href@noop {} {\
  (\bibinfo {year} {2020}{\natexlab{a}})},\ \Eprint
  {http://arxiv.org/abs/2005.13419} {arXiv:2005.13419 [hep-ex]} \BibitemShut
  {NoStop}%
\bibitem [{\citenamefont {Eichten}\ and\ \citenamefont
  {Liu}(2017)}]{Eichten:2017ual}%
  \BibitemOpen
  \bibfield  {author} {\bibinfo {author} {\bibfnamefont {E.}~\bibnamefont
  {Eichten}}\ and\ \bibinfo {author} {\bibfnamefont {Z.}~\bibnamefont {Liu}},\
  }\href@noop {} {\  (\bibinfo {year} {2017})},\ \Eprint
  {http://arxiv.org/abs/1709.09605} {arXiv:1709.09605 [hep-ph]} \BibitemShut
  {NoStop}%
\bibitem [{\citenamefont {Aaij}\ \emph {et~al.}(2018)\citenamefont {Aaij} \emph
  {et~al.}}]{Aaij:2018zrb}%
  \BibitemOpen
  \bibfield  {author} {\bibinfo {author} {\bibfnamefont {R.}~\bibnamefont
  {Aaij}} \emph {et~al.} (\bibinfo {collaboration} {LHCb}),\ }\href {\doibase
  10.1007/JHEP10(2018)086} {\bibfield  {journal} {\bibinfo  {journal} {JHEP}\
  }\textbf {\bibinfo {volume} {10}},\ \bibinfo {pages} {086} (\bibinfo {year}
  {2018})},\ \Eprint {http://arxiv.org/abs/1806.09707} {arXiv:1806.09707
  [hep-ex]} \BibitemShut {NoStop}%
\bibitem [{\citenamefont {Aaij}\ \emph
  {et~al.}(2020{\natexlab{b}})\citenamefont {Aaij} \emph {et~al.}}]{1804391}%
  \BibitemOpen
  \bibfield  {author} {\bibinfo {author} {\bibfnamefont {R.}~\bibnamefont
  {Aaij}} \emph {et~al.} (\bibinfo {collaboration} {LHCb}),\ }\href@noop {} {\
  (\bibinfo {year} {2020}{\natexlab{b}})},\ \Eprint
  {http://arxiv.org/abs/2006.16957} {arXiv:2006.16957 [hep-ex]} \BibitemShut
  {NoStop}%
\bibitem [{\citenamefont {Chen}\ \emph {et~al.}(2017)\citenamefont {Chen},
  \citenamefont {Chen}, \citenamefont {Liu}, \citenamefont {Steele},\ and\
  \citenamefont {Zhu}}]{Chen:2016jxd}%
  \BibitemOpen
  \bibfield  {author} {\bibinfo {author} {\bibfnamefont {W.}~\bibnamefont
  {Chen}}, \bibinfo {author} {\bibfnamefont {H.-X.}\ \bibnamefont {Chen}},
  \bibinfo {author} {\bibfnamefont {X.}~\bibnamefont {Liu}}, \bibinfo {author}
  {\bibfnamefont {T.}~\bibnamefont {Steele}}, \ and\ \bibinfo {author}
  {\bibfnamefont {S.-L.}\ \bibnamefont {Zhu}},\ }\href {\doibase
  10.1016/j.physletb.2017.08.034} {\bibfield  {journal} {\bibinfo  {journal}
  {Phys. Lett. B}\ }\textbf {\bibinfo {volume} {773}},\ \bibinfo {pages} {247}
  (\bibinfo {year} {2017})},\ \Eprint {http://arxiv.org/abs/1605.01647}
  {arXiv:1605.01647 [hep-ph]} \BibitemShut {NoStop}%
\bibitem [{\citenamefont {Liu}\ \emph {et~al.}(2019)\citenamefont {Liu},
  \citenamefont {Lü}, \citenamefont {Zhong},\ and\ \citenamefont
  {Zhao}}]{Liu:2019zuc}%
  \BibitemOpen
  \bibfield  {author} {\bibinfo {author} {\bibfnamefont {M.-S.}\ \bibnamefont
  {Liu}}, \bibinfo {author} {\bibfnamefont {Q.-F.}\ \bibnamefont {Lü}},
  \bibinfo {author} {\bibfnamefont {X.-H.}\ \bibnamefont {Zhong}}, \ and\
  \bibinfo {author} {\bibfnamefont {Q.}~\bibnamefont {Zhao}},\ }\href {\doibase
  10.1103/PhysRevD.100.016006} {\bibfield  {journal} {\bibinfo  {journal}
  {Phys. Rev. D}\ }\textbf {\bibinfo {volume} {100}},\ \bibinfo {pages}
  {016006} (\bibinfo {year} {2019})},\ \Eprint
  {http://arxiv.org/abs/1901.02564} {arXiv:1901.02564 [hep-ph]} \BibitemShut
  {NoStop}%
\bibitem [{\citenamefont {Karliner}\ \emph {et~al.}(2017)\citenamefont
  {Karliner}, \citenamefont {Nussinov},\ and\ \citenamefont
  {Rosner}}]{Karliner:2016zzc}%
  \BibitemOpen
  \bibfield  {author} {\bibinfo {author} {\bibfnamefont {M.}~\bibnamefont
  {Karliner}}, \bibinfo {author} {\bibfnamefont {S.}~\bibnamefont {Nussinov}},
  \ and\ \bibinfo {author} {\bibfnamefont {J.~L.}\ \bibnamefont {Rosner}},\
  }\href {\doibase 10.1103/PhysRevD.95.034011} {\bibfield  {journal} {\bibinfo
  {journal} {Phys. Rev. D}\ }\textbf {\bibinfo {volume} {95}},\ \bibinfo
  {pages} {034011} (\bibinfo {year} {2017})},\ \Eprint
  {http://arxiv.org/abs/1611.00348} {arXiv:1611.00348 [hep-ph]} \BibitemShut
  {NoStop}%
\bibitem [{\citenamefont {Berezhnoy}\ \emph {et~al.}(2012)\citenamefont
  {Berezhnoy}, \citenamefont {Luchinsky},\ and\ \citenamefont
  {Novoselov}}]{Berezhnoy:2011xn}%
  \BibitemOpen
  \bibfield  {author} {\bibinfo {author} {\bibfnamefont {A.}~\bibnamefont
  {Berezhnoy}}, \bibinfo {author} {\bibfnamefont {A.}~\bibnamefont
  {Luchinsky}}, \ and\ \bibinfo {author} {\bibfnamefont {A.}~\bibnamefont
  {Novoselov}},\ }\href {\doibase 10.1103/PhysRevD.86.034004} {\bibfield
  {journal} {\bibinfo  {journal} {Phys. Rev. D}\ }\textbf {\bibinfo {volume}
  {86}},\ \bibinfo {pages} {034004} (\bibinfo {year} {2012})},\ \Eprint
  {http://arxiv.org/abs/1111.1867} {arXiv:1111.1867 [hep-ph]} \BibitemShut
  {NoStop}%
\bibitem [{\citenamefont {Wu}\ \emph {et~al.}(2018)\citenamefont {Wu},
  \citenamefont {Liu}, \citenamefont {Chen}, \citenamefont {Liu},\ and\
  \citenamefont {Zhu}}]{Wu:2016vtq}%
  \BibitemOpen
  \bibfield  {author} {\bibinfo {author} {\bibfnamefont {J.}~\bibnamefont
  {Wu}}, \bibinfo {author} {\bibfnamefont {Y.-R.}\ \bibnamefont {Liu}},
  \bibinfo {author} {\bibfnamefont {K.}~\bibnamefont {Chen}}, \bibinfo {author}
  {\bibfnamefont {X.}~\bibnamefont {Liu}}, \ and\ \bibinfo {author}
  {\bibfnamefont {S.-L.}\ \bibnamefont {Zhu}},\ }\href {\doibase
  10.1103/PhysRevD.97.094015} {\bibfield  {journal} {\bibinfo  {journal} {Phys.
  Rev. D}\ }\textbf {\bibinfo {volume} {97}},\ \bibinfo {pages} {094015}
  (\bibinfo {year} {2018})},\ \Eprint {http://arxiv.org/abs/1605.01134}
  {arXiv:1605.01134 [hep-ph]} \BibitemShut {NoStop}%
\bibitem [{\citenamefont {Wang}(2017)}]{Wang:2017jtz}%
  \BibitemOpen
  \bibfield  {author} {\bibinfo {author} {\bibfnamefont {Z.-G.}\ \bibnamefont
  {Wang}},\ }\href {\doibase 10.1140/epjc/s10052-017-4997-0} {\bibfield
  {journal} {\bibinfo  {journal} {Eur. Phys. J. C}\ }\textbf {\bibinfo {volume}
  {77}},\ \bibinfo {pages} {432} (\bibinfo {year} {2017})},\ \Eprint
  {http://arxiv.org/abs/1701.04285} {arXiv:1701.04285 [hep-ph]} \BibitemShut
  {NoStop}%
\bibitem [{\citenamefont {Wang}\ and\ \citenamefont {Di}(2019)}]{Wang:2018poa}%
  \BibitemOpen
  \bibfield  {author} {\bibinfo {author} {\bibfnamefont {Z.-G.}\ \bibnamefont
  {Wang}}\ and\ \bibinfo {author} {\bibfnamefont {Z.-Y.}\ \bibnamefont {Di}},\
  }\href {\doibase 10.5506/APhysPolB.50.1335} {\bibfield  {journal} {\bibinfo
  {journal} {Acta Phys. Polon. B}\ }\textbf {\bibinfo {volume} {50}},\ \bibinfo
  {pages} {1335} (\bibinfo {year} {2019})},\ \Eprint
  {http://arxiv.org/abs/1807.08520} {arXiv:1807.08520 [hep-ph]} \BibitemShut
  {NoStop}%
\bibitem [{\citenamefont {Reinders}\ \emph {et~al.}(1985)\citenamefont
  {Reinders}, \citenamefont {Rubinstein},\ and\ \citenamefont
  {Yazaki}}]{Reinders:1984sr}%
  \BibitemOpen
  \bibfield  {author} {\bibinfo {author} {\bibfnamefont {L.}~\bibnamefont
  {Reinders}}, \bibinfo {author} {\bibfnamefont {H.}~\bibnamefont
  {Rubinstein}}, \ and\ \bibinfo {author} {\bibfnamefont {S.}~\bibnamefont
  {Yazaki}},\ }\href {\doibase 10.1016/0370-1573(85)90065-1} {\bibfield
  {journal} {\bibinfo  {journal} {Phys. Rept.}\ }\textbf {\bibinfo {volume}
  {127}},\ \bibinfo {pages} {1} (\bibinfo {year} {1985})}\BibitemShut {NoStop}%
\bibitem [{\citenamefont {Heller}\ and\ \citenamefont
  {Tjon}(1985)}]{Heller:1985cb}%
  \BibitemOpen
  \bibfield  {author} {\bibinfo {author} {\bibfnamefont {L.}~\bibnamefont
  {Heller}}\ and\ \bibinfo {author} {\bibfnamefont {J.}~\bibnamefont {Tjon}},\
  }\href {\doibase 10.1103/PhysRevD.32.755} {\bibfield  {journal} {\bibinfo
  {journal} {Phys. Rev. D}\ }\textbf {\bibinfo {volume} {32}},\ \bibinfo
  {pages} {755} (\bibinfo {year} {1985})}\BibitemShut {NoStop}%
\bibitem [{\citenamefont {Anwar}\ \emph {et~al.}(2018)\citenamefont {Anwar},
  \citenamefont {Ferretti}, \citenamefont {Guo}, \citenamefont {Santopinto},\
  and\ \citenamefont {Zou}}]{Anwar:2017toa}%
  \BibitemOpen
  \bibfield  {author} {\bibinfo {author} {\bibfnamefont {M.~N.}\ \bibnamefont
  {Anwar}}, \bibinfo {author} {\bibfnamefont {J.}~\bibnamefont {Ferretti}},
  \bibinfo {author} {\bibfnamefont {F.-K.}\ \bibnamefont {Guo}}, \bibinfo
  {author} {\bibfnamefont {E.}~\bibnamefont {Santopinto}}, \ and\ \bibinfo
  {author} {\bibfnamefont {B.-S.}\ \bibnamefont {Zou}},\ }\href {\doibase
  10.1140/epjc/s10052-018-6073-9} {\bibfield  {journal} {\bibinfo  {journal}
  {Eur. Phys. J. C}\ }\textbf {\bibinfo {volume} {78}},\ \bibinfo {pages} {647}
  (\bibinfo {year} {2018})},\ \Eprint {http://arxiv.org/abs/1710.02540}
  {arXiv:1710.02540 [hep-ph]} \BibitemShut {NoStop}%
\bibitem [{\citenamefont {Esposito}\ and\ \citenamefont
  {Polosa}(2018)}]{Esposito:2018cwh}%
  \BibitemOpen
  \bibfield  {author} {\bibinfo {author} {\bibfnamefont {A.}~\bibnamefont
  {Esposito}}\ and\ \bibinfo {author} {\bibfnamefont {A.~D.}\ \bibnamefont
  {Polosa}},\ }\href {\doibase 10.1140/epjc/s10052-018-6269-z} {\bibfield
  {journal} {\bibinfo  {journal} {Eur. Phys. J. C}\ }\textbf {\bibinfo {volume}
  {78}},\ \bibinfo {pages} {782} (\bibinfo {year} {2018})},\ \Eprint
  {http://arxiv.org/abs/1807.06040} {arXiv:1807.06040 [hep-ph]} \BibitemShut
  {NoStop}%
\bibitem [{\citenamefont {Ader}\ \emph {et~al.}(1982)\citenamefont {Ader},
  \citenamefont {Richard},\ and\ \citenamefont {Taxil}}]{Ader:1981db}%
  \BibitemOpen
  \bibfield  {author} {\bibinfo {author} {\bibfnamefont {J.}~\bibnamefont
  {Ader}}, \bibinfo {author} {\bibfnamefont {J.}~\bibnamefont {Richard}}, \
  and\ \bibinfo {author} {\bibfnamefont {P.}~\bibnamefont {Taxil}},\ }\href
  {\doibase 10.1103/PhysRevD.25.2370} {\bibfield  {journal} {\bibinfo
  {journal} {Phys. Rev. D}\ }\textbf {\bibinfo {volume} {25}},\ \bibinfo
  {pages} {2370} (\bibinfo {year} {1982})}\BibitemShut {NoStop}%
\bibitem [{\citenamefont {Zouzou}\ \emph {et~al.}(1986)\citenamefont {Zouzou},
  \citenamefont {Silvestre-Brac}, \citenamefont {Gignoux},\ and\ \citenamefont
  {Richard}}]{Zouzou:1986qh}%
  \BibitemOpen
  \bibfield  {author} {\bibinfo {author} {\bibfnamefont {S.}~\bibnamefont
  {Zouzou}}, \bibinfo {author} {\bibfnamefont {B.}~\bibnamefont
  {Silvestre-Brac}}, \bibinfo {author} {\bibfnamefont {C.}~\bibnamefont
  {Gignoux}}, \ and\ \bibinfo {author} {\bibfnamefont {J.}~\bibnamefont
  {Richard}},\ }\href {\doibase 10.1007/BF01557611} {\bibfield  {journal}
  {\bibinfo  {journal} {Z. Phys. C}\ }\textbf {\bibinfo {volume} {30}},\
  \bibinfo {pages} {457} (\bibinfo {year} {1986})}\BibitemShut {NoStop}%
\bibitem [{\citenamefont {Lloyd}\ and\ \citenamefont
  {Vary}(2004)}]{Lloyd:2003yc}%
  \BibitemOpen
  \bibfield  {author} {\bibinfo {author} {\bibfnamefont {R.~J.}\ \bibnamefont
  {Lloyd}}\ and\ \bibinfo {author} {\bibfnamefont {J.~P.}\ \bibnamefont
  {Vary}},\ }\href {\doibase 10.1103/PhysRevD.70.014009} {\bibfield  {journal}
  {\bibinfo  {journal} {Phys. Rev. D}\ }\textbf {\bibinfo {volume} {70}},\
  \bibinfo {pages} {014009} (\bibinfo {year} {2004})},\ \Eprint
  {http://arxiv.org/abs/hep-ph/0311179} {arXiv:hep-ph/0311179} \BibitemShut
  {NoStop}%
\bibitem [{\citenamefont {Barnea}\ \emph {et~al.}(2006)\citenamefont {Barnea},
  \citenamefont {Vijande},\ and\ \citenamefont {Valcarce}}]{Barnea:2006sd}%
  \BibitemOpen
  \bibfield  {author} {\bibinfo {author} {\bibfnamefont {N.}~\bibnamefont
  {Barnea}}, \bibinfo {author} {\bibfnamefont {J.}~\bibnamefont {Vijande}}, \
  and\ \bibinfo {author} {\bibfnamefont {A.}~\bibnamefont {Valcarce}},\ }\href
  {\doibase 10.1103/PhysRevD.73.054004} {\bibfield  {journal} {\bibinfo
  {journal} {Phys. Rev. D}\ }\textbf {\bibinfo {volume} {73}},\ \bibinfo
  {pages} {054004} (\bibinfo {year} {2006})},\ \Eprint
  {http://arxiv.org/abs/hep-ph/0604010} {arXiv:hep-ph/0604010} \BibitemShut
  {NoStop}%
\bibitem [{\citenamefont {Richard}\ \emph {et~al.}(2018)\citenamefont
  {Richard}, \citenamefont {Valcarce},\ and\ \citenamefont
  {Vijande}}]{Richard:2018yrm}%
  \BibitemOpen
  \bibfield  {author} {\bibinfo {author} {\bibfnamefont {J.-M.}\ \bibnamefont
  {Richard}}, \bibinfo {author} {\bibfnamefont {A.}~\bibnamefont {Valcarce}}, \
  and\ \bibinfo {author} {\bibfnamefont {J.}~\bibnamefont {Vijande}},\ }\href
  {\doibase 10.1103/PhysRevC.97.035211} {\bibfield  {journal} {\bibinfo
  {journal} {Phys. Rev. C}\ }\textbf {\bibinfo {volume} {97}},\ \bibinfo
  {pages} {035211} (\bibinfo {year} {2018})},\ \Eprint
  {http://arxiv.org/abs/1803.06155} {arXiv:1803.06155 [hep-ph]} \BibitemShut
  {NoStop}%
\bibitem [{\citenamefont {Richard}\ \emph {et~al.}(2017)\citenamefont
  {Richard}, \citenamefont {Valcarce},\ and\ \citenamefont
  {Vijande}}]{Richard:2017vry}%
  \BibitemOpen
  \bibfield  {author} {\bibinfo {author} {\bibfnamefont {J.-M.}\ \bibnamefont
  {Richard}}, \bibinfo {author} {\bibfnamefont {A.}~\bibnamefont {Valcarce}}, \
  and\ \bibinfo {author} {\bibfnamefont {J.}~\bibnamefont {Vijande}},\ }\href
  {\doibase 10.1103/PhysRevD.95.054019} {\bibfield  {journal} {\bibinfo
  {journal} {Phys. Rev. D}\ }\textbf {\bibinfo {volume} {95}},\ \bibinfo
  {pages} {054019} (\bibinfo {year} {2017})},\ \Eprint
  {http://arxiv.org/abs/1703.00783} {arXiv:1703.00783 [hep-ph]} \BibitemShut
  {NoStop}%
\bibitem [{\citenamefont {Vijande}\ \emph {et~al.}(2009)\citenamefont
  {Vijande}, \citenamefont {Valcarce},\ and\ \citenamefont
  {Barnea}}]{Vijande:2009kj}%
  \BibitemOpen
  \bibfield  {author} {\bibinfo {author} {\bibfnamefont {J.}~\bibnamefont
  {Vijande}}, \bibinfo {author} {\bibfnamefont {A.}~\bibnamefont {Valcarce}}, \
  and\ \bibinfo {author} {\bibfnamefont {N.}~\bibnamefont {Barnea}},\ }\href
  {\doibase 10.1103/PhysRevD.79.074010} {\bibfield  {journal} {\bibinfo
  {journal} {Phys. Rev. D}\ }\textbf {\bibinfo {volume} {79}},\ \bibinfo
  {pages} {074010} (\bibinfo {year} {2009})},\ \Eprint
  {http://arxiv.org/abs/0903.2949} {arXiv:0903.2949 [hep-ph]} \BibitemShut
  {NoStop}%
\bibitem [{\citenamefont {Debastiani}\ and\ \citenamefont
  {Navarra}(2019)}]{Debastiani:2017msn}%
  \BibitemOpen
  \bibfield  {author} {\bibinfo {author} {\bibfnamefont {V.}~\bibnamefont
  {Debastiani}}\ and\ \bibinfo {author} {\bibfnamefont {F.}~\bibnamefont
  {Navarra}},\ }\href {\doibase 10.1088/1674-1137/43/1/013105} {\bibfield
  {journal} {\bibinfo  {journal} {Chin. Phys. C}\ }\textbf {\bibinfo {volume}
  {43}},\ \bibinfo {pages} {013105} (\bibinfo {year} {2019})},\ \Eprint
  {http://arxiv.org/abs/1706.07553} {arXiv:1706.07553 [hep-ph]} \BibitemShut
  {NoStop}%
\bibitem [{\citenamefont {Chen}(2019{\natexlab{a}})}]{Chen:2019dvd}%
  \BibitemOpen
  \bibfield  {author} {\bibinfo {author} {\bibfnamefont {X.}~\bibnamefont
  {Chen}},\ }\href {\doibase 10.1140/epja/i2019-12807-2} {\bibfield  {journal}
  {\bibinfo  {journal} {Eur. Phys. J. A}\ }\textbf {\bibinfo {volume} {55}},\
  \bibinfo {pages} {106} (\bibinfo {year} {2019}{\natexlab{a}})},\ \Eprint
  {http://arxiv.org/abs/1902.00008} {arXiv:1902.00008 [hep-ph]} \BibitemShut
  {NoStop}%
\bibitem [{\citenamefont {Chen}(2019{\natexlab{b}})}]{Chen:2019vrj}%
  \BibitemOpen
  \bibfield  {author} {\bibinfo {author} {\bibfnamefont {X.}~\bibnamefont
  {Chen}},\ }\href {\doibase 10.1103/PhysRevD.100.094009} {\bibfield  {journal}
  {\bibinfo  {journal} {Phys. Rev. D}\ }\textbf {\bibinfo {volume} {100}},\
  \bibinfo {pages} {094009} (\bibinfo {year} {2019}{\natexlab{b}})},\ \Eprint
  {http://arxiv.org/abs/1908.08811} {arXiv:1908.08811 [hep-ph]} \BibitemShut
  {NoStop}%
\bibitem [{\citenamefont {Chen}(2020)}]{Chen:2020lgj}%
  \BibitemOpen
  \bibfield  {author} {\bibinfo {author} {\bibfnamefont {X.}~\bibnamefont
  {Chen}},\ }\href@noop {} {\  (\bibinfo {year} {2020})},\ \Eprint
  {http://arxiv.org/abs/2001.06755} {arXiv:2001.06755 [hep-ph]} \BibitemShut
  {NoStop}%
\bibitem [{\citenamefont {Wang}\ \emph {et~al.}(2019)\citenamefont {Wang},
  \citenamefont {Meng},\ and\ \citenamefont {Zhu}}]{Wang:2019rdo}%
  \BibitemOpen
  \bibfield  {author} {\bibinfo {author} {\bibfnamefont {G.-J.}\ \bibnamefont
  {Wang}}, \bibinfo {author} {\bibfnamefont {L.}~\bibnamefont {Meng}}, \ and\
  \bibinfo {author} {\bibfnamefont {S.-L.}\ \bibnamefont {Zhu}},\ }\href
  {\doibase 10.1103/PhysRevD.100.096013} {\bibfield  {journal} {\bibinfo
  {journal} {Phys. Rev. D}\ }\textbf {\bibinfo {volume} {100}},\ \bibinfo
  {pages} {096013} (\bibinfo {year} {2019})},\ \Eprint
  {http://arxiv.org/abs/1907.05177} {arXiv:1907.05177 [hep-ph]} \BibitemShut
  {NoStop}%
\bibitem [{\citenamefont {Yang}\ \emph
  {et~al.}(2020{\natexlab{a}})\citenamefont {Yang}, \citenamefont {Ping},
  \citenamefont {He},\ and\ \citenamefont {Wang}}]{Yang:2020rih}%
  \BibitemOpen
  \bibfield  {author} {\bibinfo {author} {\bibfnamefont {G.}~\bibnamefont
  {Yang}}, \bibinfo {author} {\bibfnamefont {J.}~\bibnamefont {Ping}}, \bibinfo
  {author} {\bibfnamefont {L.}~\bibnamefont {He}}, \ and\ \bibinfo {author}
  {\bibfnamefont {Q.}~\bibnamefont {Wang}},\ }\href@noop {} {\  (\bibinfo
  {year} {2020}{\natexlab{a}})},\ \Eprint {http://arxiv.org/abs/2006.13756}
  {arXiv:2006.13756 [hep-ph]} \BibitemShut {NoStop}%
\bibitem [{\citenamefont {Bedolla}\ \emph {et~al.}(2019)\citenamefont
  {Bedolla}, \citenamefont {Ferretti}, \citenamefont {Roberts},\ and\
  \citenamefont {Santopinto}}]{Bedolla:2019zwg}%
  \BibitemOpen
  \bibfield  {author} {\bibinfo {author} {\bibfnamefont {M.~A.}\ \bibnamefont
  {Bedolla}}, \bibinfo {author} {\bibfnamefont {J.}~\bibnamefont {Ferretti}},
  \bibinfo {author} {\bibfnamefont {C.~D.}\ \bibnamefont {Roberts}}, \ and\
  \bibinfo {author} {\bibfnamefont {E.}~\bibnamefont {Santopinto}},\
  }\href@noop {} {\  (\bibinfo {year} {2019})},\ \Eprint
  {http://arxiv.org/abs/1911.00960} {arXiv:1911.00960 [hep-ph]} \BibitemShut
  {NoStop}%
\bibitem [{\citenamefont {Hughes}\ \emph {et~al.}(2018)\citenamefont {Hughes},
  \citenamefont {Eichten},\ and\ \citenamefont {Davies}}]{Hughes:2017xie}%
  \BibitemOpen
  \bibfield  {author} {\bibinfo {author} {\bibfnamefont {C.}~\bibnamefont
  {Hughes}}, \bibinfo {author} {\bibfnamefont {E.}~\bibnamefont {Eichten}}, \
  and\ \bibinfo {author} {\bibfnamefont {C.~T.~H.}\ \bibnamefont {Davies}},\
  }\href {\doibase 10.1103/PhysRevD.97.054505} {\bibfield  {journal} {\bibinfo
  {journal} {Phys. Rev. D}\ }\textbf {\bibinfo {volume} {97}},\ \bibinfo
  {pages} {054505} (\bibinfo {year} {2018})},\ \Eprint
  {http://arxiv.org/abs/1710.03236} {arXiv:1710.03236 [hep-lat]} \BibitemShut
  {NoStop}%
\bibitem [{\citenamefont {Bai}\ \emph {et~al.}(2019)\citenamefont {Bai},
  \citenamefont {Lu},\ and\ \citenamefont {Osborne}}]{Bai:2016int}%
  \BibitemOpen
  \bibfield  {author} {\bibinfo {author} {\bibfnamefont {Y.}~\bibnamefont
  {Bai}}, \bibinfo {author} {\bibfnamefont {S.}~\bibnamefont {Lu}}, \ and\
  \bibinfo {author} {\bibfnamefont {J.}~\bibnamefont {Osborne}},\ }\href
  {\doibase 10.1016/j.physletb.2019.134930} {\bibfield  {journal} {\bibinfo
  {journal} {Phys. Lett. B}\ }\textbf {\bibinfo {volume} {798}},\ \bibinfo
  {pages} {134930} (\bibinfo {year} {2019})},\ \Eprint
  {http://arxiv.org/abs/1612.00012} {arXiv:1612.00012 [hep-ph]} \BibitemShut
  {NoStop}%
\bibitem [{\citenamefont {Silvestre-Brac}(1996)}]{SilvestreBrac:1996bg}%
  \BibitemOpen
  \bibfield  {author} {\bibinfo {author} {\bibfnamefont {B.}~\bibnamefont
  {Silvestre-Brac}},\ }\href {\doibase 10.1007/s006010050028} {\bibfield
  {journal} {\bibinfo  {journal} {Few Body Syst.}\ }\textbf {\bibinfo {volume}
  {20}},\ \bibinfo {pages} {1} (\bibinfo {year} {1996})}\BibitemShut {NoStop}%
\bibitem [{\citenamefont {Bali}\ \emph {et~al.}(2005)\citenamefont {Bali},
  \citenamefont {Neff}, \citenamefont {Duessel}, \citenamefont {Lippert},\ and\
  \citenamefont {Schilling}}]{Bali:2005fu}%
  \BibitemOpen
  \bibfield  {author} {\bibinfo {author} {\bibfnamefont {G.~S.}\ \bibnamefont
  {Bali}}, \bibinfo {author} {\bibfnamefont {H.}~\bibnamefont {Neff}}, \bibinfo
  {author} {\bibfnamefont {T.}~\bibnamefont {Duessel}}, \bibinfo {author}
  {\bibfnamefont {T.}~\bibnamefont {Lippert}}, \ and\ \bibinfo {author}
  {\bibfnamefont {K.}~\bibnamefont {Schilling}} (\bibinfo {collaboration}
  {SESAM}),\ }\href {\doibase 10.1103/PhysRevD.71.114513} {\bibfield  {journal}
  {\bibinfo  {journal} {Phys. Rev. D}\ }\textbf {\bibinfo {volume} {71}},\
  \bibinfo {pages} {114513} (\bibinfo {year} {2005})},\ \Eprint
  {http://arxiv.org/abs/hep-lat/0505012} {arXiv:hep-lat/0505012} \BibitemShut
  {NoStop}%
\bibitem [{\citenamefont {Semay}\ and\ \citenamefont
  {Silvestre-Brac}(1994)}]{Semay:1994ht}%
  \BibitemOpen
  \bibfield  {author} {\bibinfo {author} {\bibfnamefont {C.}~\bibnamefont
  {Semay}}\ and\ \bibinfo {author} {\bibfnamefont {B.}~\bibnamefont
  {Silvestre-Brac}},\ }\href {\doibase 10.1007/BF01413104} {\bibfield
  {journal} {\bibinfo  {journal} {Z. Phys. C}\ }\textbf {\bibinfo {volume}
  {61}},\ \bibinfo {pages} {271} (\bibinfo {year} {1994})}\BibitemShut
  {NoStop}%
\bibitem [{\citenamefont {Hammond}\ \emph {et~al.}(1994)\citenamefont
  {Hammond}, \citenamefont {Lester},\ and\ \citenamefont
  {Reynolds}}]{Hammond:1994bk}%
  \BibitemOpen
  \bibfield  {author} {\bibinfo {author} {\bibfnamefont {B.}~\bibnamefont
  {Hammond}}, \bibinfo {author} {\bibfnamefont {W.}~\bibnamefont {Lester}}, \
  and\ \bibinfo {author} {\bibfnamefont {P.}~\bibnamefont {Reynolds}},\
  }\href@noop {} {\emph {\bibinfo {title} {Monte Carlo Methods in ab Initio
  Quantum Chemistry}}}\ (\bibinfo  {publisher} {World Scientific},\ \bibinfo
  {address} {Singapore},\ \bibinfo {year} {1994})\BibitemShut {NoStop}%
\bibitem [{\citenamefont {Foulkes}\ \emph {et~al.}(2001)\citenamefont
  {Foulkes}, \citenamefont {Mitas}, \citenamefont {Needs},\ and\ \citenamefont
  {Rajagopal}}]{Foulkes:2001zz}%
  \BibitemOpen
  \bibfield  {author} {\bibinfo {author} {\bibfnamefont {W.}~\bibnamefont
  {Foulkes}}, \bibinfo {author} {\bibfnamefont {L.}~\bibnamefont {Mitas}},
  \bibinfo {author} {\bibfnamefont {R.}~\bibnamefont {Needs}}, \ and\ \bibinfo
  {author} {\bibfnamefont {G.}~\bibnamefont {Rajagopal}},\ }\href {\doibase
  10.1103/RevModPhys.73.33} {\bibfield  {journal} {\bibinfo  {journal} {Rev.
  Mod. Phys.}\ }\textbf {\bibinfo {volume} {73}},\ \bibinfo {pages} {33}
  (\bibinfo {year} {2001})}\BibitemShut {NoStop}%
\bibitem [{\citenamefont {Nightingale}\ and\ \citenamefont
  {Umrigar}(2014)}]{Nightingale:2014bk}%
  \BibitemOpen
  \bibfield  {author} {\bibinfo {author} {\bibfnamefont {M.}~\bibnamefont
  {Nightingale}}\ and\ \bibinfo {author} {\bibfnamefont {C.~J.}\ \bibnamefont
  {Umrigar}},\ }\href@noop {} {\emph {\bibinfo {title} {Quantum Monte Carlo
  Methods in Physics and Chemistry}}}\ (\bibinfo  {publisher} {Springer},\
  \bibinfo {address} {Vienna},\ \bibinfo {year} {2014})\BibitemShut {NoStop}%
\bibitem [{\citenamefont {Schmidt}\ and\ \citenamefont
  {Ceperley}(1992)}]{Schmidt1992}%
  \BibitemOpen
  \bibfield  {author} {\bibinfo {author} {\bibfnamefont {K.~E.}\ \bibnamefont
  {Schmidt}}\ and\ \bibinfo {author} {\bibfnamefont {D.~M.}\ \bibnamefont
  {Ceperley}},\ }\enquote {\bibinfo {title} {Monte carlo techniques for quantum
  fluids, solids and droplets},}\ in\ \href {\doibase
  10.1007/978-3-662-02855-1_7} {\emph {\bibinfo {booktitle} {The Monte Carlo
  Method in Condensed Matter Physics}}},\ \bibinfo {editor} {edited by\
  \bibinfo {editor} {\bibfnamefont {K.}~\bibnamefont {Binder}}}\ (\bibinfo
  {publisher} {Springer Berlin Heidelberg},\ \bibinfo {address} {Berlin,
  Heidelberg},\ \bibinfo {year} {1992})\ pp.\ \bibinfo {pages}
  {205--248}\BibitemShut {NoStop}%
\bibitem [{\citenamefont {Ceperley}(1995)}]{Ceperley:1995zz}%
  \BibitemOpen
  \bibfield  {author} {\bibinfo {author} {\bibfnamefont {D.}~\bibnamefont
  {Ceperley}},\ }\href {\doibase 10.1103/RevModPhys.67.279} {\bibfield
  {journal} {\bibinfo  {journal} {Rev. Mod. Phys.}\ }\textbf {\bibinfo {volume}
  {67}},\ \bibinfo {pages} {279} (\bibinfo {year} {1995})}\BibitemShut
  {NoStop}%
\bibitem [{\citenamefont {Carlson}\ \emph {et~al.}(2003)\citenamefont
  {Carlson}, \citenamefont {Chang}, \citenamefont {Pandharipande},\ and\
  \citenamefont {Schmidt}}]{Carlson:2003zz}%
  \BibitemOpen
  \bibfield  {author} {\bibinfo {author} {\bibfnamefont {J.}~\bibnamefont
  {Carlson}}, \bibinfo {author} {\bibfnamefont {S.-Y.}\ \bibnamefont {Chang}},
  \bibinfo {author} {\bibfnamefont {V.}~\bibnamefont {Pandharipande}}, \ and\
  \bibinfo {author} {\bibfnamefont {K.}~\bibnamefont {Schmidt}},\ }\href
  {\doibase 10.1103/PhysRevLett.91.050401} {\bibfield  {journal} {\bibinfo
  {journal} {Phys. Rev. Lett.}\ }\textbf {\bibinfo {volume} {91}},\ \bibinfo
  {pages} {050401} (\bibinfo {year} {2003})},\ \Eprint
  {http://arxiv.org/abs/physics/0303094} {arXiv:physics/0303094} \BibitemShut
  {NoStop}%
\bibitem [{\citenamefont {Giorgini}\ \emph {et~al.}(2008)\citenamefont
  {Giorgini}, \citenamefont {Pitaevskii},\ and\ \citenamefont
  {Stringari}}]{Giorgini:2008zz}%
  \BibitemOpen
  \bibfield  {author} {\bibinfo {author} {\bibfnamefont {S.}~\bibnamefont
  {Giorgini}}, \bibinfo {author} {\bibfnamefont {L.~P.}\ \bibnamefont
  {Pitaevskii}}, \ and\ \bibinfo {author} {\bibfnamefont {S.}~\bibnamefont
  {Stringari}},\ }\href {\doibase 10.1103/RevModPhys.80.1215} {\bibfield
  {journal} {\bibinfo  {journal} {Rev. Mod. Phys.}\ }\textbf {\bibinfo {volume}
  {80}},\ \bibinfo {pages} {1215} (\bibinfo {year} {2008})},\ \Eprint
  {http://arxiv.org/abs/0706.3360} {arXiv:0706.3360 [cond-mat.other]}
  \BibitemShut {NoStop}%
\bibitem [{\citenamefont {De~Soto}\ \emph {et~al.}(2014)\citenamefont
  {De~Soto}, \citenamefont {Carbonell-Coronado},\ and\ \citenamefont
  {Gordillo}}]{PhysRevA.89.023633}%
  \BibitemOpen
  \bibfield  {author} {\bibinfo {author} {\bibfnamefont {F.}~\bibnamefont
  {De~Soto}}, \bibinfo {author} {\bibfnamefont {C.}~\bibnamefont
  {Carbonell-Coronado}}, \ and\ \bibinfo {author} {\bibfnamefont {M.~C.}\
  \bibnamefont {Gordillo}},\ }\href {\doibase 10.1103/PhysRevA.89.023633}
  {\bibfield  {journal} {\bibinfo  {journal} {Phys. Rev. A}\ }\textbf {\bibinfo
  {volume} {89}},\ \bibinfo {pages} {023633} (\bibinfo {year}
  {2014})}\BibitemShut {NoStop}%
\bibitem [{\citenamefont {Carbonell-Coronado}\ \emph
  {et~al.}(2016)\citenamefont {Carbonell-Coronado}, \citenamefont {Soto},\ and\
  \citenamefont {Gordillo}}]{Carbonell_Coronado_2016}%
  \BibitemOpen
  \bibfield  {author} {\bibinfo {author} {\bibfnamefont {C.}~\bibnamefont
  {Carbonell-Coronado}}, \bibinfo {author} {\bibfnamefont {F.~D.}\ \bibnamefont
  {Soto}}, \ and\ \bibinfo {author} {\bibfnamefont {M.~C.}\ \bibnamefont
  {Gordillo}},\ }\href {\doibase 10.1088/1367-2630/18/2/025015} {\bibfield
  {journal} {\bibinfo  {journal} {New Journal of Physics}\ }\textbf {\bibinfo
  {volume} {18}},\ \bibinfo {pages} {025015} (\bibinfo {year}
  {2016})}\BibitemShut {NoStop}%
\bibitem [{\citenamefont {Lomnitz-Adler}\ \emph {et~al.}(1981)\citenamefont
  {Lomnitz-Adler}, \citenamefont {Pandharipande},\ and\ \citenamefont
  {Smith}}]{Lomnitz-Adler:1981dmh}%
  \BibitemOpen
  \bibfield  {author} {\bibinfo {author} {\bibfnamefont {J.}~\bibnamefont
  {Lomnitz-Adler}}, \bibinfo {author} {\bibfnamefont {V.}~\bibnamefont
  {Pandharipande}}, \ and\ \bibinfo {author} {\bibfnamefont {R.}~\bibnamefont
  {Smith}},\ }\href {\doibase 10.1016/0375-9474(81)90642-4} {\bibfield
  {journal} {\bibinfo  {journal} {Nucl. Phys. A}\ }\textbf {\bibinfo {volume}
  {361}},\ \bibinfo {pages} {399} (\bibinfo {year} {1981})}\BibitemShut
  {NoStop}%
\bibitem [{\citenamefont {Carlson}(1987)}]{Carlson:1987zz}%
  \BibitemOpen
  \bibfield  {author} {\bibinfo {author} {\bibfnamefont {J.}~\bibnamefont
  {Carlson}},\ }\href {\doibase 10.1103/PhysRevC.36.2026} {\bibfield  {journal}
  {\bibinfo  {journal} {Phys. Rev. C}\ }\textbf {\bibinfo {volume} {36}},\
  \bibinfo {pages} {2026} (\bibinfo {year} {1987})}\BibitemShut {NoStop}%
\bibitem [{\citenamefont {Carlson}(1988)}]{Carlson:1988zz}%
  \BibitemOpen
  \bibfield  {author} {\bibinfo {author} {\bibfnamefont {J.}~\bibnamefont
  {Carlson}},\ }\href {\doibase 10.1103/PhysRevC.38.1879} {\bibfield  {journal}
  {\bibinfo  {journal} {Phys. Rev. C}\ }\textbf {\bibinfo {volume} {38}},\
  \bibinfo {pages} {1879} (\bibinfo {year} {1988})}\BibitemShut {NoStop}%
\bibitem [{\citenamefont {Lovato}\ \emph {et~al.}(2013)\citenamefont {Lovato},
  \citenamefont {Gandolfi}, \citenamefont {Butler}, \citenamefont {Carlson},
  \citenamefont {Lusk}, \citenamefont {Pieper},\ and\ \citenamefont
  {Schiavilla}}]{Lovato:2013cua}%
  \BibitemOpen
  \bibfield  {author} {\bibinfo {author} {\bibfnamefont {A.}~\bibnamefont
  {Lovato}}, \bibinfo {author} {\bibfnamefont {S.}~\bibnamefont {Gandolfi}},
  \bibinfo {author} {\bibfnamefont {R.}~\bibnamefont {Butler}}, \bibinfo
  {author} {\bibfnamefont {J.}~\bibnamefont {Carlson}}, \bibinfo {author}
  {\bibfnamefont {E.}~\bibnamefont {Lusk}}, \bibinfo {author} {\bibfnamefont
  {S.~C.}\ \bibnamefont {Pieper}}, \ and\ \bibinfo {author} {\bibfnamefont
  {R.}~\bibnamefont {Schiavilla}},\ }\href {\doibase
  10.1103/PhysRevLett.111.092501} {\bibfield  {journal} {\bibinfo  {journal}
  {Phys. Rev. Lett.}\ }\textbf {\bibinfo {volume} {111}},\ \bibinfo {pages}
  {092501} (\bibinfo {year} {2013})},\ \Eprint {http://arxiv.org/abs/1305.6959}
  {arXiv:1305.6959 [nucl-th]} \BibitemShut {NoStop}%
\bibitem [{\citenamefont {Lovato}\ \emph {et~al.}(2014)\citenamefont {Lovato},
  \citenamefont {Gandolfi}, \citenamefont {Carlson}, \citenamefont {Pieper},\
  and\ \citenamefont {Schiavilla}}]{Lovato:2014eva}%
  \BibitemOpen
  \bibfield  {author} {\bibinfo {author} {\bibfnamefont {A.}~\bibnamefont
  {Lovato}}, \bibinfo {author} {\bibfnamefont {S.}~\bibnamefont {Gandolfi}},
  \bibinfo {author} {\bibfnamefont {J.}~\bibnamefont {Carlson}}, \bibinfo
  {author} {\bibfnamefont {S.~C.}\ \bibnamefont {Pieper}}, \ and\ \bibinfo
  {author} {\bibfnamefont {R.}~\bibnamefont {Schiavilla}},\ }\href {\doibase
  10.1103/PhysRevLett.112.182502} {\bibfield  {journal} {\bibinfo  {journal}
  {Phys. Rev. Lett.}\ }\textbf {\bibinfo {volume} {112}},\ \bibinfo {pages}
  {182502} (\bibinfo {year} {2014})},\ \Eprint {http://arxiv.org/abs/1401.2605}
  {arXiv:1401.2605 [nucl-th]} \BibitemShut {NoStop}%
\bibitem [{\citenamefont {Lovato}\ \emph {et~al.}(2015)\citenamefont {Lovato},
  \citenamefont {Gandolfi}, \citenamefont {Carlson}, \citenamefont {Pieper},\
  and\ \citenamefont {Schiavilla}}]{Lovato:2015qka}%
  \BibitemOpen
  \bibfield  {author} {\bibinfo {author} {\bibfnamefont {A.}~\bibnamefont
  {Lovato}}, \bibinfo {author} {\bibfnamefont {S.}~\bibnamefont {Gandolfi}},
  \bibinfo {author} {\bibfnamefont {J.}~\bibnamefont {Carlson}}, \bibinfo
  {author} {\bibfnamefont {S.~C.}\ \bibnamefont {Pieper}}, \ and\ \bibinfo
  {author} {\bibfnamefont {R.}~\bibnamefont {Schiavilla}},\ }\href {\doibase
  10.1103/PhysRevC.91.062501} {\bibfield  {journal} {\bibinfo  {journal} {Phys.
  Rev. C}\ }\textbf {\bibinfo {volume} {91}},\ \bibinfo {pages} {062501}
  (\bibinfo {year} {2015})},\ \Eprint {http://arxiv.org/abs/1501.01981}
  {arXiv:1501.01981 [nucl-th]} \BibitemShut {NoStop}%
\bibitem [{\citenamefont {Schmidt}\ and\ \citenamefont
  {Fantoni}(1999)}]{Schmidt:1999lik}%
  \BibitemOpen
  \bibfield  {author} {\bibinfo {author} {\bibfnamefont {K.}~\bibnamefont
  {Schmidt}}\ and\ \bibinfo {author} {\bibfnamefont {S.}~\bibnamefont
  {Fantoni}},\ }\href {\doibase 10.1016/S0370-2693(98)01522-6} {\bibfield
  {journal} {\bibinfo  {journal} {Phys. Lett. B}\ }\textbf {\bibinfo {volume}
  {446}},\ \bibinfo {pages} {99} (\bibinfo {year} {1999})}\BibitemShut
  {NoStop}%
\bibitem [{\citenamefont {Carlson}\ \emph
  {et~al.}(1983{\natexlab{a}})\citenamefont {Carlson}, \citenamefont {Kogut},\
  and\ \citenamefont {Pandharipande}}]{Carlson:1982xi}%
  \BibitemOpen
  \bibfield  {author} {\bibinfo {author} {\bibfnamefont {J.}~\bibnamefont
  {Carlson}}, \bibinfo {author} {\bibfnamefont {J.~B.}\ \bibnamefont {Kogut}},
  \ and\ \bibinfo {author} {\bibfnamefont {V.}~\bibnamefont {Pandharipande}},\
  }\href {\doibase 10.1103/PhysRevD.27.233} {\bibfield  {journal} {\bibinfo
  {journal} {Phys. Rev. D}\ }\textbf {\bibinfo {volume} {27}},\ \bibinfo
  {pages} {233} (\bibinfo {year} {1983}{\natexlab{a}})}\BibitemShut {NoStop}%
\bibitem [{\citenamefont {Carlson}\ \emph
  {et~al.}(1983{\natexlab{b}})\citenamefont {Carlson}, \citenamefont {Kogut},\
  and\ \citenamefont {Pandharipande}}]{Carlson:1983rw}%
  \BibitemOpen
  \bibfield  {author} {\bibinfo {author} {\bibfnamefont {J.}~\bibnamefont
  {Carlson}}, \bibinfo {author} {\bibfnamefont {J.}~\bibnamefont {Kogut}}, \
  and\ \bibinfo {author} {\bibfnamefont {V.}~\bibnamefont {Pandharipande}},\
  }\href {\doibase 10.1103/PhysRevD.28.2807} {\bibfield  {journal} {\bibinfo
  {journal} {Phys. Rev. D}\ }\textbf {\bibinfo {volume} {28}},\ \bibinfo
  {pages} {2807} (\bibinfo {year} {1983}{\natexlab{b}})}\BibitemShut {NoStop}%
\bibitem [{\citenamefont {Isgur}\ and\ \citenamefont
  {Karl}(1978)}]{Isgur:1978xj}%
  \BibitemOpen
  \bibfield  {author} {\bibinfo {author} {\bibfnamefont {N.}~\bibnamefont
  {Isgur}}\ and\ \bibinfo {author} {\bibfnamefont {G.}~\bibnamefont {Karl}},\
  }\href {\doibase 10.1103/PhysRevD.18.4187} {\bibfield  {journal} {\bibinfo
  {journal} {Phys. Rev. D}\ }\textbf {\bibinfo {volume} {18}},\ \bibinfo
  {pages} {4187} (\bibinfo {year} {1978})}\BibitemShut {NoStop}%
\bibitem [{\citenamefont {Isgur}\ and\ \citenamefont
  {Karl}(1979{\natexlab{a}})}]{Isgur:1978wd}%
  \BibitemOpen
  \bibfield  {author} {\bibinfo {author} {\bibfnamefont {N.}~\bibnamefont
  {Isgur}}\ and\ \bibinfo {author} {\bibfnamefont {G.}~\bibnamefont {Karl}},\
  }\href {\doibase 10.1103/PhysRevD.19.2653} {\bibfield  {journal} {\bibinfo
  {journal} {Phys. Rev. D}\ }\textbf {\bibinfo {volume} {19}},\ \bibinfo
  {pages} {2653} (\bibinfo {year} {1979}{\natexlab{a}})},\ \bibinfo {note}
  {[Erratum: Phys.Rev.D 23, 817 (1981)]}\BibitemShut {NoStop}%
\bibitem [{\citenamefont {Isgur}\ and\ \citenamefont
  {Karl}(1979{\natexlab{b}})}]{Isgur:1979be}%
  \BibitemOpen
  \bibfield  {author} {\bibinfo {author} {\bibfnamefont {N.}~\bibnamefont
  {Isgur}}\ and\ \bibinfo {author} {\bibfnamefont {G.}~\bibnamefont {Karl}},\
  }\href {\doibase 10.1103/PhysRevD.20.1191} {\bibfield  {journal} {\bibinfo
  {journal} {Phys. Rev. D}\ }\textbf {\bibinfo {volume} {20}},\ \bibinfo
  {pages} {1191} (\bibinfo {year} {1979}{\natexlab{b}})}\BibitemShut {NoStop}%
\bibitem [{\citenamefont {Capstick}\ and\ \citenamefont
  {Isgur}(1985)}]{Capstick:1986bm}%
  \BibitemOpen
  \bibfield  {author} {\bibinfo {author} {\bibfnamefont {S.}~\bibnamefont
  {Capstick}}\ and\ \bibinfo {author} {\bibfnamefont {N.}~\bibnamefont
  {Isgur}},\ }\href {\doibase 10.1103/PhysRevD.34.2809} {\bibfield  {journal}
  {\bibinfo  {journal} {AIP Conf. Proc.}\ }\textbf {\bibinfo {volume} {132}},\
  \bibinfo {pages} {267} (\bibinfo {year} {1985})}\BibitemShut {NoStop}%
\bibitem [{\citenamefont {Kalos}\ \emph {et~al.}(1974)\citenamefont {Kalos},
  \citenamefont {Levesque},\ and\ \citenamefont {Verlet}}]{PhysRevA.9.2178}%
  \BibitemOpen
  \bibfield  {author} {\bibinfo {author} {\bibfnamefont {M.~H.}\ \bibnamefont
  {Kalos}}, \bibinfo {author} {\bibfnamefont {D.}~\bibnamefont {Levesque}}, \
  and\ \bibinfo {author} {\bibfnamefont {L.}~\bibnamefont {Verlet}},\ }\href
  {\doibase 10.1103/PhysRevA.9.2178} {\bibfield  {journal} {\bibinfo  {journal}
  {Phys. Rev. A}\ }\textbf {\bibinfo {volume} {9}},\ \bibinfo {pages} {2178}
  (\bibinfo {year} {1974})}\BibitemShut {NoStop}%
\bibitem [{\citenamefont {Boronat}\ and\ \citenamefont
  {Casulleras}(1994)}]{PhysRevB.49.8920}%
  \BibitemOpen
  \bibfield  {author} {\bibinfo {author} {\bibfnamefont {J.}~\bibnamefont
  {Boronat}}\ and\ \bibinfo {author} {\bibfnamefont {J.}~\bibnamefont
  {Casulleras}},\ }\href {\doibase 10.1103/PhysRevB.49.8920} {\bibfield
  {journal} {\bibinfo  {journal} {Phys. Rev. B}\ }\textbf {\bibinfo {volume}
  {49}},\ \bibinfo {pages} {8920} (\bibinfo {year} {1994})}\BibitemShut
  {NoStop}%
\bibitem [{\citenamefont {S\'anchez-Baena}\ \emph {et~al.}(2018)\citenamefont
  {S\'anchez-Baena}, \citenamefont {Boronat},\ and\ \citenamefont
  {Mazzanti}}]{PhysRevA.98.053632}%
  \BibitemOpen
  \bibfield  {author} {\bibinfo {author} {\bibfnamefont {J.}~\bibnamefont
  {S\'anchez-Baena}}, \bibinfo {author} {\bibfnamefont {J.}~\bibnamefont
  {Boronat}}, \ and\ \bibinfo {author} {\bibfnamefont {F.}~\bibnamefont
  {Mazzanti}},\ }\href {\doibase 10.1103/PhysRevA.98.053632} {\bibfield
  {journal} {\bibinfo  {journal} {Phys. Rev. A}\ }\textbf {\bibinfo {volume}
  {98}},\ \bibinfo {pages} {053632} (\bibinfo {year} {2018})}\BibitemShut
  {NoStop}%
\bibitem [{\citenamefont {Casulleras}\ and\ \citenamefont
  {Boronat}(1995)}]{Casulleras_1995}%
  \BibitemOpen
  \bibfield  {author} {\bibinfo {author} {\bibfnamefont {J.}~\bibnamefont
  {Casulleras}}\ and\ \bibinfo {author} {\bibfnamefont {J.}~\bibnamefont
  {Boronat}},\ }\href {\doibase 10.1103/physrevb.52.3654} {\bibfield  {journal}
  {\bibinfo  {journal} {Physical Review B}\ }\textbf {\bibinfo {volume} {52}},\
  \bibinfo {pages} {3654–3661} (\bibinfo {year} {1995})}\BibitemShut
  {NoStop}%
\bibitem [{\citenamefont {Krotscheck}\ and\ \citenamefont
  {Navarro}(2002)}]{Boronat:2002book}%
  \BibitemOpen
  \bibfield  {author} {\bibinfo {author} {\bibfnamefont {E.}~\bibnamefont
  {Krotscheck}}\ and\ \bibinfo {author} {\bibfnamefont {J.}~\bibnamefont
  {Navarro}},\ }\href {https://www.worldscientific.com/doi/abs/10.1142/4718}
  {\emph {\bibinfo {title} {Microscopic Approaches to Quantum Liquids in
  Confined Geometries}}}\ (\bibinfo  {publisher} {World Scientific},\ \bibinfo
  {year} {2002})\ \Eprint
  {http://arxiv.org/abs/https://www.worldscientific.com/doi/pdf/10.1142/4718}
  {https://www.worldscientific.com/doi/pdf/10.1142/4718} \BibitemShut {NoStop}%
\bibitem [{\citenamefont {Pineda}\ and\ \citenamefont
  {Segovia}(2013)}]{Pineda:2013lta}%
  \BibitemOpen
  \bibfield  {author} {\bibinfo {author} {\bibfnamefont {A.}~\bibnamefont
  {Pineda}}\ and\ \bibinfo {author} {\bibfnamefont {J.}~\bibnamefont
  {Segovia}},\ }\href {\doibase 10.1103/PhysRevD.87.074024} {\bibfield
  {journal} {\bibinfo  {journal} {Phys. Rev. D}\ }\textbf {\bibinfo {volume}
  {87}},\ \bibinfo {pages} {074024} (\bibinfo {year} {2013})},\ \Eprint
  {http://arxiv.org/abs/1302.3528} {arXiv:1302.3528 [hep-ph]} \BibitemShut
  {NoStop}%
\bibitem [{\citenamefont {Peset}\ \emph {et~al.}(2016)\citenamefont {Peset},
  \citenamefont {Pineda},\ and\ \citenamefont {Stahlhofen}}]{Peset:2015vvi}%
  \BibitemOpen
  \bibfield  {author} {\bibinfo {author} {\bibfnamefont {C.}~\bibnamefont
  {Peset}}, \bibinfo {author} {\bibfnamefont {A.}~\bibnamefont {Pineda}}, \
  and\ \bibinfo {author} {\bibfnamefont {M.}~\bibnamefont {Stahlhofen}},\
  }\href {\doibase 10.1007/JHEP05(2016)017} {\bibfield  {journal} {\bibinfo
  {journal} {JHEP}\ }\textbf {\bibinfo {volume} {05}},\ \bibinfo {pages} {017}
  (\bibinfo {year} {2016})},\ \Eprint {http://arxiv.org/abs/1511.08210}
  {arXiv:1511.08210 [hep-ph]} \BibitemShut {NoStop}%
\bibitem [{\citenamefont {Yang}\ \emph
  {et~al.}(2020{\natexlab{b}})\citenamefont {Yang}, \citenamefont {Ping},
  \citenamefont {Ortega},\ and\ \citenamefont {Segovia}}]{Yang:2019lsg}%
  \BibitemOpen
  \bibfield  {author} {\bibinfo {author} {\bibfnamefont {G.}~\bibnamefont
  {Yang}}, \bibinfo {author} {\bibfnamefont {J.}~\bibnamefont {Ping}}, \bibinfo
  {author} {\bibfnamefont {P.~G.}\ \bibnamefont {Ortega}}, \ and\ \bibinfo
  {author} {\bibfnamefont {J.}~\bibnamefont {Segovia}},\ }\href {\doibase
  10.1088/1674-1137/44/2/023102} {\bibfield  {journal} {\bibinfo  {journal}
  {Chin. Phys. C}\ }\textbf {\bibinfo {volume} {44}},\ \bibinfo {pages}
  {023102} (\bibinfo {year} {2020}{\natexlab{b}})},\ \Eprint
  {http://arxiv.org/abs/1904.10166} {arXiv:1904.10166 [hep-ph]} \BibitemShut
  {NoStop}%
\bibitem [{\citenamefont {Aaij}\ \emph {et~al.}(2015)\citenamefont {Aaij} \emph
  {et~al.}}]{Aaij:2015tga}%
  \BibitemOpen
  \bibfield  {author} {\bibinfo {author} {\bibfnamefont {R.}~\bibnamefont
  {Aaij}} \emph {et~al.} (\bibinfo {collaboration} {LHCb}),\ }\href {\doibase
  10.1103/PhysRevLett.115.072001} {\bibfield  {journal} {\bibinfo  {journal}
  {Phys. Rev. Lett.}\ }\textbf {\bibinfo {volume} {115}},\ \bibinfo {pages}
  {072001} (\bibinfo {year} {2015})},\ \Eprint
  {http://arxiv.org/abs/1507.03414} {arXiv:1507.03414 [hep-ex]} \BibitemShut
  {NoStop}%
\bibitem [{\citenamefont {Aaij}\ \emph {et~al.}(2019)\citenamefont {Aaij} \emph
  {et~al.}}]{lhcb:2019pc}%
  \BibitemOpen
  \bibfield  {author} {\bibinfo {author} {\bibfnamefont {R.}~\bibnamefont
  {Aaij}} \emph {et~al.} (\bibinfo {collaboration} {LHCb}),\ }\href {\doibase
  10.1103/PhysRevLett.122.222001} {\bibfield  {journal} {\bibinfo  {journal}
  {Phys. Rev. Lett.}\ }\textbf {\bibinfo {volume} {122}},\ \bibinfo {pages}
  {222001} (\bibinfo {year} {2019})},\ \Eprint
  {http://arxiv.org/abs/1904.03947} {arXiv:1904.03947 [hep-ex]} \BibitemShut
  {NoStop}%
\bibitem [{\citenamefont {Clement}(2017)}]{Clement:2016vnl}%
  \BibitemOpen
  \bibfield  {author} {\bibinfo {author} {\bibfnamefont {H.}~\bibnamefont
  {Clement}},\ }\href {\doibase 10.1016/j.ppnp.2016.12.004} {\bibfield
  {journal} {\bibinfo  {journal} {Prog. Part. Nucl. Phys.}\ }\textbf {\bibinfo
  {volume} {93}},\ \bibinfo {pages} {195} (\bibinfo {year} {2017})},\ \Eprint
  {http://arxiv.org/abs/1610.05591} {arXiv:1610.05591 [nucl-ex]} \BibitemShut
  {NoStop}%
\bibitem [{\citenamefont {Iwasaki}(1975)}]{Iwasaki:1975pv}%
  \BibitemOpen
  \bibfield  {author} {\bibinfo {author} {\bibfnamefont {Y.}~\bibnamefont
  {Iwasaki}},\ }\href {\doibase 10.1143/PTP.54.492} {\bibfield  {journal}
  {\bibinfo  {journal} {Prog. Theor. Phys.}\ }\textbf {\bibinfo {volume}
  {54}},\ \bibinfo {pages} {492} (\bibinfo {year} {1975})}\BibitemShut
  {NoStop}%
\end{thebibliography}%

\end{document}